\documentclass[traditabstract]{aa} 
\usepackage[pdftex]{graphicx}
\usepackage{breqn}
\usepackage[flushleft]{threeparttable}

\def\la{\lower.5ex\hbox{$\; \buildrel < \over \sim \;$}}
\def\ga{\lower.5ex\hbox{$\; \buildrel > \over \sim \;$}}

\def\sigmaturb{\sigma_{\rm disp}}
\def\vdisp{\sigma_{\rm disp}^\star}
\def\vtherm{c_{\rm s}}
\def\vturbb{v_{\rm turb,3D}}
\def\vturb{v_{\rm turb}}
\def\ldriv{l_{\rm driv}}
\def\sigmastar{\dot{\Sigma}_\star}
\def\mdot{\dot{\rm M}}
\def\phiv{\phi_{\rm{v}}}
\def\rhosfr{\dot{\rho}_\star}
\def\tauffcl{t_{\rm ff,cl}}
\def\tauturbcl{t_{\rm turb,cl}}

\begin{document}

   \title{Deciphering the radio star formation correlation on kpc scales}
   \subtitle{II. The integrated infrared - radio continuum and star formation - radio continuum correlations}

   \author{B.~Vollmer\inst{1}, M.~Soida\inst{2} \and J.~Dallant\inst{1}}

   \institute{Universit\'e de Strasbourg, CNRS, Observatoire astronomique de Strasbourg, UMR 7550, F-67000 Strasbourg, France \and
          Astronomical Observatory, Jagiellonian University, Krak\'ow, Poland}

   \date{Received ; accepted }


  \abstract
{
Given the multiple energy loss mechanisms of cosmic ray electrons in galaxies, the tightness of the infrared - radio continuum correlation
is surprising.  Since the radio continuum emission at GHz frequencies is optically thin, this offers the opportunity to
obtain unbiased star formation rates from radio continuum flux density measurements.
The calorimeter theory can naturally explain the tightness of the FIR - radio correlation but makes predictions, which do
not agree with observations. Non-calorimeter models often have to involve a conspiracy to maintain the tightness of the 
FIR - radio correlation. We extended the analytical model of galactic disks of Vollmer et al. (2017) by including a simplified prescription for
the synchrotron emissivity. The galactic gas disks of local spiral galaxies, low-z starburst galaxies, high-z 
main sequence starforming, and high-z starburst galaxies are treated as turbulent clumpy accretion disks. 
The magnetic field strength is determined by the equipartition between the turbulent kinetic and the magnetic energy densities.
Our fiducial model, which neither includes galactic winds nor CR electron secondaries, reproduces the
observed radio continuum SEDs of most ($\sim 70$\,\%) of the galaxies.
Except for the local spiral galaxies, fast galactic winds can potentially make the conflicting models agree with observations.
The observed IR - radio correlations are reproduced by the model within $2\sigma$ of the joint uncertainty of model and data for all datasets.
The model agrees with the observed SFR - radio correlations within $\sim 4\sigma$.
Energy equipartition between the CR particles and the magnetic field only approximately holds in our models of main sequence starforming galaxies.
If a CR electron calorimeter is assumed, the slope of the IR - radio correlation flattens significantly.
Inverse Compton (IC) losses are not dominant in the starburst galaxies because in these galaxies not only the gas density but
also the turbulent velocity dispersion is higher than in normally starforming galaxies. 
Equipartition between the turbulent kinetic and magnetic field energy densities then leads to very high magnetic field strengths
and very short synchrotron timescales. The exponents of our model SFR - radio correlations at $150$~MHz and $1.4$~GHz 
are very close to one.
}


   \authorrunning{Vollmer et al.}
   \titlerunning{The integrated infrared - radio continuum correlation}

   \maketitle
%

\section{Introduction\label{sec:introduction}}

One of the tightest correlations in astronomy is the relation between the integrated radio continuum (synchrotron) and the far-infrared (FIR)
emission (Helou et al. 1985, Condon 1992, Mauch \& Sadler  2007, Yun et al. 2001, Bell 2003, Farrah et al. 2003, Appleton et al. 2004, 
Kovacs et al. 2006, Murphy et al. 2009, Sargent et al. 2010, Jarvis et al. 2010, Basu et al. 2015, Magnelli et al. 2015, Delhaize et al. 2017,
Read et al. 2018, Thomson et al. 2019, Algera et al. 2020, Molnar et al. 2021, Delvecchio et al. 2021). 
It holds over five orders of magnitude in various types of galaxies, including starbursts.
The common interpretation of the correlation is that both emission types are proportional to star formation:
the radio emission via (i) the cosmic ray (CR) source term caused by supernova explosions and the turbulent amplification of
the small-scale magnetic field (small-scale dynamo e.g., Schleicher \& Beck 2013) and (ii) the far-infrared emission via
the dust heating, mainly through massive stars. 

In local galaxies the correlation between the star formation rate and the radio continuum emission is as tight as the correlation involving the
far-infrared emission (Bell 2003, Murphy et al. 2011, Heesen et al. 2014, Boselli et al. 2015, Li et al. 2016, Brown et al. 2017, G\"urkan et al. 2018, 
Wang et al. 2019, Heesen et al. 2019, Smith et al. 2021). 
Since the radio continuum emission at GHz frequencies is optically thin, this offers the opportunity to
obtain unbiased star formation rates from radio continuum flux density measurements (e.g., Davies et al. 2017). A possible contribution of an
active galactic nucleus (AGN) has to be recognized and subtracted, if possible. Whereas the exponents, close to unity, and normalizations of the
FIR and star formation - radio continuum correlations are well studied, the detailed physics that lead to these relations are
only broadly understood. 

Radio continuum emission observed at frequencies below a few GHz is usually dominated by synchrotron emission, which
is emitted by CR electrons with relativistic velocities that spiral around galactic magnetic fields.
The magnetic field can be regular, meaning structured on large-scales (kpc), or tangled on small-scales via turbulent motions.
The turbulent magnetic field has an isotropic and an anisotropic component.
The total magnetic field $B$ is the quadratic sum of the ordered and turbulent magnetic
field components. The ordered magnetic field includes the large-scale regular magnetic field, and the anisotropic small-scale magnetic field.
Anisotropic small-scale magnetic fields can be produced by a large-scale gas and associated magnetic field compression.

The energy loss caused by synchrotron emission depends on the magnetic field strength and the electron energy or Lorentz factor $\gamma$:
\begin{equation}
\frac{d\,E}{d\,t}=b(E)=3.1 \times 10^{-18} {\rm GeV\,sec^{-1}} \times \big(\frac{E}{\rm GeV} \big)^2 \big( \frac{B}{10~\mu{\rm G}} \big)^2\ ,
\end{equation}
where $B=10~\mu$G is the typical magnetic field strength in local spiral galaxies (e.g., Beck 2015).
A CR electron with energy $E$ emits most of its energy at a critical frequency $\nu_{\rm c}$ where
\begin{equation}
\nu_{\rm c}=1.6 \times 10^{-1} \big(\frac{B}{10~\mu {\rm G}} \big) \big(\frac{E}{\rm GeV}  \big)^2 \sin \alpha ~{\rm GHz}\ ,
\end{equation}
where $\alpha$ is the pitch angle of the particle’s path with respect to the magnetic field direction.
The timescale for synchrotron emission is 
\begin{equation}
\label{eq:synch}
t_{\rm sync}=\frac{E}{b(E)} \simeq 4.5 \times 10^7 \big( \frac{B}{10~\mu{\rm G}} \big)^{-3/2} \big(\frac{\nu}{\rm GHz}\big)^{-1/2}~{\rm yr}\ .
\end{equation}
For the calculation of the mean energy of CR electrons we use the mean frequency calculated via the expectation value of $x=\nu/\nu_{\rm c}$
\begin{equation}
\frac{\nu_{\rm s}}{\nu_{\rm c}}=\int x\,G(x)\,E^{-2.3} {\rm d}E/(\int G(x)\,E^{-2.3} {\rm d}E)=0.85
\end{equation}
with the synchrotron kernel $G(x)$ given by Eq.D3 of Aharonian et al. (2010). This yields
\begin{equation}
\label{eq:meanfreq}
\nu_{\rm s}=1.3 \times 10^{-1} \big(\frac{B}{10~\mu {\rm G}} \big) \big(\frac{E}{\rm GeV}  \big)^2 ~{\rm GHz}\ .
\end{equation}

Cosmic ray particles are mainly produced in supernova shocks via Fermi acceleration. 
However, the relativistic electrons do not stay at the location of their creation. They propagate either via diffusion,
or by streaming with the Alfv\'en velocity. In addition, CR electrons can be transported into the halo
by advection, meaning a galactic wind. During the transport process, the CR electron loses energy via synchrotron emission.
The associated diffusion--advection--loss equation for the CR electron density $n$ reads as
\begin{dmath}
\label{eq:diffeq}
\frac{\partial n}{\partial t} = D {\nabla}^2 n + \frac{\partial}{\partial E} \big( b(E) n(E) \big) - 
(\vec{u}+\vec{v}) \nabla n + \frac{p}{3} \frac{\partial n}{\partial p} \nabla \vec{u} + Q(E) - \frac{n}{t_{\rm loss}}\ ,
\end{dmath}
where $D$ is the diffusion coefficient, $E$ the CR electron energy, $\vec{u}$ the advective flow velocity, 
$\vec{v}$ the streaming velocity, $p$ the CR electron pressure,
$Q$ the source term, and $b(E)$ the energy loss rate.
The first part of the right-hand side is the diffusion term, followed by the synchrotron loss, advection, streaming, adiabatic energy gain or loss,
and the source terms. The advection and adiabatic terms are only important for the transport in a vertical direction. 
Energy can be lost via inverse Compton radiation, bremsstrahlung, pion or ionization energy loss, and most importantly synchrotron emission
(e.g., Murphy 2009, Lacki et al. 2010).

V\"olk (1989) developed a calorimeter theory, assuming that galaxies CR electrons lose their energy
before escaping galaxies, with most of the energy radiated as synchrotron radio emission. In addition, galaxies are assumed to
be optically thick to UV light from massive young stars, which is absorbed by dust and re-radiated in the FIR.
The calorimeter theory can naturally explain the tightness of the FIR - radio correlation.
In the Milky Way, however, the inferred diffusive escape time is shorter than the typical estimated synchrotron cooling time 
casting doubt on the validity of the electron calorimeter assumption. In addition, calorimeter theory predicts a spectral index
$\alpha \sim -1$ ($S_{\nu} \propto \nu^{\alpha}$), which is in conflict with the observed spectral indices of $\alpha \sim -0.7$ to $-0.8$  
for normal galaxies (Vollmer et al. 2005, 2010).

On the other hand, non-calorimeter models (Helou \& Bicay 1993, Niklas \& Beck 1997, Murphy 2009, Lacki et al. 2010)
often have to invoke ``conspiracy'' to maintain the tightness of the FIR - radio correlation. 
Murphy (2009) stated that to keep a fixed ratio between the FIR and non-thermal radio continuum emission of
a normal star-forming galaxy, whose CR electrons typically lose most of their energy to synchrotron
radiation and inverse Compton scattering, requires a nearly constant ratio between galaxy magnetic field
and radiation field energy densities. Lacki et al. (2010) found that the correlation is caused by a combination of the efficient 
cooling of CR electrons (calorimetry) in starbursts and a conspiracy of several factors. For lower surface density galaxies, the 
decreasing radio emission caused by CR escape is balanced by the decreasing FIR emission caused by the low effective UV
dust opacity. In starbursts, bremsstrahlung, ionization, and inverse Compton cooling decrease the radio emission,
but they are countered by secondary electrons/positrons and the dependence of synchrotron frequency on energy,
both of which increase the radio emission. Lacki et al. (2010) predicted spectral exponents $\alpha$, which were significantly steeper
than those derived from observations of normal galaxies.

Vollmer et al. (2017) developed an analytical 1D model of turbulent clumpy star-forming galactic disks
and applied it to well defined samples of local spiral galaxies, ultraluminous infrared galaxies (ULIRGs),
high-z star-forming galaxies, and high-z starburst galaxies. The model has a large-scale part (gas surface density, volume density, disk height, 
turbulent driving length scale, velocity dispersion, gas viscosity, volume filling factor, and molecular fraction), which 
is governed by vertical pressure equilibrium, the Toomre $Q$ parameter, conservation of the turbulent energy flux, a relation between the
gas viscosity and the gas surface density, a star-formation recipe, and a simple closed-box model for the gas metallicity,
and a small-scale part (non-self-gravitating and self-gravitating gas clouds) governed by turbulent scaling relations.
The model yields radial profiles of molecular line and infrared emission. The global metallicities, total infrared luminosities and dust 
spectral energy distributions (SEDs), dust temperature, CO luminosities and spectral line energy density distributions (SLEDs) 
of the four galaxy samples could be reproduced by the model. 

In this work we added a recipe for the non-thermal radio continuum emission of the galactic disks and compare the results to available radio 
and infrared observations of starforming galaxies at various redshifts. The recipe includes (i) energy equipartition between the
turbulent kinetic energy of the gas and the magnetic field and (ii) CR energy loss terms as described in Murphy (2009) and Lacki et al. (2010).
The Lacki et al. (2010) model assumes a gas surface density $\Sigma_{\rm g}$ and scale height. 
Moreover, the star formation rate per area is given by a Schmidt-Kennicutt law
$\dot\Sigma_* \propto \Sigma_{\rm g}^{1.4}$ and the strength of the magnetic field is linked to the gas surface density via a power law.
The advantage of our analytical model is that it gives access to radial profiles of the gas and star formation volume densities and 
to the gas velocity dispersion. With these quantities the CR electron source term and cooling times can be evaluated. 
The infrared and radio luminosities of a given galaxy are directly calculated by the model.

\section{The analytical model \label{sec:model}}

The theory of clumpy gas disks (Vollmer \& Beckert 2003, Vollmer \& Leroy 2011, Vollmer et al. 2017) provides the large-scale and small-scale properties of galactic gas
disks. Large-scale properties considered are the gas surface density, density, disk height, turbulent driving length scale, velocity dispersion, 
gas viscosity, volume filling factor, and molecular fraction. Small-scale properties are the mass, size, density,
turbulent, free-fall, and molecular formation timescales of the most massive self-gravitating gas clouds. These quantities depend on the stellar 
surface density, the angular velocity, the disk radius $R$, and three additional parameters, which are the
Toomre parameter $Q$ of the gas, the mass accretion rate $\dot{M}$, and the ratio $\delta$ between the driving length scale of turbulence and the
cloud size. The large-scale part of the model disk is governed by vertical pressure equilibrium, the Toomre $Q$ parameter,
conservation of the turbulent energy flux via $\dot{M}$, 
a relation between the gas viscosity and the gas surface density, a star-formation recipe, 
and a simple closed-box model for the gas metallicity. We used the modified version of the large-scale model presented in Vollmer et al. (2021),
which is presented in Appendix~\ref{sec:model1},
with a constant relating supernova energy input to star formation $\xi=9.2 \times 10^{-6}$~pc$^2$yr$^{-2}$. 
This modified version treats the turbulent scaling relations in a physically more consistent way than the old version.
The model equations are equivalent 
to those of Vollmer et al. (2017) with $\xi=4.6 \times 10^{-6}$~pc$^2$yr$^{-2}$ (see Appendix~\ref{sec:model1}). 
The factor of two between the constants
is within the uncertainties of the underlying observations, the Galactic star formation rate, supernova explosion rate, and the 
fraction of supernova energy injected into ISM turbulence. We verified that both descriptions of the large-scale lead to 
comparable results within the uncertainties. We used a constant $Q$ parameter for all galaxies except for
NGC~628, NGC~3198, NGC~3351, NGC~5055, NGC~5194, and NGC~7331 where we assumed the radial profiles of Vollmer \& Leroy (2011), 
which increase towards the galaxy centers.

The small-scale part is divided into two parts according to gas density: non-self-gravitating and self-gravitating gas clouds. 
The mass fraction at a given density is determined by a density probability distribution involving the overdensity and the Mach number. 
Both density regimes are governed by different observed scale relations. The dense gas clouds are mechanically heated by turbulence. In addition, 
they are heated by CRs. The gas temperature of the molecular gas is calculated through the equilibrium between gas heating and cooling 
via molecular line emission (CO, H$_2$, H$_2$O). No photodissociation regions were included in the model.

For the calculations of the model IR emission we refer to Sects.~2.1.3 and 8 of Vollmer et al. (2017).
In short, the dust is heated by the interstellar UV and optical radiation field. 
We assume that the UV radiation is emitted by young massive stars whose surface density is proportional 
to the star-formation rate per unit area $\dot{\Sigma_*}$. The optical light stems from the majority of disks stars.
The contributions of each component was chosen such that the normalizations of $\dot{\Sigma_*}$ and stellar mass
surface density $\Sigma_*$ are set by observations of the ISRF at the solar radius.
In addition, the local Galactic star-formation rate is assumed to be $\dot{\Sigma_*}=6.7 \times 10^{-10}~{\rm M}_{\odot}{\rm pc}^{-2}{\rm yr}^{-1}$.
The model does not assume an explicit initial mass function (IMF).
In the presence of dust and gas, the interstellar radiation field is attenuated.
For this attenuation we adopted the mean extinction of a sphere of constant density.
We assumed a dust mass absorption coefficient of the following form:
\begin{equation}
\label{eq:kappa}
\kappa(\lambda)=\kappa_0\,(\lambda_0/\lambda)^{\beta}\ ,
\end{equation}
with $\lambda_0=250~\mu$m, $\kappa_0=0.48~{\rm m}^2{\rm kg}^{-1}$ (Dale et al. 2012), and a gas-to-dust ratio of $M_{\rm gas}/M_{\rm dust}=\frac{Z}{Z_{\odot}} \times 100$ 
(including helium; R{\'e}my-Ruyer et al. 2014). We allowed for energy transfer between dust and gas due to collisions
The dust temperature of a gas cloud of given density and size illuminated by the local mean radiation field is calculated by solving
the equilibrium between radiative heating and cooling and the heat transfer between gas and dust. 
The IR emission of the diffuse warm neutral medium is taken into account.

The model inputs are the rotation curve and the radial profiles of the stellar mass surface density and the Toomre $Q$ parameter.
The constant mass accretion rate $\dot{M}$ is determined by the integrated star formation rate $\dot{M}_*$ of the galaxy:
for a given Toomre $Q$ parameter a higher star formation rate leads to a higher turbulent velocity dispersion,
which in turn leads to a higher turbulent viscosity and thus a higher $\dot{M}$ (see Appendix~ of Vollmer \& Leroy 2011).
The model results are radial profiles of the large- and small-scale properties of the galactic disk, the molecular line emission, and the infrared 
emission at multiple wavelengths.

The free-free radio continuum emission from electrons in H{\sc ii} regions around ionizing young, massive stars is expected to be closely 
connected to the warm dust emission that is heated by the same stars (e.g., Condon 1992). Although the cosmic ray electrons
responsible for synchrotron emission also originate from supernova remnants located within star-formation regions, the 
synchrotron-IR correlation is not as tight as the 
free-free-IR correlation locally, as a result of the propagation of CRes from their places of birth (e.g., Tabatabaei et al. 2013).
The smallest scale on which the synchrotron-IR correlation holds is approximately the propagation length of cosmic ray
electrons, which is $\ga 0.5$~kpc at $\nu=5$~GHz and $\ga 1$~kpc at $\nu=1.4$~GHz in massive local spiral galaxies 
(e.g., Tabatabaei et al. 2013, Vollmer et al. 2020). Therefore, only the large-scale part detailed in Appendix~\ref{sec:model1} was used
for the calculations of the model radio continuum emission.

We assumed a stationary CR electron density distribution ($\partial n/\partial t$=0; Eq.~\ref{eq:diffeq}). 
The CR electrons are transported into the halo through diffusion or advection
where they lose their energy via adiabatic losses or where the energy loss through synchrotron emission is so small that the
emitted radio continuum emission cannot be detected. Furthermore, we assumed that the source term of CR electrons is
proportional to the star formation rate per unit volume $\dot{\rho}_*$. For the energy distribution of the cosmic
ray electrons the standard assumption is a power law with index $q$, which leads to a power law of the radio continuum spectrum
with index $-(q-1)/2$ (e.g., Beck).

Under these assumptions the synchrotron emissivity is given by the density per unit energy interval of the primary cosmic ray 
electrons, where $E$ is the energy and $n_0 \propto \dot{\rho}_* t_{\rm eff}$
\begin{equation}
\label{eq:emissivity}
\epsilon_\nu {\rm d}\nu \propto \dot{\rho}_* t_{\rm eff} E^{-q} \frac{E}{t_{\rm sync}} {\rm d}E\ .
\end{equation}
This is equivalent to the approach of Werhahn et al. (2021b) who set the CR proton luminosity proportional to the
star formation rate and the primary CR electron luminosity proportional to the proton luminosity if
fixed shapes of the primary electron and proton energy spectra are assumed.

The effective lifetime of synchrotron emitting CR electrons $t_{\rm eff}$ is given by
\begin{equation}
\frac{1}{t_{\rm eff}}=\frac{1}{t_{\rm sync}}+\frac{1}{t_{\rm diff}}+\frac{1}{t_{\rm wind}}+\frac{1}{t_{\rm brems}}+\frac{1}{t_{\rm IC}}+\frac{1}{t_{\rm ion}}\ .
\end{equation}
For the characteristic timescales we follow the prescriptions of Lacki et al. (2010).
The diffusion timescale based on observations of beryllium isotope ratios at the Solar Circle (Connell 1998, Webber et al. 2003) is
\begin{equation}
\label{eq:tdiffe}
t_{\rm diff}=26/\sqrt{E/3 {\rm GeV}}~{\rm Myr},
\end{equation}
where the mean energy $E$ is calculated via the mean synchrotron frequency of Eq.~\ref{eq:meanfreq}.
The CR escape time through advection by galactic winds is 
\begin{equation}
\label{eq:wind}
t_{\rm wind }=1 \,\frac{H}{\sqrt{2} v_{\rm rot}}~{\rm Myr},
\end{equation}
where $H$ is the disk height in pc and $v_{\rm rot}$ the rotation velocity in km\,s$^{-1}$.
This timescale is drastically increased if the star formation surface density is lower than the Heckman limit of 
$\dot{\Sigma}_*=10^{-7}$~M$_{\odot}$yr$^{-1}$pc$^{-2}$ (Heckman 2002).
The characteristic time for bremsstrahlung is
\begin{equation} 
t_{\rm brems}=37\, (\frac{n}{{\rm cm}^{-3}})^{-1}~{\rm Myr},
\end{equation}
that for inverse Compton energy losses is
\begin{equation}
t_{\rm IC}=180\, (\frac{B}{10~\mu{\rm G}})^{\frac{1}{2}}(\nu_{\rm GHz})^{-\frac{1}{2}}(\frac{U}{10^{-12}~{\rm erg\,cm}^{-3}})^{-1}~{\rm Myr}
\end{equation}
where $U$ is the interstellar radiation field. The ionization energy loss timescale is
\begin{equation}
t_{\rm ion}=210\, (\frac{B}{10~\mu{\rm G}})^{-\frac{1}{2}}(\nu_{\rm GHz})^{\frac{1}{2}}(\frac{n}{{\rm cm}^{-3}})^{-1}\ .
\end{equation}
The magnetic field strength $B$ is calculated under the assumption of energy equipartition between the turbulent kinetic energy of the gas
and the magnetic field:
\begin{equation}
\label{eq:Bmag2}
\frac{B^2}{8 \pi}=\frac{1}{2}\rho v_{\rm turb}^2\ ,
\end{equation}
where $\rho$ is the total midplane density of the gas and $v_{\rm turb}$ its turbulent velocity dispersion.

Secondary CR electrons can be produced via collisions between the ISM and CR protons.
The proton lifetime to pion losses (Mannheim \& Schlickeiser 1994) is
\begin{equation}
t_{\pi}=50\,(\frac{n}{{\rm cm}^{-3}})^{-1}~{\rm Myr}.
\end{equation}
The effective lifetime of CR protons is given by
\begin{equation}
\frac{1}{t_{\rm eff,p}}=\frac{1}{t_{\rm wind}}+\frac{1}{t_{\rm diff,p}}\ ,
\end{equation}
where the proton diffusion timescale is $\sqrt{16}$ times shorter than the CR electron diffusion timescale (Appendix B3 of Werhahn et al. 2021a).
The CR electron secondary fraction is given by
\begin{equation}
\label{eq:secondaries}
\eta_{\rm sec}=\frac{1}{2} (1+\frac{t_{\pi}}{t_{\rm eff,p}})
\end{equation}
(Werhahn et al. 2021a).
In models that include CR electron secondaries the CR electron density is multiplied by $(1+\eta_{\rm sec})$.

With $\nu=C B E^2$ the synchrotron emissivity of Eq.~\ref{eq:emissivity} becomes
\begin{equation}
\label{eq:emfinal}
\epsilon_{\nu}= \xi \dot{\rho}_* \frac{t_{\rm eff}(\nu)}{t_{\rm sync}(\nu)} B^{\frac{q}{2}-1} \nu^{-\frac{q}{2}}\ .
\end{equation}
The constant is $C=e/(2 \pi m_{\rm e}^2 c^2)$.
With the cosmic ray electron density $n_0$ and Eq.~\ref{eq:synch} the classical expression 
$\epsilon_{\nu} \propto n_0 B^{(q+1)/2} \nu^{(1-q)/2}$ is recovered.
The factor $\xi$ was chosen such that the radio-IR correlations measured by Yun et al. (2001) and Molnar et al. (2021) are 
reproduced within $2 \sigma$ (Fig.~\ref{fig:galaxies_FRC_vrotDifferentForPhibbs_nosecnowind_3}).
Since it was not possible to exactly match the correlation offsets of Yun et al. (2001) and Molnar et al. (2021) at the
same time, our choice represents the best compromise (last column of Table~\ref{tab:corrs}).
We assume $q=2$ as our fiducial model but also investigated the case of $q=2.3$.
The gas density $\rho$, turbulent gas velocity dispersion $v_{\rm turb}$, and interstellar radiation field $U$ are 
directly taken from the analytical model of Vollmer et al. (2017).

Following Tsang (2007) and Beck \& Krause (2005) the synchrotron emissivity is given by 
\begin{equation}
\label{eq:emtsang}
\epsilon_{\nu=}=a(s) \alpha_{\rm f} n_0 h \nu_{\rm L} \big(\frac{\nu}{\nu_{\rm L}}\big)^{-(q-1)/2}\ ,
\end{equation}
with $a(s)=3^{q/2}/(4 \pi(q+1))\,\Gamma((3s+19)/12)\,\Gamma((3s-1)/12)$, $\alpha_{\rm f}$ is the fine structure constant,
$h$ the Planck constant, and $\nu_{\rm L}$ the Larmor frequency.
Furthermore, the number density of relativistic electrons in the interval of Lorentz factor $\gamma$ to $\gamma + {\rm d}\gamma$
is $n_0 \gamma^{-q} {\rm d}\gamma$. From the combination of Eq.~\ref{eq:emfinal} and Eq.~\ref{eq:emtsang}
we calculated the total number density of CR electrons $n_{\rm CRe}=\int_{\gamma_1}^{\gamma_2} n_0 \gamma^{-q} {\rm d}\gamma$.
The integration limits correspond to CR electron energies of $E_1=1$~GeV and $E_2=100$~GeV. In addition, we used $q=2$.

The synchrotron luminosity was calculated via 
\begin{equation}
\label{eq:lnu8}
L_{\nu}=8 \pi^2 \int \epsilon_{\nu} \frac{(1-\exp(-\tau))}{\tau}\,H\,R\,{\rm d}R\ ,
\end{equation}
where $\tau=\tau_{\rm ff}+\tau_{\rm sync}$ is the optical depth caused by free-free and synchrotron self-absorption.
We also calculated the synchrotron luminosity using the thickness of the thin starforming disk $l_{\rm driv}$ instead of the
height of the gas disk $H$. It turned out that in this case we had to increase the normalization $\xi$ by $0.1$~dex to reproduce the
radio-IR correlations measured by Yun et al. (2001) and Molnar et al. (2021). Moreover, the model slopes of these
correlations increased by $0.1$ (e.g., from $1.0$ to $1.1$). On the other hand, the radio SEDs of NGC~628 and NGC~3184 are
better reproduced by the model using $l_{\rm driv}$ than by the model using $H$ as disk thickness. We did not use the
thickness of the gas disk ($2 \times H$) because we did not want to deviate too much from $I_{\nu} \propto \dot{\Sigma}_* = \dot{\rho}_* l_{\rm driv}$
(see Appendix~\ref{sec:model1}) in the case of an electron calorimeter (Fig.~\ref{fig:galaxies_FRC_vrotDifferentForPhibbs_calorimeter}), 
where $I_{\nu}$ is the specific intensity and $\dot{\Sigma}_*$ and $\dot{\rho}_*$ are
the star formation rate per unit area and unit volume, respectively.
Since most spiral galaxies host a thick disk of radio continuum emission (Krause et al. 2018), we think that using a vertical integration
length larger than the thickness of the thin star-forming disk is appropriate.

For the free-free absorption we used
\begin{equation} 
\tau_{\rm ff}=4.5 \times 10^{-9} (\frac{n}{{\rm cm}^{-3}})\,(\frac{l_{\rm driv}}{1~{\rm pc}})\,\nu_{\rm GHz}^{-2.1}\ ,
\end{equation}
where the height of the star-forming disk is assumed to be of the order of the turbulent driving lengthscale.
For the optical depth of synchrotron self-absorption we used the formalism described by Tsang (2007).

Optically thin thermal emission was added according to the recipe of Murphy et al. (2012)
\begin{equation}
\big(\frac{L_{\nu}^{\rm ff}}{{\rm erg}\,s^{-1} {\rm Hz}^{-1}}\big)= 2.33 \times 10^{27} \big( \frac{T_{\rm e}}{10^4\ {\rm K}} \big)^{0.45} \big( \frac{\nu}{\rm GHz} \big)^{-0.1} \big(\frac{\rm SFR}{{\rm M}_{\odot}{\rm yr}^{-1}}\big)
\end{equation}
with an electron temperature of $T_{\rm e}=8000$~K.

For the galaxies at high redshifts the IC losses off the cosmic microwave background (CMB) are taken into account
via the inverse Compton equivalent magnetic field:
\begin{equation}
U(z)=U+\frac{(3.25~\mu{\rm G} (1+z)^2)^2}{8 \pi}\ .
\end{equation}

\section{The galaxy samples \label{sec:samples}}

Most galaxies form stars at a rate, which is proportional to their stellar mass.  The tight relation between star 
formation and stellar mass is called the main sequence of star forming galaxies, in place from redshift $\sim$0 up to $\sim4$
(e.g., Speagle et al. 2014). Galaxies with much higher star-formation rates than predicted by the main sequence are
called starburst galaxies. Two of our four galaxy samples are made of main sequence galaxies (local spirals and
high-z starforming galaxies) and the other two are starburst samples (low-z starbursts/ULIRGs and high-z starbursts).
We note that high-z and dusty starburst galaxies with star-formation rates higher than $200$~M$_{\odot}$yr$^{-1}$ are usually 
called sub-mm galaxies (e.g., Bothwell et al. 2013).

The sample of local spiral galaxies (Table~\ref{tab:gleroy}) with masses in excess of $10^{10}$~M$_{\odot}$
is taken from Leroy et al. (2008).  The gas masses were derived from IRAM 30 m CO(2-1)
HERACLES (Leroy et al. 2009) and VLA H{\sc i} THINGS (Walter et al. 2008) data. The star-formation rate was derived from
Spitzer MIR and GALEX UV data (Leroy et al. 2008). The total infrared (TIR) luminosities are taken from Dale et al. (2012).

The low-z starburst/ULIRG sample (Table~\ref{tab:gulirg}) was taken from Downes \& Solomon (1998). These authors derived the spatial extent, 
rotation velocity, gas mass, and dynamical mass $M_{\rm dyn}$ for local ULIRGs from PdB interferometric CO-line observations.
The total infrared luminosities were taken from Graci\'a-Carpio et al. (2008).  The star formation 
rates were derived by applying a conversion factor of $\dot{M}_*/L_{\rm TIR}=1.7 \times 10^{-10}$~M$_{\odot}$yr$^{-1}$L$_{\odot}^{-1}$.

The high-z star-forming sample (Table~\ref{tab:gphibbs}) was taken from PHIBSS (Tacconi et al. 2013), the IRAM PdB high-z blue 
sequence CO(3-2) survey of the molecular gas properties in massive, main-sequence star-forming galaxies at $z=1$-$1.5$. For our
purpose, we only took the disk galaxies from PHIBSS based on their kinematical and structural properties (Tacconi et al. 2013).
The vast majority of the sample galaxies belong to the star-formation
main sequence. Only four out of $42$ galaxies can be qualified as starburst galaxies.
Their star-formation rates are based on the sum of the observed UV-
and IR-luminosities, or an extinction-corrected H$\alpha$ luminosity. The quoted TIR luminosities were 
derived from spectral energy distributions (for wavelengths $\leq 70$~$\mu$m) by Barro et al. (2011).
Following Vollmer et al. (2017), we assumed flat rotation curves for galactic radii $R > 0.5$~kpc 
($v_{\rm rot}=v_{\rm max} \left(1-\exp(-R/0.1~{\rm kpc})\right)$). This assumption led to an acceptable agreement
between the model and observed CO flux densities.

The high-z starburst galaxy sample (Table~\ref{tab:gbzk}) was drawn from Genzel et al. (2010). 
The total infrared luminosities are based on the $850$~$\mu$m flux densities (Genzel et al. 2010). 
For eight out of ten galaxies observationally derived TIR luminosities are available 
(Kovacs et al. 2006, Valiante et al. 2009, Chapman et al. (2010), Magnelli et al. 2012).

Many of the starburst galaxies are interaction-induced mergers. For these galaxies a disk model might be questionable.
However, since the two rotating nuclear disks of the prototypical local starburst galaxy Arp~220 are resolved by ALMA (Scoville et al. 2017) and 
these disks are sources of intense radio continuum emission (Rovilos et al. 2002), we think that a disk model is
appropriate for these systems.

For all model calculations we used a scaling between the driving length scale and the size of the largest self-gravitating structures
of $\delta=5$ (Eq.~\ref{larson}). Our model results are not sensitive to a variation of $\delta$ by a factor of $2$ (Liz\'ee et al. 2022).
The Toomre $Q$ parameters were chosen such that the model CO luminosities match the observed CO luminosities (Vollmer et al. 2017).
The mass accretion rate was set by the observed star-formation rate. Vollmer et al. (2017) estimated the overall model uncertainties
to be $\sim 0.3$~dex.

\section{The L$_{\rm TIR}$-SFR conversion factor \label{sec:sfrircorr}}

The total infrared luminosity is frequently used to estimate the star formation rate of galaxies (Kennicutt 1998).
The TIR luminosity to SFR conversion factor ($\dot{M}_*/L_{\rm TIR}$) depends on how efficiently stellar light
is absorbed by dust and re-radiated in the IR (e.g., Inoue et al. 2000). The dust can be heated by ionizing UV emission of a young
stellar population or non-ionizing emission of an older stellar population. 
Since the light of starburst galaxies is dominated by the youngest stellar populations
(see Fig.~5 of Madau \& Dickinson 2014), whereas the light of older stellar populations significantly contributes
to dust heating in the main sequence starforming galaxies, one expects higher $\dot{M}_*/L_{\rm TIR}$ for starburst galaxies than for
main sequence starforming galaxies.
Indeed, Rowlands et al. (2014) showed that the TIR luminosity to SFR conversion factor
can be significantly lower for galaxies with $L_{\rm TIR} < 3 \times 10^{11}$~L$_{\odot}$ than for galaxies with $L_{\rm TIR} > 3 \times 10^{11}$~L$_{\odot}$
(their Fig.~7), where it corresponds to the Kennicutt (1998) value. These authors stated that ``galaxies with
a significant contribution to the infrared luminosity from the diffuse ISM (mostly powered by stars older than $10$~Myr) lie further from
the Kennicutt et al. (1998) relation.''

In the framework of theoretically derived TIR luminosities, $\dot{M}_*/L_{\rm TIR}$ depends on the assumed 
initial mass function (IMF). 
A Salpeter IMF (Kennicutt 1998) leads to $\dot{M}_*/L_{\rm TIR}=1.7 \times 10^{-10}$~M$_{\odot}$yr$^{-1}$L$_{\odot}^{-1}$, 
a Kroupa et al. (1993) IMF to a lower conversion factor of $\dot{M}_*/L_{\rm TIR}=1.1 \times 10^{-10}$~M$_{\odot}$yr$^{-1}$L$_{\odot}^{-1}$,
and a Chabrier (2003) IMF to $\dot{M}_*/L_{\rm TIR}=1.0 \times 10^{-10}$~M$_{\odot}$yr$^{-1}$L$_{\odot}^{-1}$.
Whereas Genzel et al. (2010) and Tacconi et al. (2013) assumed a Chabrier IMF, Gracia-Carpio et al. (2008) 
assumed a Salpeter IMF.
If the star formation rate is derived by a combination of $24$~$\mu$m and FUV or H$\alpha$ emission, $\dot{M}_*/L_{\rm TIR}$
can be calculated with the observed TIR luminosity.

\begin{figure}
  \centering
  \resizebox{\hsize}{!}{\includegraphics{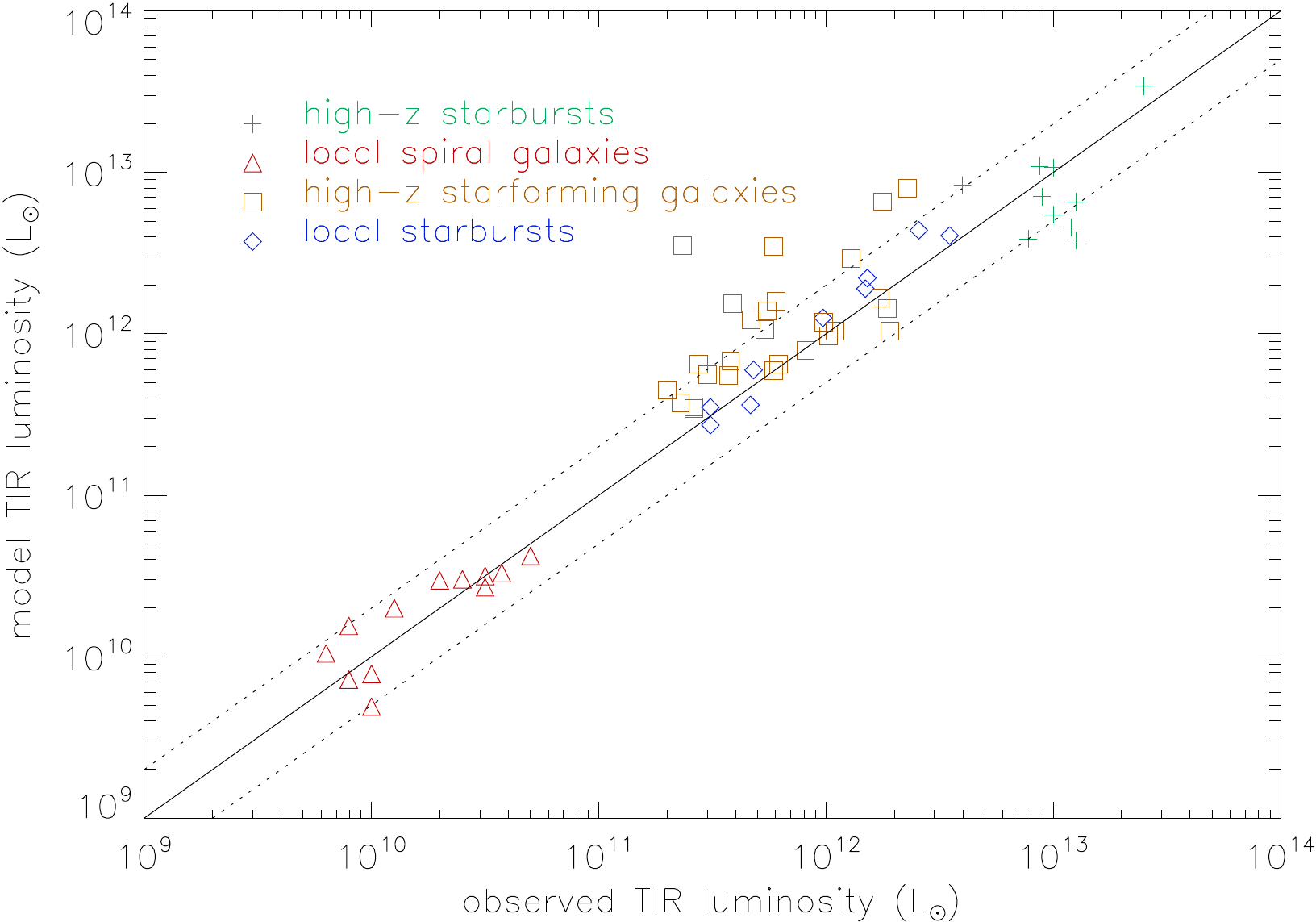}}
  \caption{Model TIR luminosity as a function of the TIR luminosity derived from observations. The solid line corresponds to the one to one correlation.
    The dotted lines are located at distances of $\pm 0.3$~dex from the solid line. 
  \label{fig:galaxies_FRC_vrotDifferentForPhibbs_TIR}}
\end{figure}

The TIR luminosity to SFR conversion factor of the local spiral galaxies calculated in this way is 
$\dot{M}_*/L_{\rm TIR}=0.9 \times 10^{-10}$~M$_{\odot}$yr$^{-1}$L$_{\odot}^{-1}$ with an uncertainty of $30$\,\%.
Our model reproduces the observed TIR luminosities within a factor of two (Fig.~\ref{fig:galaxies_FRC_vrotDifferentForPhibbs_TIR}).
It turned out that the star formation rates of the local starburst galaxies calculated with 
$\dot{M}_*/L_{\rm TIR}=1.7 \times 10^{-10}$~M$_{\odot}$yr$^{-1}$L$_{\odot}^{-1}$ lead to model IR spectral density distributions and TIR luminosities, 
which are consistent with observations (Fig.~\ref{fig:galaxies_FRC_vrotDifferentForPhibbs_TIR} and \ref{fig:IRspectra_spirals}).
The mean conversion factor of the high-z starforming galaxies calculated with the TIR luminosities of Barro et al. (2011) is 
$\dot{M}_*/L_{\rm TIR}=1.4 \times 10^{-10}$~M$_{\odot}$yr$^{-1}$L$_{\odot}^{-1}$ with an uncertainty of a factor of two.
We realized that we had to enhance the star-formation rates of the high-z starburst galaxies by a factor of two to reproduce the observed IR 
SEDs (Fig.~\ref{fig:IRspectra_smm}) and the observed CO luminosities (Fig.~10 of Vollmer et al. 2017).
For the maximum observed TIR luminosities (Table~\ref{tab:gbzk}) and the enhanced SFRs the conversion factor is 
$\dot{M}_*/L_{\rm TIR}=(1.7 \pm 0.4) \times 10^{-10}$~M$_{\odot}$yr$^{-1}$L$_{\odot}^{-1}$.
For the observationally derived TIR luminosities of Genzel et al. (2010) and the enhanced SFRs the conversion factor is 
$\dot{M}_*/L_{\rm TIR}=2 \times 10^{-10}$~M$_{\odot}$yr$^{-1}$L$_{\odot}^{-1}$.

\begin{figure}
  \centering
  \resizebox{\hsize}{!}{\includegraphics{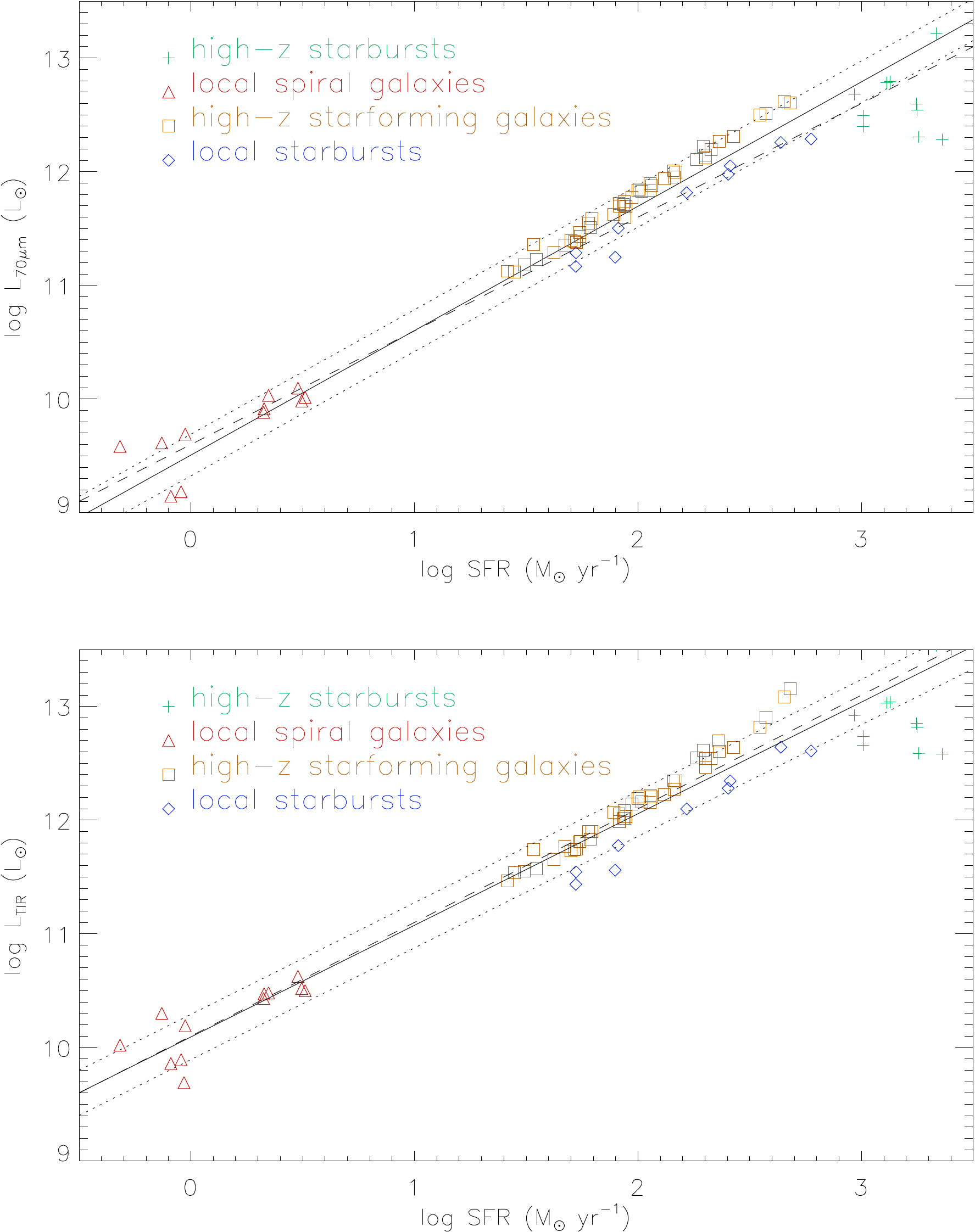}}
  \caption{Upper panel: model SFR - $70$~$\mu$m correlation. Lower panel: model SFR - TIR correlation. The solid and dotted lines  represent 
    an outlier-resistant linear regression and its uncertainty. The dashed line corresponds to a linear correlation.
  \label{fig:galaxies_FRC_vrotDifferentForPhibbs_nosecnowind_2}}
\end{figure}

The model SFR-IR correlations are presented in Fig.~\ref{fig:galaxies_FRC_vrotDifferentForPhibbs_nosecnowind_2}.
The IR luminosities are measured at $70$~$\mu$m and between $8$~$\mu$m and $1000$~$\mu$m (total IR, TIR).
We note that there are two high-z starburst galaxies that have significantly lower IR luminosities than the
majority of the high-z starburst galaxies.
The slopes of both log-log correlations are close to unity, i.e. the correlation is close to linear. 
The slope of the log(SFR) - log($70$~$\mu$m) correlation is $1.08 \pm 0.05$, that of the log(SFR) - log(TIR) correlation is $0.97 \pm 0.04$.

At a given star-formation rate the $70$~$\mu$m luminosities of the low-z starburst galaxies are closer to those
of the high-z starforming galaxies than for the TIR luminosities. One might expect the opposite trend because of
the higher dust temperatures of the low-z starburst galaxies ($< T_{\rm dust}>=44$~K; Vollmer et al. 2017) compared to the high-z starforming galaxies
($< T_{\rm dust}>=31$~K; Vollmer et al. 2017). However, the inspection of the two associated modified Planck curves normalized by the total
IR luminosity corroborated our result.

The monochromatic and total IR luminosities of the high-z starforming galaxies are about 50\,\% higher than
those of the low-z starbursts at the same SFR.  The model TIR luminosity to SFR conversion factors are 
($0.9 \pm 0.4$, $1.5 \pm 0.4$, $0.7 \pm 0.2$, and $2.4 \pm 1.7$)$\times 10^{-10}$~M$_{\odot}$yr$^{-1}$L$_{\odot}^{-1}$ for the local
spiral, local starburst, high-z starforming, and high-z starburst galaxies, respectively.
The starburst galaxies thus show significantly higher TIR luminosity to SFR conversion factors than the main sequence
starforming galaxies. This trend is consistent with the conversion factors based on the observed TIR luminosities
for the local galaxies. On the other hand, the model conversion factor of the high-z starforming galaxies is lower and that of the high-z starburst
galaxies is higher than the corresponding conversion factors based on the observationally derived TIR luminosities.

We note that the star formation rates of the high-z starforming galaxies given by Tacconi et al. (2013) lead 
to model IR spectral density distributions for $14$ out of $22$ galaxies with well-sampled VizieR SEDs, which are consistent with observations 
(Fig.~\ref{fig:IRspectra_phibbs}).
Out of the $16$ galaxies with $L_{\rm TIR,model}/L_{\rm TIR, Barro} > 1.4$, $13$ galaxies have a well sampled
VizieR IR SED. Nine out of these $13$ galaxies have model IR SEDs, which are consistent with the VizieR IR SEDs.
Within the high-z starburst sample the model IR SEDs of SMMJ123549+6215 and SMMJ123707+6214 are consistent with the VizieR IR SEDs,
whereas they are significantly lower than the VizieR SEDs for SMMJ163650+4057 and SMMJ163658+4105 (Fig.~\ref{fig:IRspectra_smm}).

We conclude that our model TIR luminosity to SFR conversion factors for the local galaxies are consistent with observations.
As expected, the conversion factor of the local spiral galaxies is lower than that of the local starburst galaxies because of
additional dust heating in the spiral galaxies by older stellar populations (Rowlands et al. 2014).
We might expect the same trend for the high-z galaxies, as predicted by our model. However, the available observationally derived TIR
luminosities lead to a common  $\dot{M}_*/L_{\rm TIR} \sim 1.3 \times 10^{-10}$~M$_{\odot}$yr$^{-1}$L$_{\odot}^{-1}$ for the high-z main sequence and
starburst galaxies.

\section{Results \label{sec:results}}

The velocities of ionized winds are typically hundreds of km\,s$^{-1}$ at large radii (several kpc; Veilleux et al. 2005, Heckman et al. 2015). 
The CR electron advection timescale is $t_{\rm adv}=L/v_{\rm wind}$, where $L$ is the height of the radio continuum emission
and $v_{\rm wind}$ is the mean velocity between $z=0$ and $z=L$. Since galactic winds are accelerating with increasing height,
the mean wind velocity up to $z=L$ is uncertain. For simplicity, we set $L=H$.
We calculated different wind models (Table~\ref{tab:models}): with a slow ($v_{\rm wind}=0.1\,\sqrt{2}\,v_{\rm rot}$), 
medium velocity ($v_{\rm wind}=\sqrt{2}\,v_{\rm rot}$), and fast wind ($v_{\rm wind}=10\,\sqrt{2}\,v_{\rm rot}$).
In addition, we calculated models with and without secondary CR electrons,
set $q=2.0$ and $2.3$ (Eq.~\ref{eq:emissivity}), and replaced equipartition between the
turbulent kinetic and magnetic energy density by (i) $B=5.3 \times (\Sigma/10~$M$_{\odot}$yr$^{-1}$)~$\mu$G (Parker limit; Lacki et al. 2010)
where $\Sigma$ is the gas surface density and (ii) 
$B=8.8/\sqrt{n/{\rm cm}^3}$~$\mu$G. The normalizations of the magnetic field strength were chosen such that the model integrated
radio continuum emission of the local spiral galaxies are close to observations.
\begin{table*}
      \caption{Models.}
         \label{tab:models}
      \[
       \begin{tabular}{llll}
        \hline
        name & ingredients \\ 
        \hline
        fiducial & $t_{\rm wind }=10 \,\frac{H}{\sqrt{2} v_{\rm rot}}~{\rm Myr}$ & no secondaries & \\
        wind & $t_{\rm wind }$ according to Eq.~\ref{eq:wind} & no secondaries & \\
        sec+wind & $t_{\rm wind }$ according to Eq.~\ref{eq:wind} & secondaries (Eq.~\ref{eq:secondaries}) & \\
        sec+fastwind & $t_{\rm wind }=0.1 \,\frac{H}{\sqrt{2} v_{\rm rot}}~{\rm Myr}$ & secondaries & \\
        exp & $t_{\rm wind }=10 \,\frac{H}{\sqrt{2} v_{\rm rot}}~{\rm Myr}$ & no secondaries & $q=2.3$ (Eq.~\ref{eq:emissivity}) \\
        Bsigma & $t_{\rm wind }=10 \,\frac{H}{\sqrt{2} v_{\rm rot}}~{\rm Myr}$ & no secondaries & $B=5.3 \times (\Sigma/10~$M$_{\odot}$pc$^{-2}$)~$\mu$G \\
        Brho & $t_{\rm wind }=10 \,\frac{H}{\sqrt{2} v_{\rm rot}}~{\rm Myr}$ & no secondaries & $B=8.8/\sqrt{n/{\rm cm}^{-3}}$~$\mu$G  \\
	\hline
        \end{tabular}
      \]
\end{table*}

Before the presentation of the integrated radio continuum spectra calculated by our model, we present the radial profiles of the
magnetic field strength, CR electron density and optical depth, and the synchrotron and energy loss timescales of the fiducial model.

\subsection{Magnetic field strength and CR density \label{sec:magcr}}
 
\begin{figure}
  \centering
  \resizebox{\hsize}{!}{\includegraphics{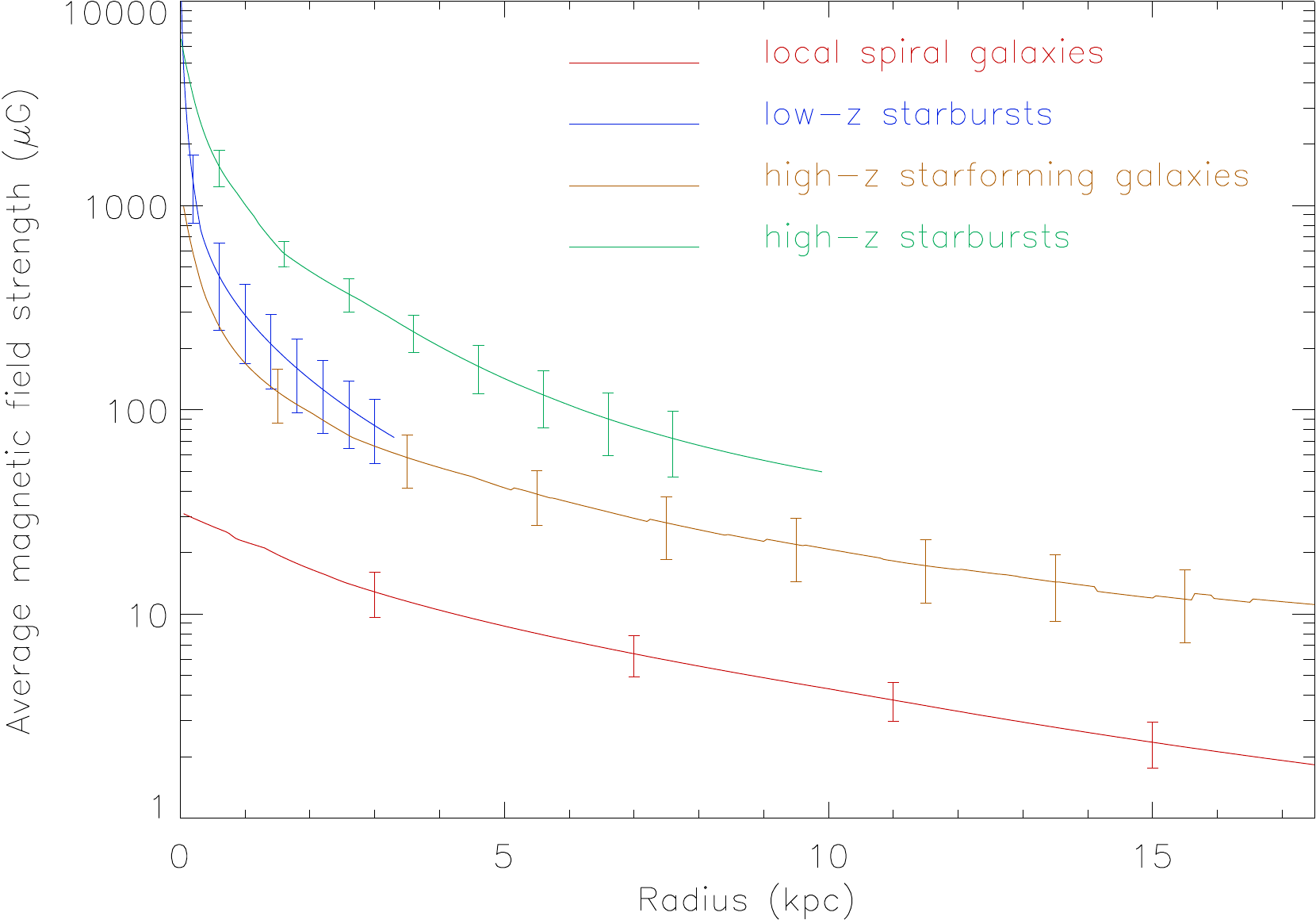}}
  \resizebox{\hsize}{!}{\includegraphics{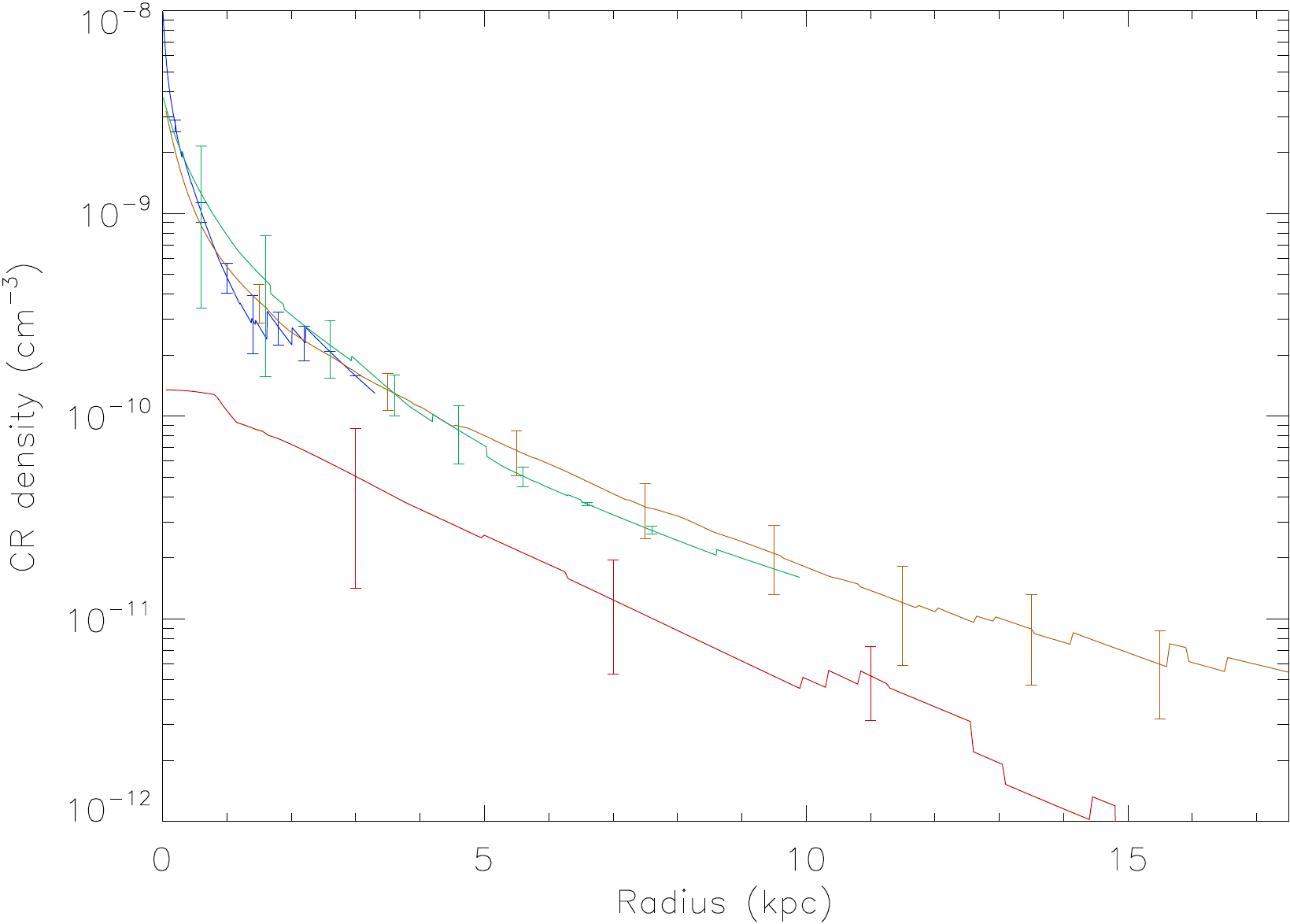}}
  \resizebox{\hsize}{!}{\includegraphics{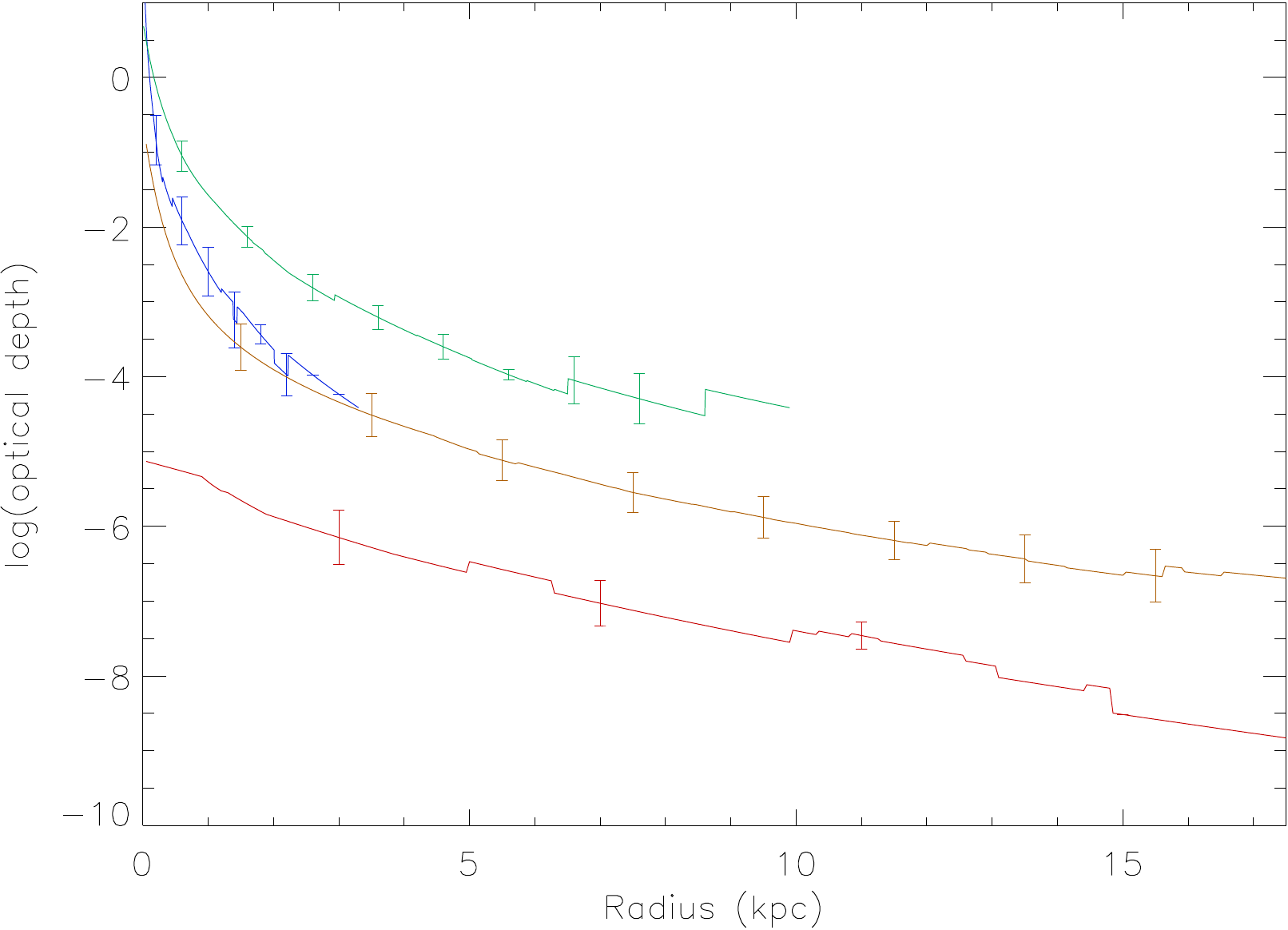}}
  \caption{Model median radial profiles of the magnetic field strength (upper panel), CR electron density (middle panel), and
    free-free optical depth at $\nu=1.4$~GHz (lower panel) for the four galaxy samples with the associated semi-interquartile ranges.
    The jumps at $R > 10$~kpc are caused by taking the median of models with the different radial sizes.
  \label{fig:average_Bmagn_all}}
\end{figure}

The median radial profiles of the magnetic field strength, CR electron density, and free-free optical depths of the four galaxy samples 
are presented in Fig.~\ref{fig:average_Bmagn_all}.
Whereas the median magnetic field strengths of the low-z starburst and high-z starforming galaxies are similar within the inner $3$~kpc,
they are three to four times higher/lower in the high-z starburst galaxies/local spiral galaxies. The magnetic field strengths in the central kpc
are about $\sim 30$~$\mu$G, $\sim 0.5$~mG, and $\sim 2$~mG in the local spirals, low-z starbursts/high-z galaxies, and high-z starburst 
galaxies. This is due to the fact that the turbulent velocity dispersion increases with the star formation rate (Eq.~\ref{starformation}) and the 
magnetic field strength is proportional to the turbulent velocity dispersion (Eq.~\ref{eq:Bmag2}).

The median radial profiles of the CR electron density at $R > 2$~kpc have approximately exponential shapes. The ratios between the
profiles of the different samples are significantly smaller than those of the profiles of the magnetic field strength.
The CR electron densities of the low-z starbursts, high-z starburst galaxies, and high-z starforming galaxies are similar, that of the local spiral galaxies 
are about a factor of three smaller at a given radius. The exponential scale lengths of the magnetic field strength and
the CR electron density are presented in Table~\ref{tab:CRscale}.
 \begin{table}[!ht]
      \caption{Scale lengths of the magnetic field and the CR electron density in kpc.}
         \label{tab:CRscale}
      \[
         \begin{tabular}{lcccc}
           \hline
             & local & low-z & high-z & high-z \\
            & spirals & starbursts & starforming & starbursts \\
           \hline
           $l_B$ & 5.3 & 1.0 & 7.0 & 2.9 \\
           $l_{n_{\rm CR}}$ & 2.8 & 0.6 & 3.6 & 2.2 \\
           \hline
           $R_{\rm in}^{\rm (a)}$ & 0.0 & 0.3 & 2.0 & 1.5 \\
           $R_{\rm out}^{\rm (a)}$ & 10.0 & 1.6 & 15.0 & 8.0 \\
           \hline
         \end{tabular}
      \]
      \begin{tablenotes}{}{}
      \item $^{\rm (a)}$ inner and outer radius for the scale length calculation in kpc.
\end{tablenotes}
\end{table}

The median radial profiles of the free-free optical depths at $\nu=1.4$~GHz of all galaxy samples are significantly smaller than unity except for the central $\sim 100$~pc.
of the low-z starbursts and high-z starburst galaxies. Therefore, free-free absorption is expected to play a role in the centers of low-z starbursts and high-z starburst galaxies.
Synchrotron self-absorption is negligible for all galaxies in all samples.

\subsection{Energy loss timescales \label{sec:elosstimes}}

\begin{figure*}
  \centering
  \resizebox{\hsize}{!}{\includegraphics{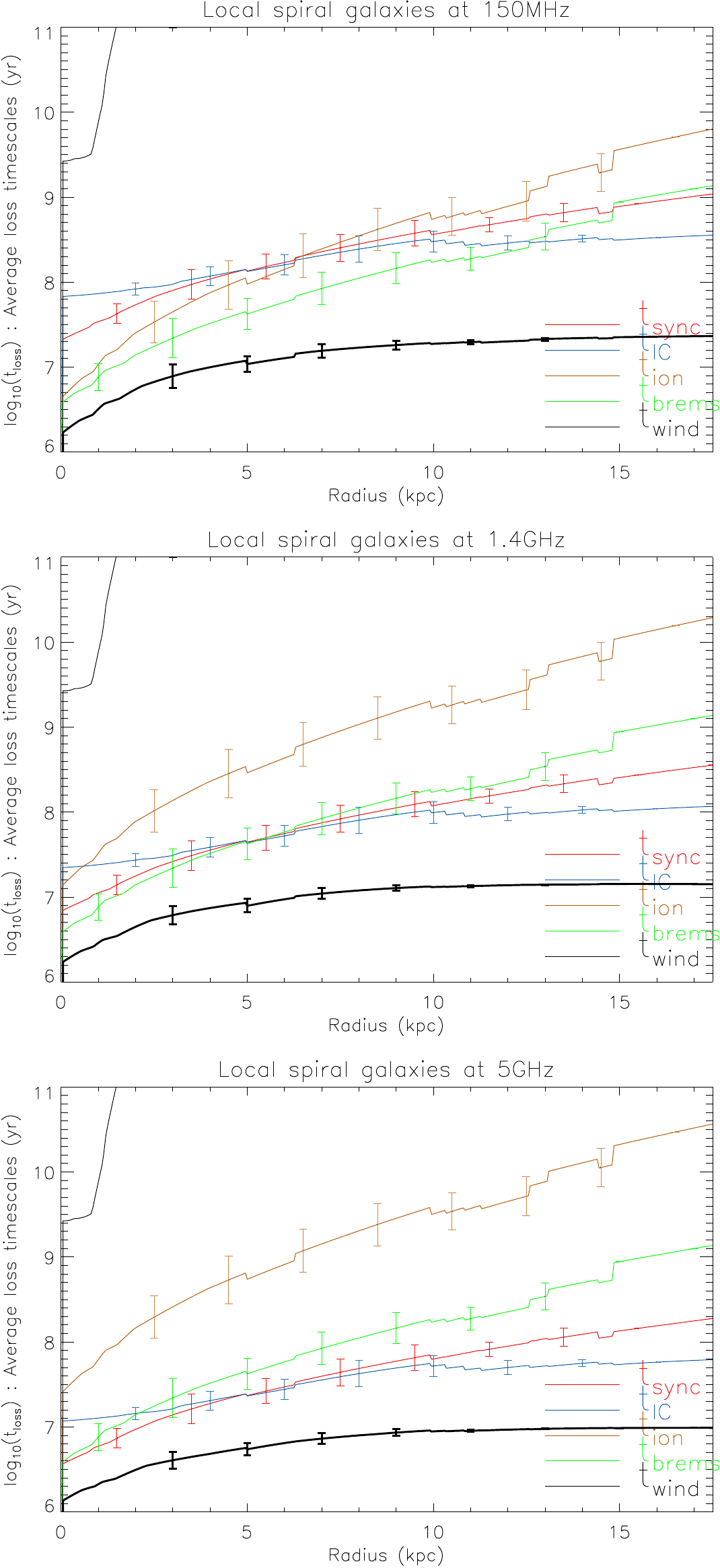}\includegraphics{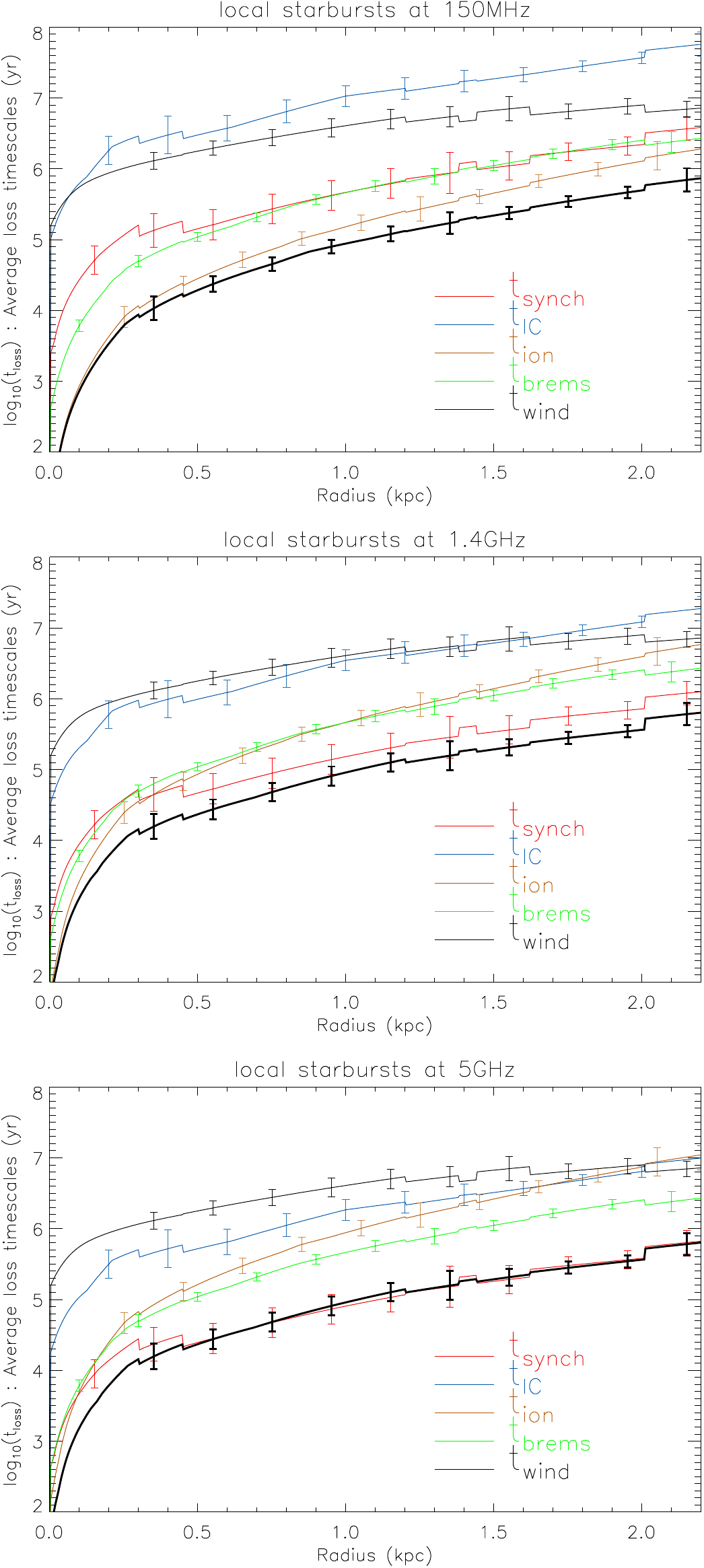}}
  \caption{Radial profiles of the different median model timescales for the local spirals (left panels) and low-z starbursts (right panels)
    at $\nu=1.4$~GHz (upper panels) and $\nu=5$~GHz (lower panels) with the associated semi-interquartile ranges.
  \label{fig:average_loss_timescales1}}
\end{figure*}

\begin{figure*}
  \centering
  \resizebox{\hsize}{!}{\includegraphics{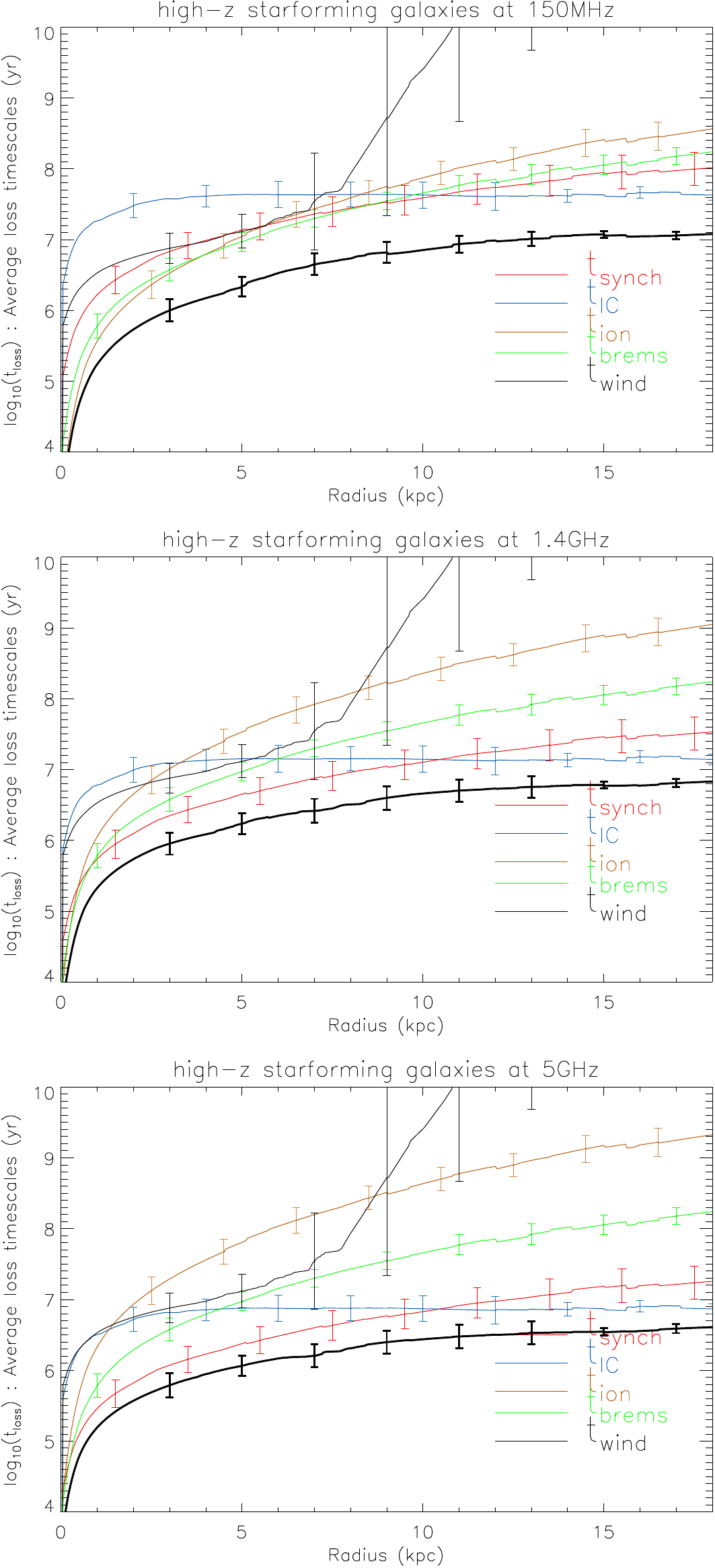}\includegraphics{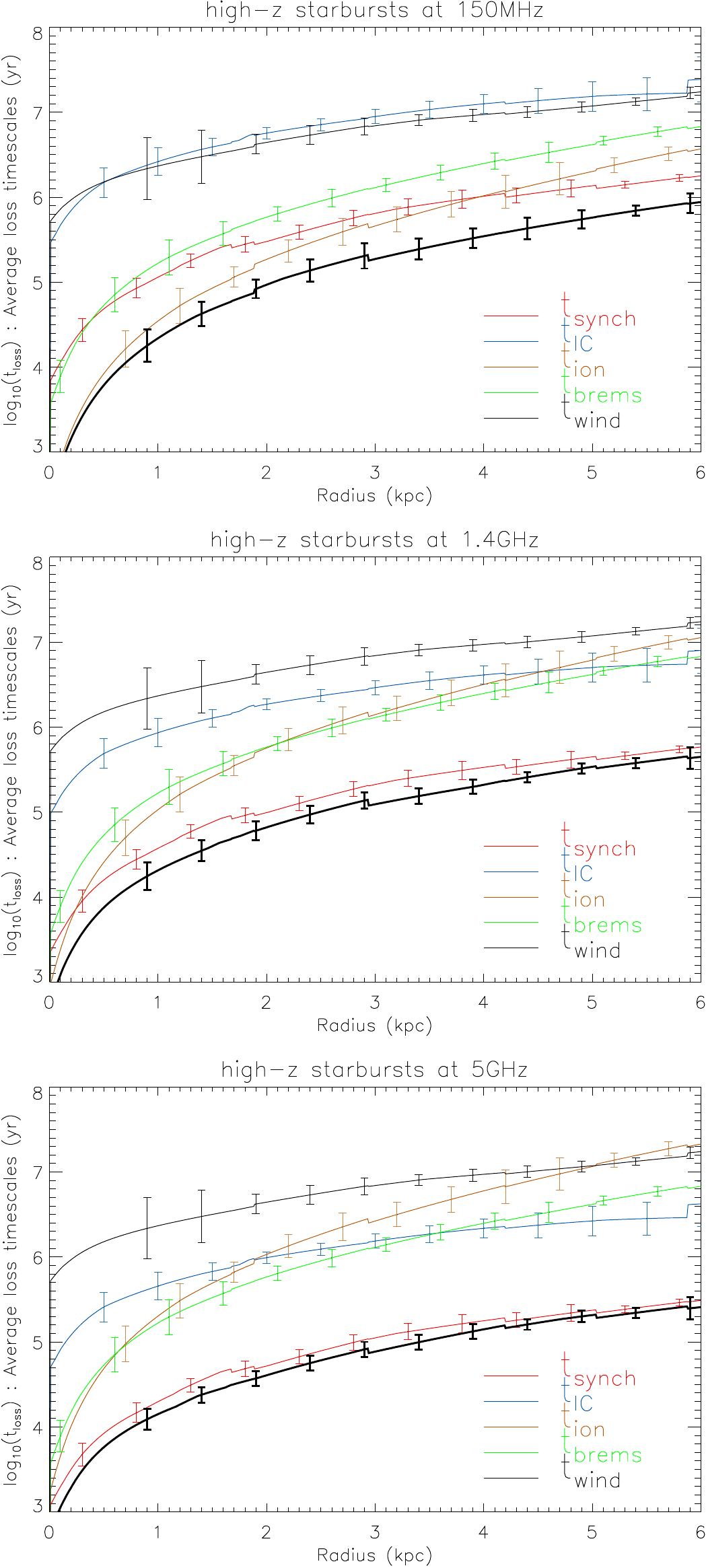}}
  \caption{Radial profiles of the different median model timescales for the high-z starforming galaxies (left panels) and high-z starburst galaxies (right panels)
    at $\nu=1.4$~GHz (upper panels) and $\nu=5$~GHz (lower panels) with the associated semi-interquartile ranges.
  \label{fig:average_loss_timescales2}}
\end{figure*}

The radial profiles of the different median model energy loss timescales (Sect.~\ref{sec:model}) at $150$~MHz, $1.4$~GHz, and
$5$~GHz for the four galaxy samples are presented in
Figs.~\ref{fig:average_loss_timescales1} and \ref{fig:average_loss_timescales2}. In the local spiral galaxies the
synchrotron, IC, and ionic timescales for $\nu=150$~MHz are similar for $3~{\rm kpc} \la R \la 9~{\rm kpc}$,
whereas the bremsstrahlung timescale is about a factor of three lower.
At $\nu=1.4$~GHz the synchrotron, IC, and bremsstrahlung timescales are similar for $3~{\rm kpc} \la R \la 9~{\rm kpc}$, whereas the
ionic timescale is about a factor of ten higher. Within the inner $3$~kpc bremsstrahlung leads to the smallest energy loss timescales.
At $\nu=5$~GHz bremsstrahlung becomes less important because of the decreasing synchrotron and IC timescales with increasing frequency.
The advection of CR electrons by galactic winds does not play a role in local spiral galaxies.

The situation is different in the low-z starbursts. Due to the high magnetic field strengths the synchrotron timescale is by far the
smallest timescale at $\nu=5$~GHz. At $\nu=1.4$~GHz, the ionic and bremsstrahlung timescales are comparable to the synchrotron
timescale in the inner few hundred parsec. A fast ($v_{\rm wind}=10\,\sqrt{2}\,v_{\rm rot}$) wind leads to timescales 
comparable to the synchrotron timescale.
Thus, fast winds are expected to significantly decrease the radio continuum emission of low-z starbursts at $\nu \la 1.4$~GHz.
At $\nu=150$~MHz the ionic timescale is a factor of about three lower than the synchrotron timescale at all radii.
The ionic timescale thus sets the CR electron energy loss timescale at this frequency.

In the high-z starforming galaxies synchrotron losses dominate within the effective radius (about half of the radial ranges
shown in Figs.~\ref{fig:average_loss_timescales1} and \ref{fig:average_loss_timescales2}) at $\nu=5$~GHz and $\nu=1.4$~GHz,
whereas ionic and bremsstrahlung losses dominate  at $\nu=150$~MHz.
Bremsstrahlung losses contribute in the centers, whereas IC losses become more and more important at larger radii. 
The latter losses dominate beyond the effective radius at all frequencies.
Medium velocity winds play an important role for the energy loss of CR electrons within the central $5$~kpc.

In the high-z starburst galaxies the magnetic field strength is so high that the synchrotron losses dominate at all radii at $\nu=5$~GHz and $\nu=1.4$~GHz.
At $\nu=150$~MHz ionic losses dominate for radii smaller than $4$~kpc. Energy losses due to galactic winds do not play any role in these galaxies. 

\subsection{IR and radio continuum SEDs \label{sec:IRradioseds}}

\begin{figure*}
  \centering
  \resizebox{15cm}{!}{\includegraphics{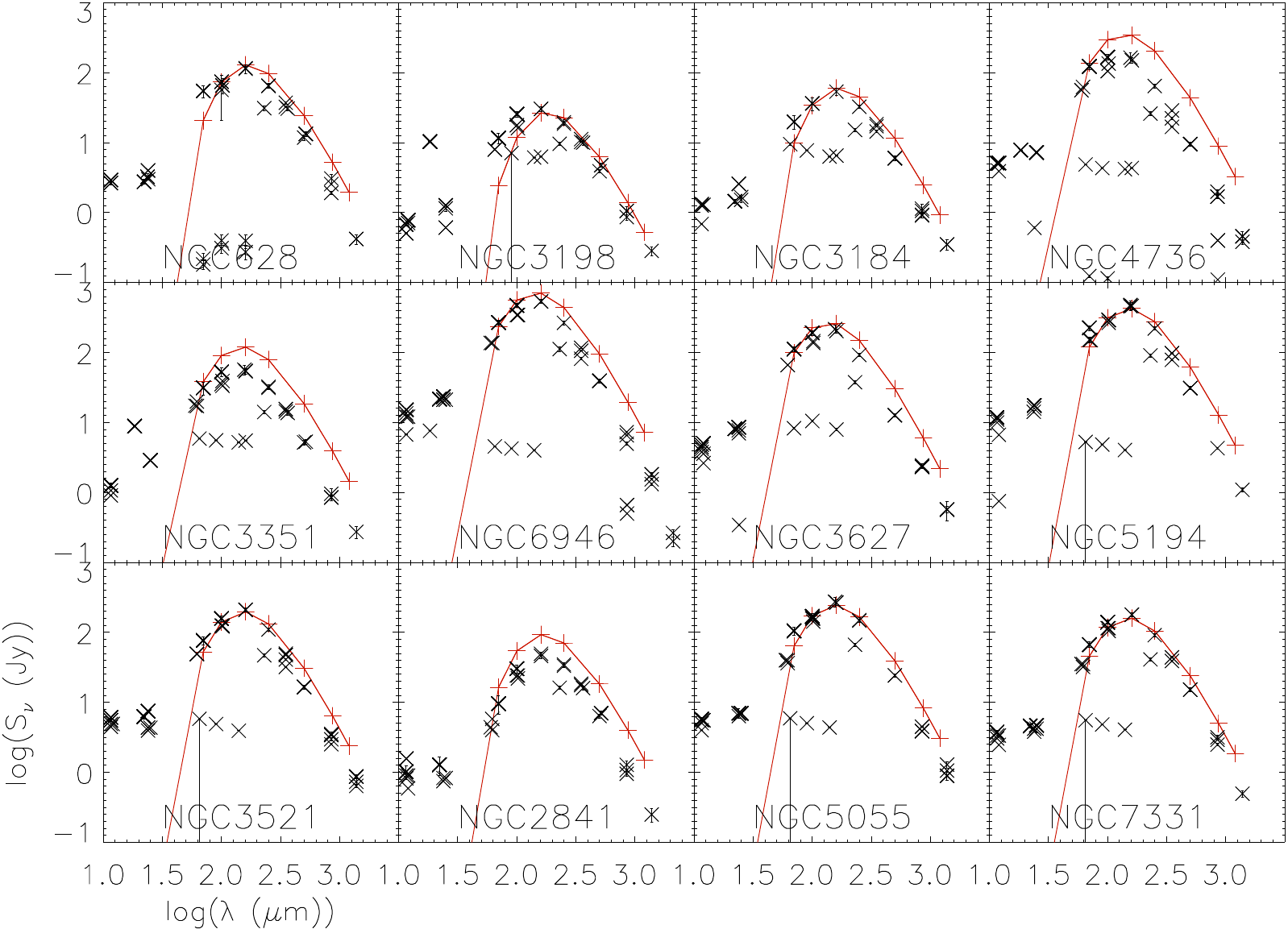}}
  \resizebox{15cm}{!}{\includegraphics{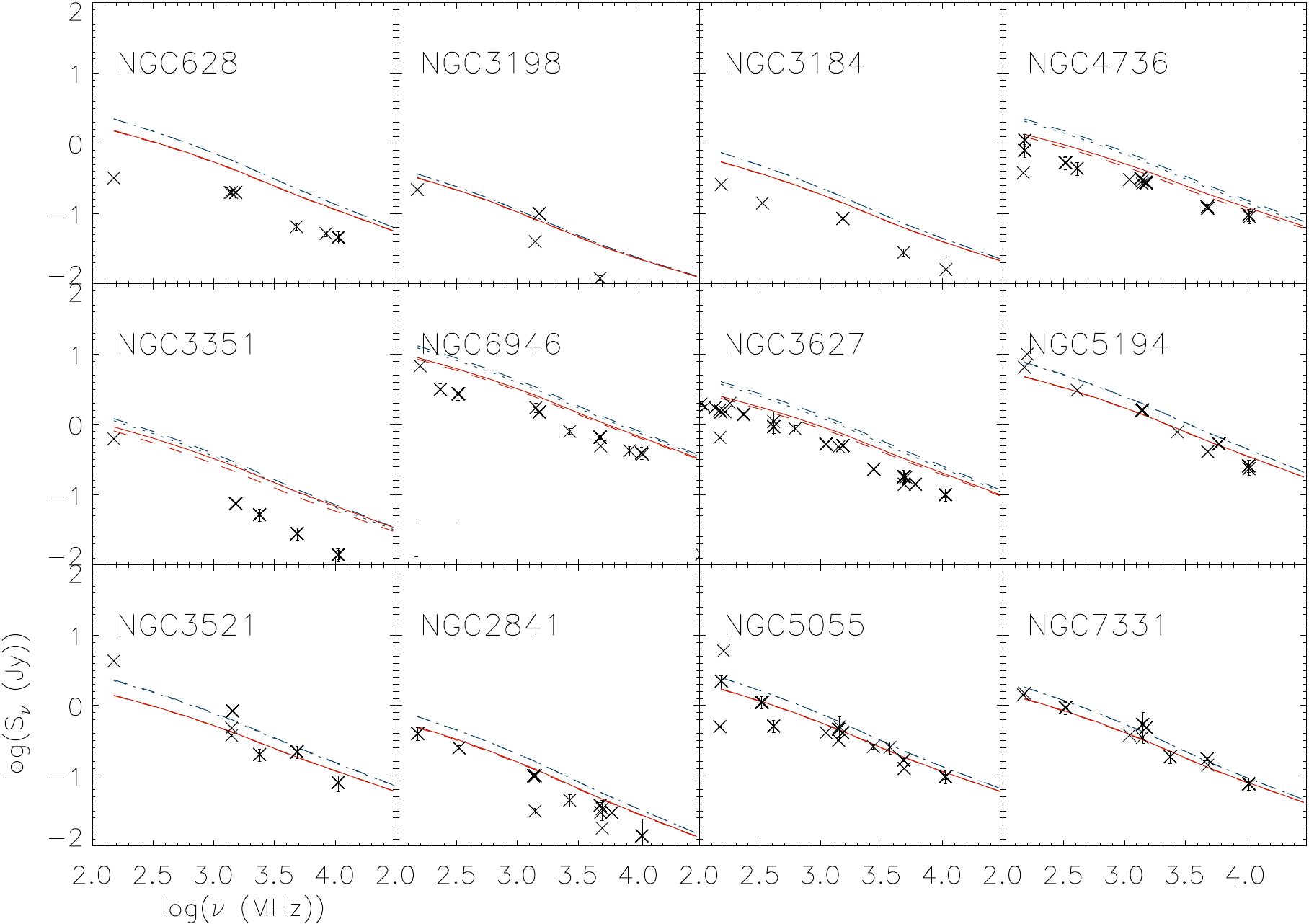}}
  \caption{Local spiral galaxies. Crosses: observations; red lines: model. Upper panels: IR SEDs. 
    Data points with significantly lower flux densities are due to measurements within smaller apertures.
    Lower panels: radio continuum SEDs.
    Solid red line: fiducial model. Dashed red line: wind model. Dashed blue line: sec+wind model. Dotted blue line: sec+fastwind model.
  \label{fig:IRspectra_spirals}}
\end{figure*}

\begin{figure*}
  \centering
  \resizebox{15cm}{!}{\includegraphics{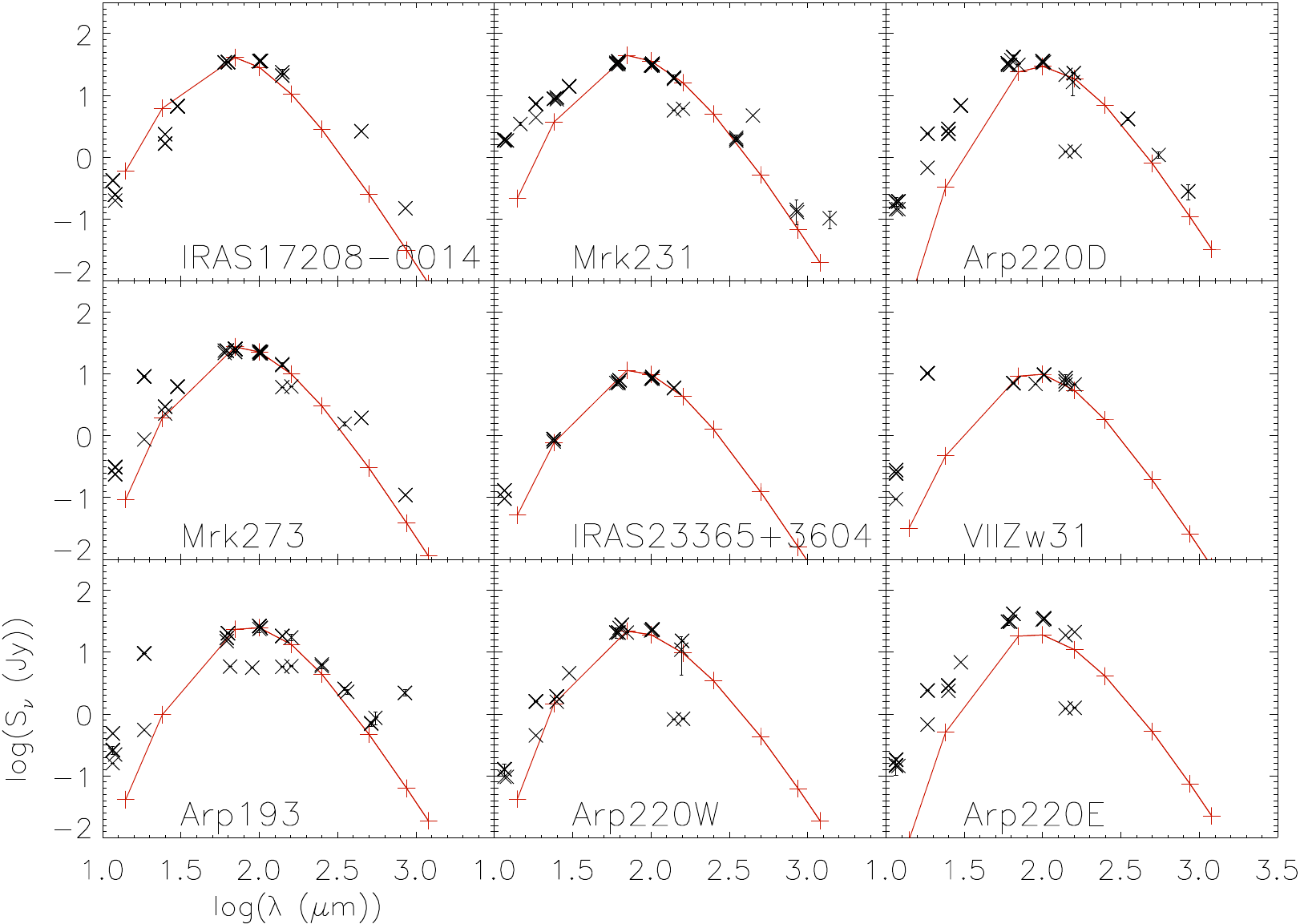}}
  \resizebox{15cm}{!}{\includegraphics{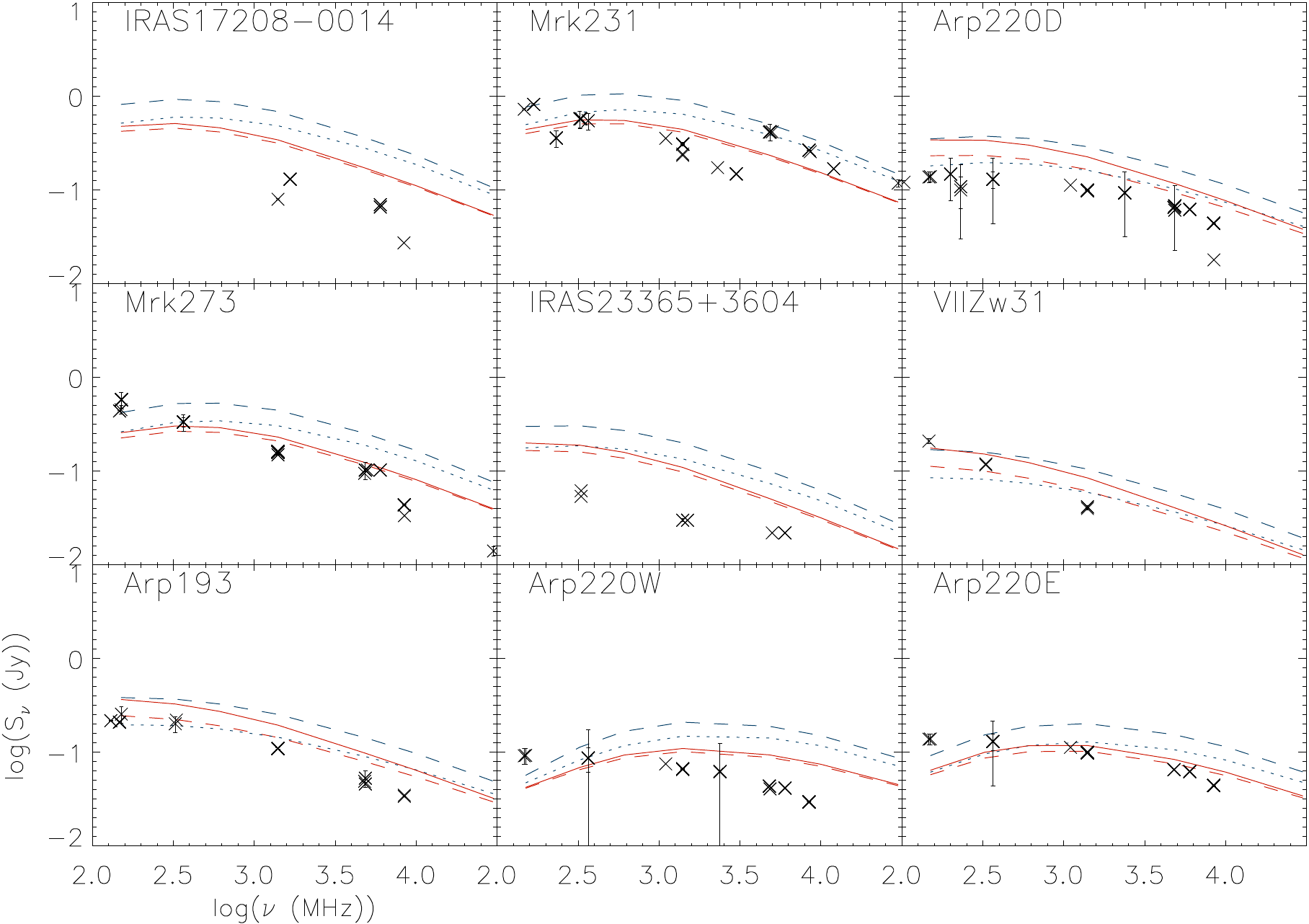}}
  \caption{Low-z starbursts.  Crosses: observations; red lines: model. Upper panels: IR SEDs. Lower panels: radio continuum SEDs.
    Solid red line: fiducial model. Dashed red line: wind model. Dashed blue line: sec+wind model. Dotted blue line: sec+fastwind model.
    Data points with significantly lower flux densities are due to measurements within smaller apertures.
  \label{fig:IRspectra_ulirgs}}
\end{figure*}

\begin{figure*}
  \centering
  \resizebox{15cm}{!}{\includegraphics{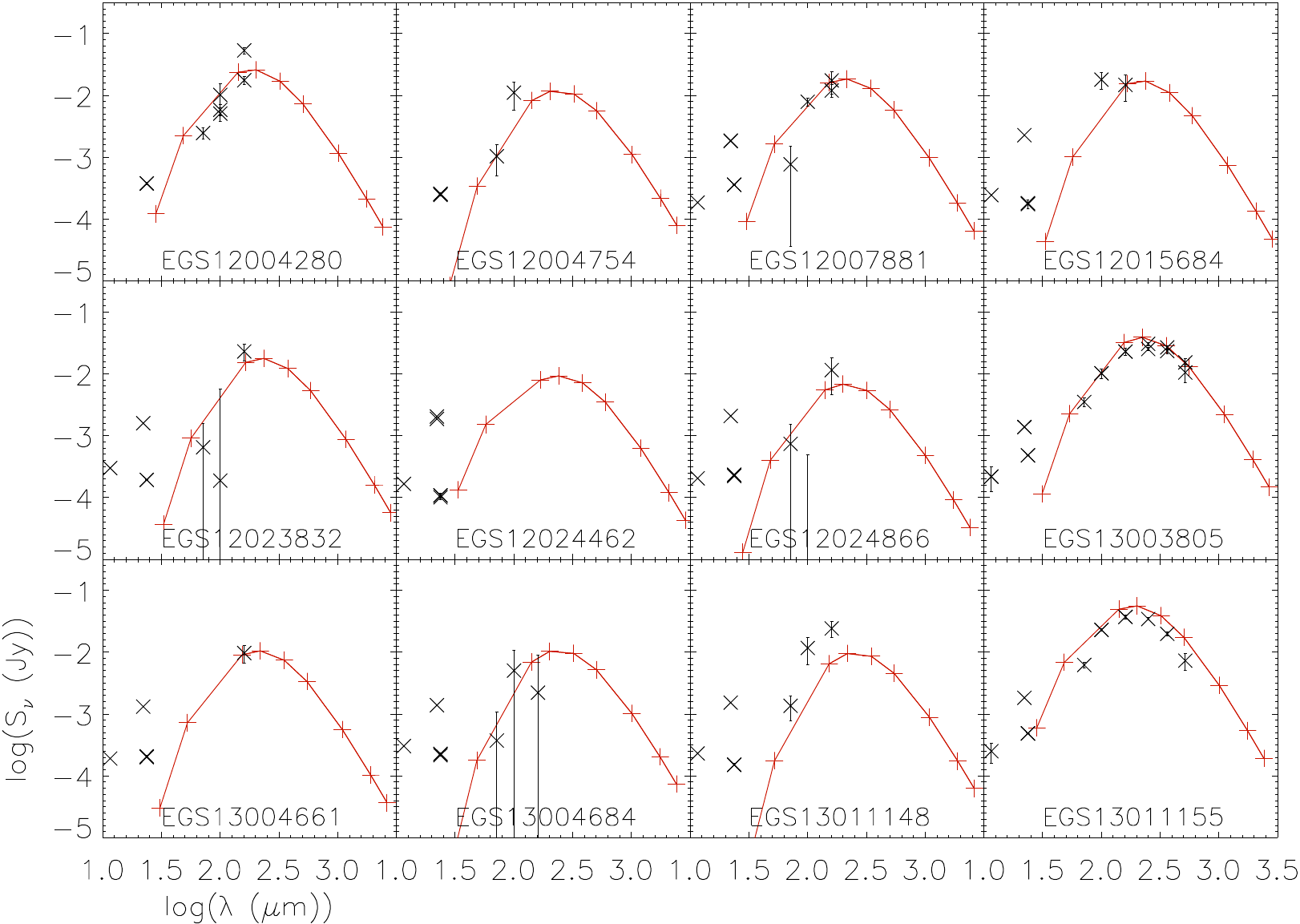}}
  \resizebox{15cm}{!}{\includegraphics{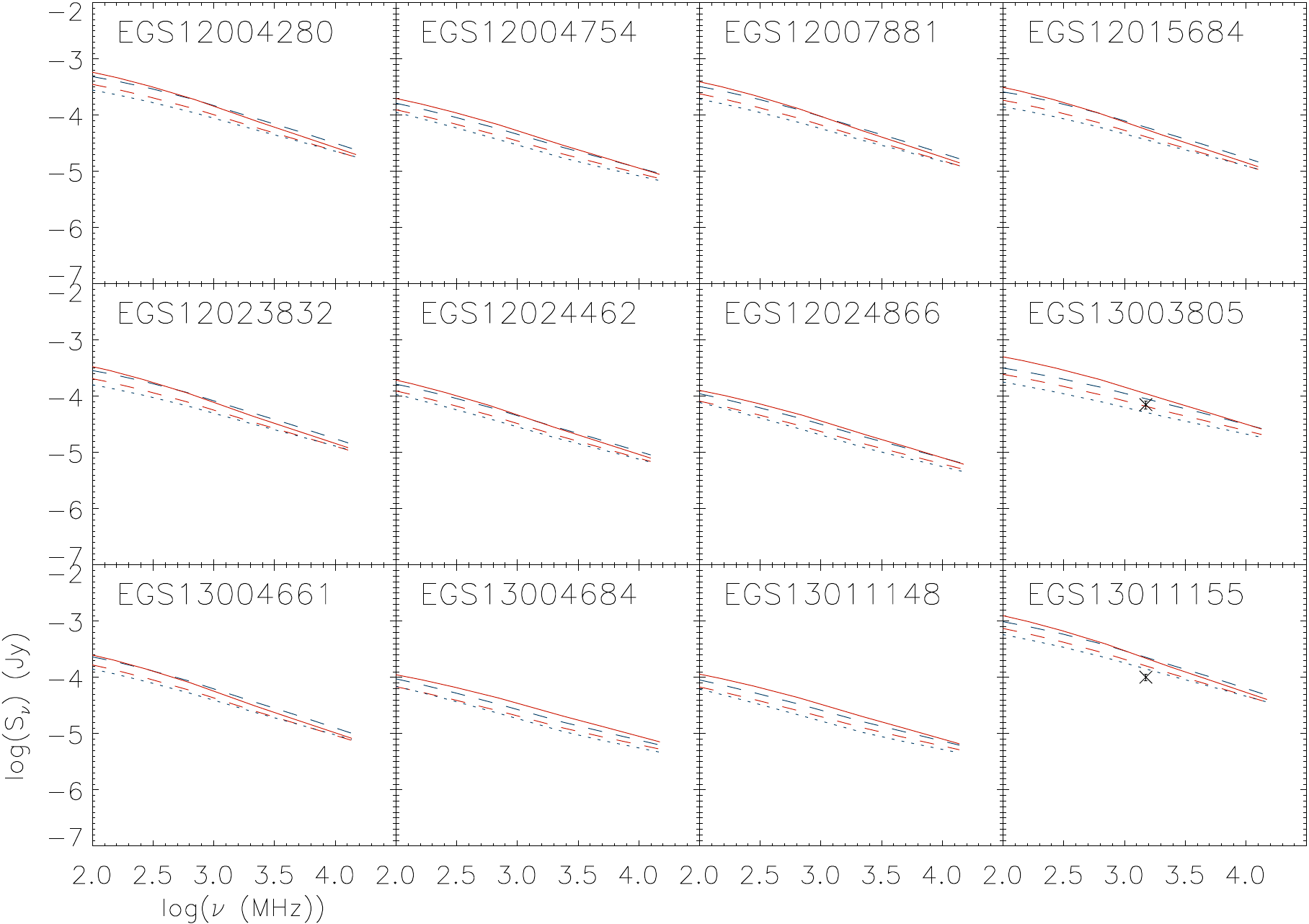}}
  \caption{High-z starforming galaxies. Crosses: observations; lines: model. Upper panels: IR SEDs. Lower panels: radio continuum SEDs.
    Solid red line: fiducial model. Dashed red line: wind model. Dashed blue line: sec+wind model. Dotted blue line: sec+fastwind model.  
    The IR and radio SEDs of the rest of the sample are presented in Fig.~\ref{fig:IRspectra_phibbs2} to \ref{fig:IRspectra_phibbs4}.
  \label{fig:IRspectra_phibbs}}
\end{figure*}

\begin{figure*}
  \centering
  \resizebox{15cm}{!}{\includegraphics{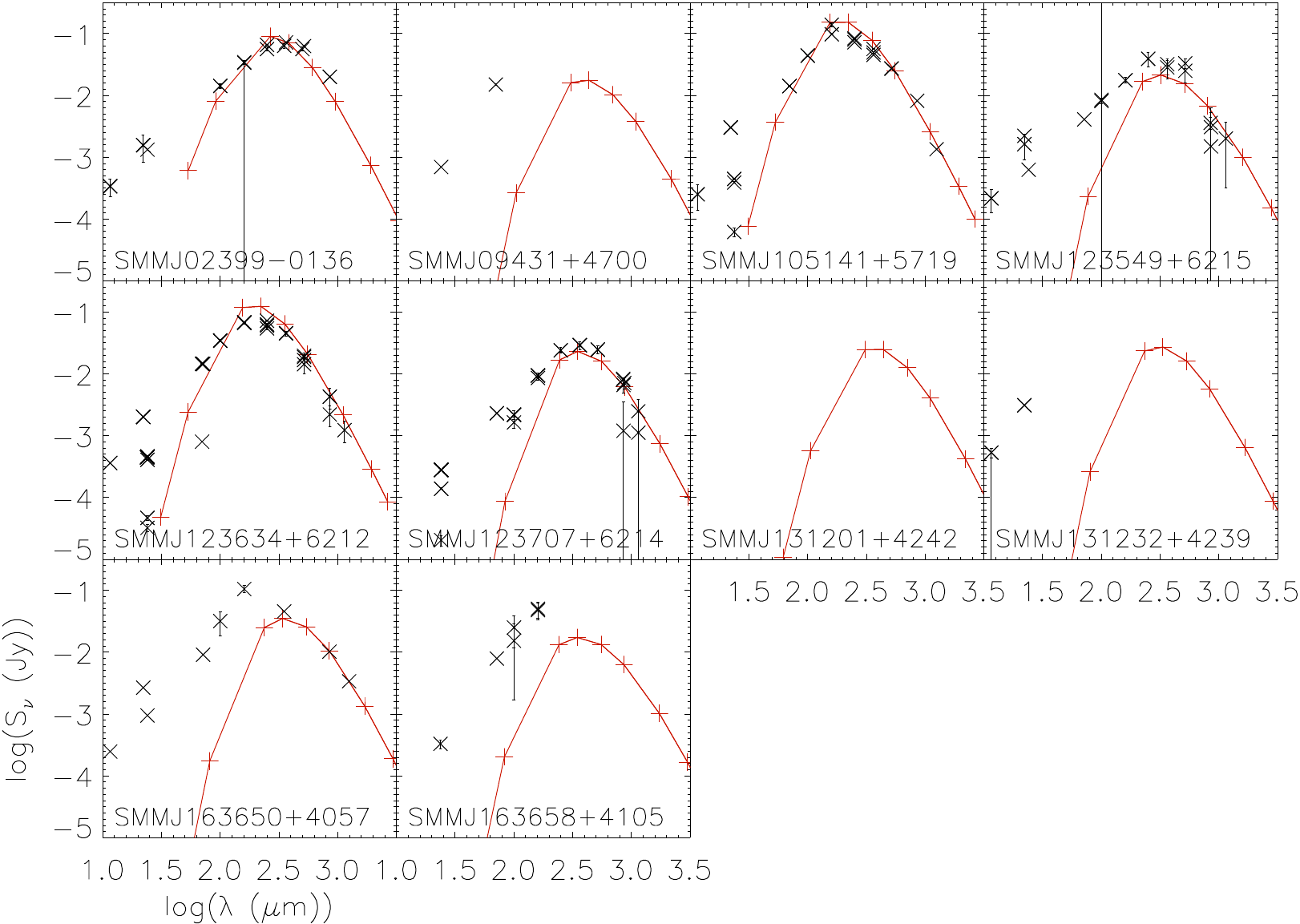}}
  \resizebox{15cm}{!}{\includegraphics{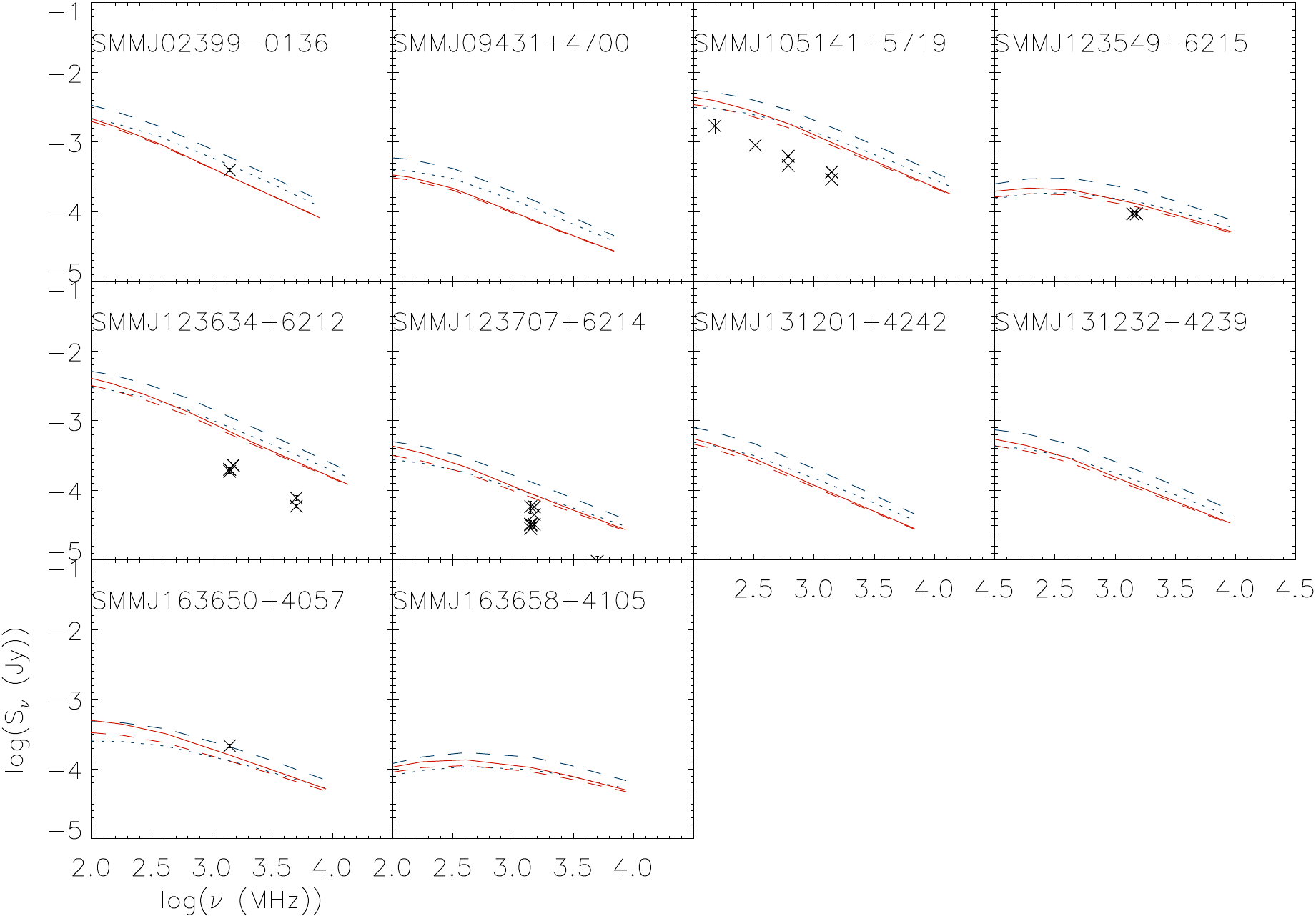}}
  \caption{Submillimeter galaxies. Crosses: observations; red lines: model. Upper panels: IR SEDs. Lower panels: radio continuum SEDs.
    Solid red line: fiducial model. Dashed red line: wind model. Dashed blue line: sec+wind model. Dotted blue line: sec+fastwind model.
  \label{fig:IRspectra_smm}}
\end{figure*}

The IR and radio continuum SEDs of the four galaxy samples are presented in Figs.~\ref{fig:IRspectra_spirals} to \ref{fig:IRspectra_smm}.
The observed IR and radio continuum flux densities were extracted from the CDS/VizieR database (Ochsenbein et al. 2000).
The comparison with Figs.~C1-C4 of Vollmer et al. (2017) show the significant increase of IR flux density measurements in the VizieR data
over the last five years. As stated in Vollmer et al. (2017) the IR SEDs of the galaxies in all samples are reproduced by the model in a 
satisfactory way.

The different model radio continuum SEDs of the local spiral galaxies (Fig.~\ref{fig:IRspectra_spirals}) are very close to each other
because neither winds nor secondary CR electrons have a significant effect on the CR electron distributions.
The models reproduce the observed radio continuum SEDs within about 50\,\%, except for NGC~628, NGC~3184 
where the model overpredicts the flux densities by a factor of two
and NGC~NGC~3351 where the model overpredicts the flux densities at $\nu > 1$~GHz by a factor of three. 
Given that the model IR SED also overestimates the
VizieR IR SED, the assumed star formation rate (Leroy et al. 2008) might be overestimated by about a factor of two.

Within the low-z starburst sample the influence of a wind on the radio continuum SED in models without secondary CR electrons is only
significant in three out of nine galaxies. The models with secondaries and a medium velocity wind always lead to
significantly higher radio continuum flux densities than observed. Overall, the model that is closest to observations is the fiducial model
with a medium velocity wind (wind; Table~\ref{tab:models}). The model radio continuum SEDs of IRAS~17208-0014 and IRAS~23365+3604
overpredict the observed SEDs by a factor of two to three.
 
Only $8$ out of $44$ high-z starforming galaxies have radio continuum flux density measurements mainly at $\nu=1.4$~GHz in VizieR.
Of these, five model flux densities are close to the observed values whereas two model flux densities are significantly higher and
one flux density is significantly lower than observed. 

Six out of ten high-z starburst galaxies have radio continuum flux density measurements in VizieR. Four model radio 
SEDs are close to observations.
The remaining two model SEDs overpredict the observed  radio continuum flux densities by a factor of $\sim 3$.

We conclude that the observed radio continuum SEDs of most of the local galaxies (spirals and low-z starbursts) and high-z 
galaxies (main sequence and starbursts) are reproduced by the fiducial model in a satisfactory way. 
On the other hand, the model significantly overpredicts the observed radio continuum SEDs of $\sim 25$\,\% of the 
low-z galaxies and $\sim 35$\,\% of the high-z galaxies.

\subsection{Alternative magnetic field strength and CR energy distribution prescriptions}
 
\begin{figure*}
  \centering
  \resizebox{15cm}{!}{\includegraphics{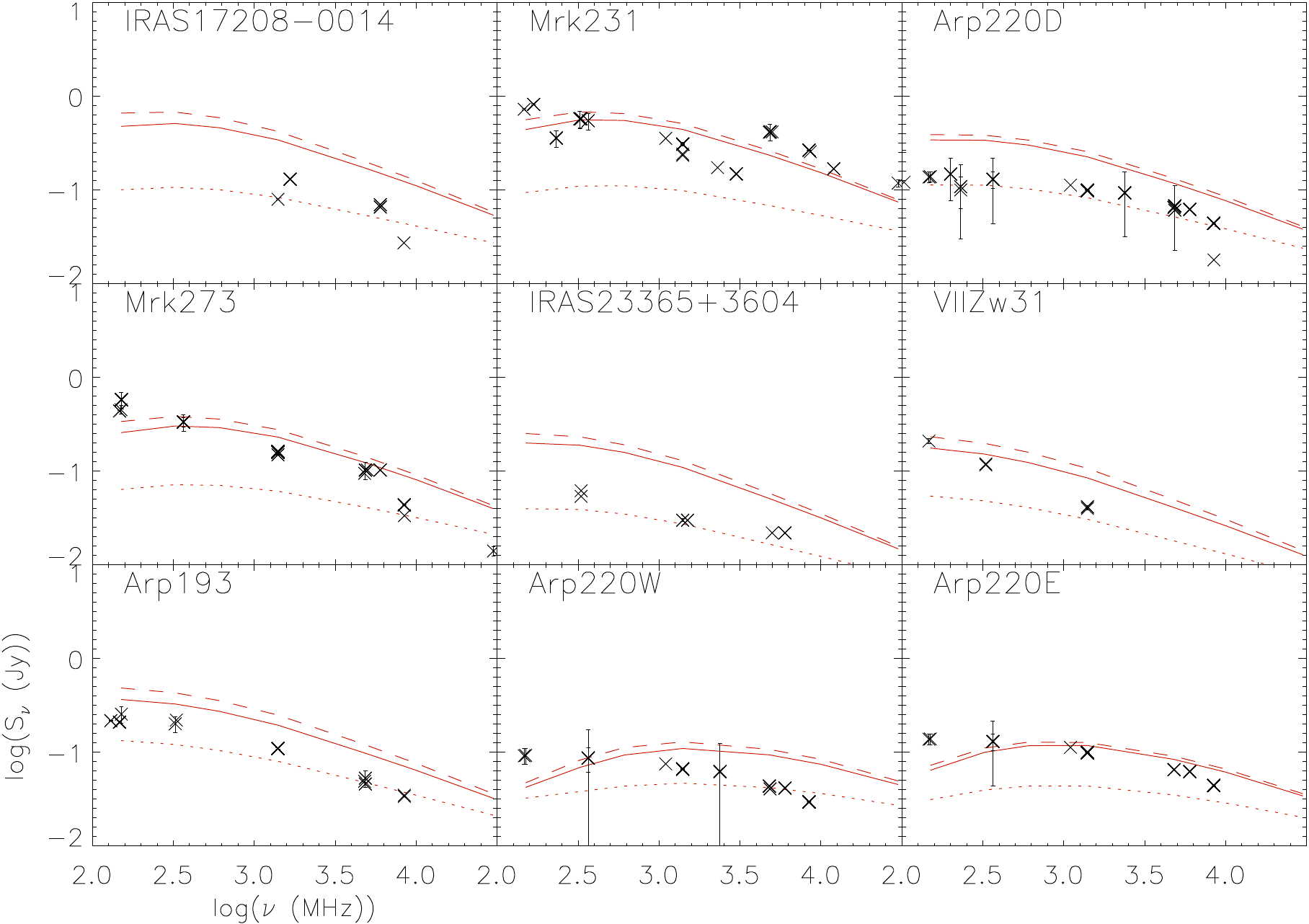}}
  \caption{Radio continuum SEDs of the low-z starbursts. Solid line: fiducial model. Dashed line: $B=5.3 \times (\Sigma/10~$M$_{\odot}$yr$^{-1}$)~$\mu$G. 
    Dotted line: $B=8.8/\sqrt{n/{\rm cm}^3}$~$\mu$G. 
  \label{fig:radiospectra_ulirgs2}}
\end{figure*}

The influence of the different recipes for the magnetic field strength can best be recognized in the radio continuum SEDs
of the low-z starburst sample (Fig.~\ref{fig:radiospectra_ulirgs2}). The radio continuum SEDs of the models involving
(i) equipartition between the turbulent kinetic and magnetic energy densities and (ii) $B=5.3 \times (\Sigma/10~$M$_{\odot}$yr$^{-1}$)~$\mu$G
are similar. Compared to equipartition, the latter recipe leads to $\sim 10$\,\% higher radio continuum flux densities.
On the other hand, the recipe $B=8.8/\sqrt{n/{\rm cm}^{-3}}$~$\mu$G leads to radio continuum flux densities, which are
significantly smaller than observed (up to a factor of ten) for five out of nine low-z starbursts.
Models of IRAS~17208-004, Arp220D, and IRAS~23365+3604 with faster winds naturally lead to better reproductions of the observed radio 
continuum SEDs. We therefore believe that the recipe involving only the gas density
does not reproduce the available observations and should be discarded.

The use of an exponent for the energy dependence of the primary CR injection of $q=2.3$ instead of $q=2.0$ leads to
$\sim 50$\,\% higher CR electron densities than those of the fiducial model and to exponents of the IR radio correlations, which are higher 
by $\sim 0.1$ compared to the exponents of the fiducial model.
Furthermore, the radio continuum SEDs become steeper and the radio continuum luminosities of the low-z starbursts, high-z starforming, 
and high-z starburst galaxies become $\sim 50$\,\% higher compared to the values of the fiducial model. Therefore, 
the $q=2.3$ models are less good in
reproducing the radio continuum emission of the low-z starburst and high-z galaxy samples.

\subsection{The IR-radio correlation \label{sec:IRradiocorrelation}}

Since our fiducial model is our preferred model, we will only show and discuss the SFR-IR, IR-radio, and SFR-radio correlations
for this model.

\begin{figure*}
  \centering
  \resizebox{\hsize}{!}{\includegraphics{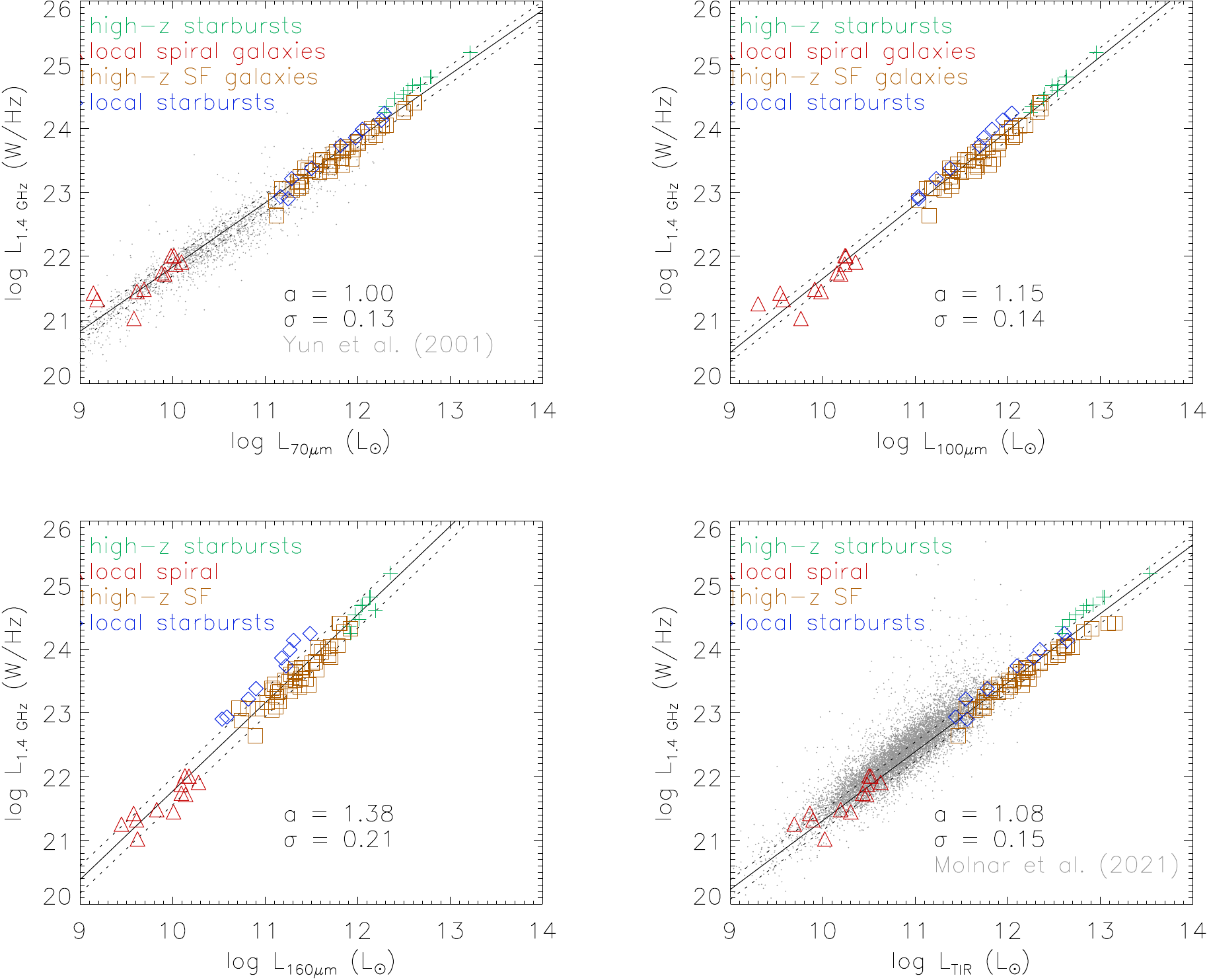}}
  \caption{Upper left: $70$~$\mu$m - $1.4$~GHz correlation. Upper right: $100$~$\mu$m - $1.4$~GHz correlation.
    Lower left: $160$~$\mu$m - $1.4$~GHz correlation. Lower right: TIR - $1.4$~GHz correlation.
    Colored symbols: model galaxies. Black solid and dotted lines: model linear regression.
    Grey dots: observations.
  \label{fig:galaxies_FRC_vrotDifferentForPhibbs_nosecnowind_3}}
\end{figure*}

The monochromatic ($70,\ 100,\ 160$~$\mu$m) and total IR - radio correlations of all four samples are shown in 
Fig.~\ref{fig:galaxies_FRC_vrotDifferentForPhibbs_nosecnowind_3}.
We calculated the slopes and offsets of the correlation by using an outlier-resistant bisector fit.
The results can be found together with the correlation scatter in Fig.~\ref{fig:galaxies_FRC_vrotDifferentForPhibbs_nosecnowind_3}.

The exponents derived from the bisector fits of the monochromatic IR - radio correlations increase with increasing wavelength from $1.00$ at $70$~$\mu$m
to $1.15$ at $100$~$\mu$m, and $1.38$ at $160$~$\mu$m. The exponent of the TIR - radio correlation is $1.08$.
These slopes are consistent with those derived with a Bayesian approach whose uncertainties are about $0.05$ (Table~\ref{tab:corrs}).
The corresponding exponents found by Molnar et al. (2021) are $1.01 \pm 0.01$, $1.05 \pm 0.09$, and $1.17 \pm 0.13$ at $60$, $100$, and $160$~$\mu$m.
That of the TIR - radio correlation is $1.11 \pm 0.01$. There is thus agreement within $0.2$, $1.0$, $1.5$, and $0.6\sigma$ of the joint 
uncertainty of model and data.
The associated scatters of the data around the power-law correlation are $\sim 0.2$~dex for all four correlations.

Furthermore, we used the Bayesian approach to linear regression with errors in both directions (Kelly et al. 2007).
We assumed uncertainties on the TIR and radio luminosities of $0.2$~dex for the local galaxies and $0.3$~dex for the 
high-z galaxies.
For a direct comparison we also calculated the slopes and offsets for the data of Yun et al. (2001; uncertainties of $0.05$~dex in both directions) 
and Molnar et al. (2021; symmetrized mean TIR luminosity uncertainties).
Since small deviations of the correlation slope lead to large deviations of the offset at $\log(L_{\rm IR})=0$, 
we decided to calculate the offsets at an infrared luminosity of $10^{10}$~L$_{\odot}$. The resulting slopes and offsets derived by the Bayesian approach
are presented in Table~\ref{tab:corrs}.
\begin{table*}[!ht]
      \caption{Correlation fits with Bayesian approach.}
         \label{tab:corrs}
      \[
         \begin{tabular}{lcccc}
           \hline
            & x-axis & y-axis & slope & offset at $L=10^{10}$~L$_{\odot}$ \\
           \hline
-           Yun et al. (2001) & 60~$\mu$m & 1.4~GHz & $0.99 \pm 0.01$ &   $21.76 \pm 0.01$ \\
           model (all samples) & 70~$\mu$m & 1.4~GHz & $0.97 \pm 0.04$ & $21.90 \pm 0.07$ \\
           model (local samples) & 70~$\mu$m & 1.4~GHz & $0.96 \pm 0.06$ & $21.93 \pm 0.08$ \\
           \hline
           Molnar et al. (2001) & TIR & 1.4~GHz & $1.07 \pm 0.01$ & $21.45 \pm 0.01$ \\
           model (all samples) & TIR & 1.4~GHz & $1.09 \pm 0.05$ & $21.31 \pm 0.09$ \\
           \hline
           Bell et al. (2003) & TIR & 1.4~GHz & $1.11 \pm 0.03$ & $21.36 \pm 0.03$ \\
           model (local samples) & TIR & 1.4~GHz & $1.12 \pm 0.08$ & $21.32 \pm 0.11$ \\
           \hline
           Basu et al. (2015) & TIR & 1.4~GHz & $1.11 \pm 0.04$ & - \\
           \hline
         \end{tabular}
      \]
\end{table*}
There is agreement between the slopes within $0.5 \sigma$ and between the offsets within $2 \sigma$ of the joint 
uncertainty of model and data for both datasets.

\begin{figure}
  \centering
  \resizebox{\hsize}{!}{\includegraphics{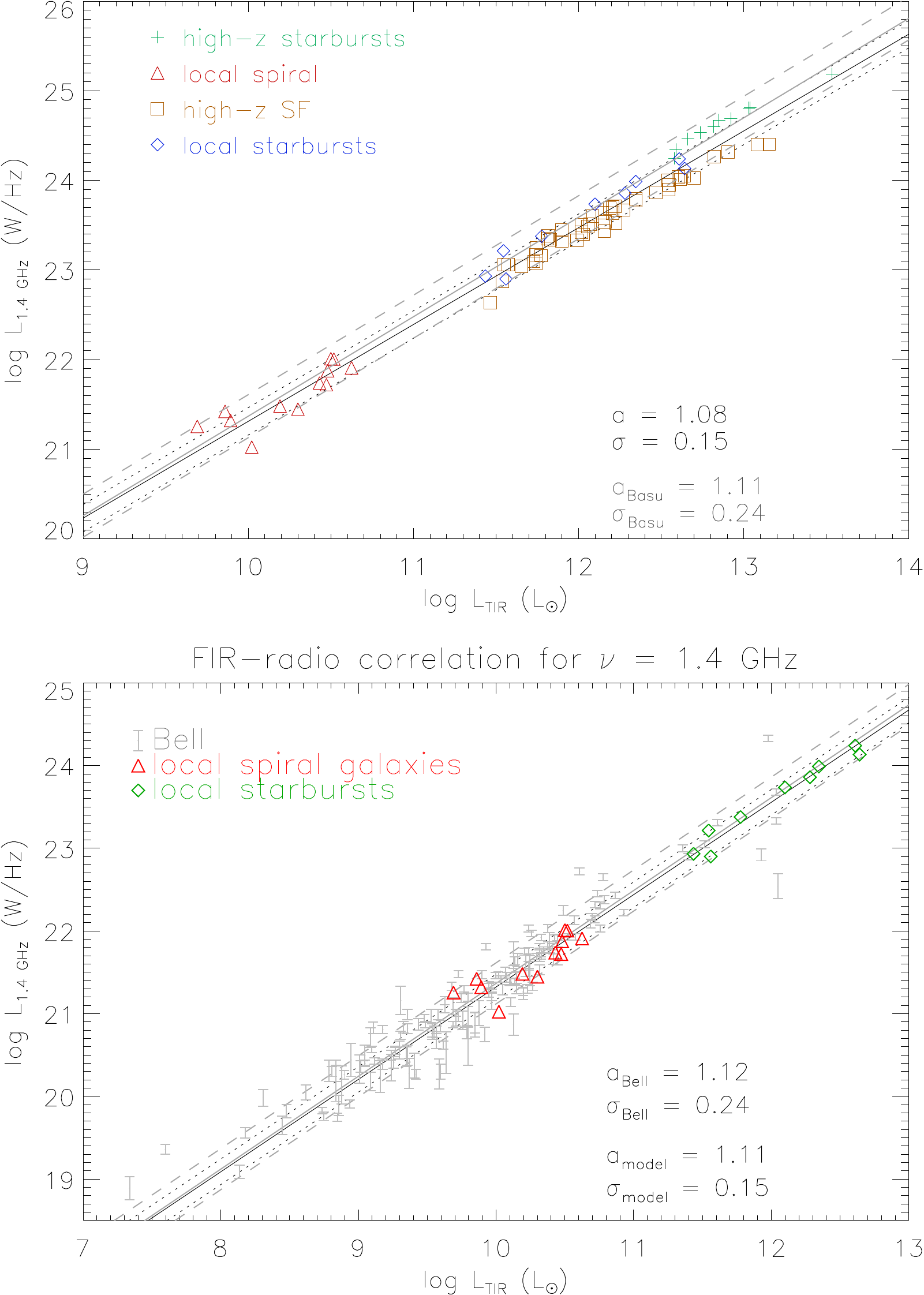}}
  \caption{TIR - $1.4$~GHz correlations. Colored symbols: model galaxies. Black solid and dotted lines:
    model linear regression. Upper panel: grey solid and dotted lines: observed linear regression (Basu et al. 2015).
    Lower panel: grey error bars: data from Bell (2003). Grey solid and dashed lines: observed linear regression (Bell 2003).
  \label{fig:galaxies_FRC_vrotDifferentForPhibbs_nosecnowind_4}}
\end{figure}

Bell (2003) assembled a diverse sample of local galaxies from the literature with far-ultraviolet (FUV), optical, infrared (IR), and 
radio luminosities and found a nearly linear radio-IR correlation. The left panel of  
Fig.~\ref{fig:galaxies_FRC_vrotDifferentForPhibbs_nosecnowind_4} shows the direct comparison between our local model galaxies
(spirals and low-z starbursts) and the compilation of Bell (2003). As before, we assumed an uncertainties of $0.2$~dex
for the model TIR and radio luminosities.
There is agreement between the slopes within $0.1 \sigma$ and between the offsets within $0.4 \sigma$ of the joint 
uncertainty of model and data (Table~\ref{tab:corrs}).

Basu et al. (2015) studied the radio - TIR correlation in starforming galaxies chosen from the PRism MUltiobject Survey up to 
redshift of $1.2$ in the XMM-LSS field employing the technique of image stacking. They found a exponent of the TIR - $1.4$~GHz
correlation of  $1.11 \pm 0.04$. The upper left panels of Fig.~\ref{fig:galaxies_FRC_vrotDifferentForPhibbs_nosecnowind_4} shows 
the direct comparison between our model galaxies (local and high-redshift) and those of Basu et al. (2015) show comparable exponents and scatters. 
There is agreement between the slopes within $0.4 \sigma$ of the joint uncertainty of model and data (Table~\ref{tab:corrs}).

The radio-FIR correlation is generally quantified via the parameter $q_{\rm IR}$ defined as $q_{\rm IR}=\log(L_{\rm IR}/L_{\rm radio})$.
Following Helou et al. (1985) we define for the bolometric case
\begin{equation}
q_{\rm IR}=\log\big(\frac{L_{\rm IR}({\rm W})}{3.75 \times 10^{12}~{\rm Hz}}\big)-\log\big(L_{\rm 1.4GHz}({\rm W\,Hz}^{-1})\big)\ .
\end{equation}
We compared the TIR luminosity integrated between $8$ and $1000$~$\mu$m,
the FIR luminosity, which is typically integrated between $40$ and $120$~$\mu$m (for a consistent comparison
with our model we integrated the model IR SEDs between $70$~$\mu$m and $160$~$\mu$m), and the monochromatic
luminosity at $70$~$\mu$m.

We compiled IR-to-radio luminosity ratios for different galaxy types from the literature (Table~\ref{tab:qtirlit} and compared them to
the values of our model $q_{\rm IR}$ (Table~\ref{tab:galq}). 
\begin{table*}[!ht]
      \caption{Galaxy samples for the calculation of IR-to-radio luminosity ratios.}
         \label{tab:qtirlit}
      \[
         \begin{tabular}{lccc}
           \hline
            sample & galaxy type & luminosity range & redshift \\
           \hline
           Jarvis et al. (2010) & local galaxies & $10^{10} \le L_{\rm TIR} \le 10^{11}$~L$_{\odot}$  & $z < 0.5$ \\
           Molnar et al. (2021) & local galaxies & $10^{10} \le L_{\rm TIR} \le 10^{12}$~L$_{\odot}$ & $z < 0.2$ \\
           Sargent et al. (2010) & local starburst galaxies & $L_{\rm TIR} > 10^{12}$~L$_{\odot}$ & $z \sim 0.1$ \\
           Farrah et al. (2003) & local starburst galaxies & $L_{\rm TIR} > 10^{12}$~L$_{\odot}$ & $z < 0.1$ \\
           Yun et al. (2001) & local galaxies & $10^{9} \le L_{60 \mu m} \le 10^{12}$~L$_{\odot}$ & $z \le 0.05$ \\
           Magnelli et al. (2015) & local galaxies & $10^{9} \la L_{\rm TIR} \la 10^{11}$~L$_{\odot}$ & $z \le 0.2$ \\
           Delhaize et al. (2017) & high-z starforming galaxies & $10^{11} \le L_{\rm TIR} \le 10^{12}$~L$_{\odot}$ & $z \sim 1$ \\
           Delvecchio et al. (2021) & high-z starforming galaxies & $10^{11} \le L_{\rm TIR} \le 5 \times 10^{12}$~L$_{\odot}$ & $z \sim 1$ \\
           Algera et al. (2020) & high-z starburst galaxies & $10^{12} \le L_{\rm TIR} \le 10^{13}$~L$_{\odot}$ & $2 \la z \la 3$ \\
           Thomson et al. (2019) & high-z starburst galaxies & & $2 \la z \la 3$ \\
           Thomson et al. (2014) & high-z starburst galaxies &  $10^{12} \le L_{\rm TIR} \le 10^{13}$~L$_{\odot}$ & $2 \la z \la 3$ \\
           Basu et al. (2015) & all & $10^{10} \le L_{\rm TIR} \le 10^{12}$~L$_{\odot}$ & $0 \le z \le 1$ \\
           \hline
           model  & local spirals &  $10^{10} \la L_{\rm TIR} \la 10^{11}$~L$_{\odot}$ & $z \sim 0$ \\
           model & low-z starburst galaxies & $10^{12} \la L_{\rm TIR} \la 10^{13}$~L$_{\odot}$ & $z \sim 0$ \\
           model & high-z starforming galaxies & $3 \times 10^{11} \la L_{\rm TIR} \la 3 \times 10^{12}$~L$_{\odot}$ & $1 \le z \le 2$ \\
           model & high-z starburst galaxies & $3 \times 10^{12} \la L_{\rm TIR} \la 10^{13}$~L$_{\odot}$ & $z \sim 2$ \\           
           \hline
         \end{tabular}
      \]
\end{table*}
We divided our samples in local galaxies (spirals and low-z starbursts) and total sample 
(spirals, low-z starbursts, high-z starforming, and high-z starburst galaxies). 
The literature samples are relatively well matched to the model samples in terms of IR luminosity and redshift ranges (Table~\ref{tab:qtirlit}.
In the cases where two groups determined $q_{\rm IR}$ independently for a given galaxy sample, the values are consistent
except for the TIR - radio correlation of high-z starburst galaxies, where the difference exceeds $0.3$~dex between the value of
Thomson et al. (2019) and that of Algera et al. (2020).  
 
As expected, the IR-to-radio luminosity ratios of the different models are comparable for the local spirals.
Moreover, the inclusion of a wind increases $q_{\rm IR}$ whereas the inclusion of secondaries decreases $q_{\rm IR}$.
The magnetic field strength recipe involving the gas surface density (model Bsigma) does not significantly change $q_{\rm IR}$
whereas the recipe involving the gas density (model Brho) leads to the highest $q_{\rm IR}$ values for all galaxy
samples except the local spirals and high-z starburst galaxies. Since these values are significantly higher than the
observed ones, we can discard model Brho. 

The TIR-to-radio luminosity ratios of our fiducial model are consistent (within $0.2$~dex or $2 \sigma$) with observations for
the local, high-z starforming galaxies, and the total sample.
The TIR-to-radio luminosity ratios of the high-z starburst galaxies is consistent within $0.5 \sigma$ with that
of Algera et al. (2020) and Thomson et al. (2019). However, it is $6 \sigma$ higher than the value found by Thomson et al. (2014).
The  model TIR-to-radio luminosity ratios of the low-z starbursts deviate from the observed value by $0.3$~dex or $3 \sigma$.
Based on the observed high $q_{\rm TIR}$ of the low-z starbursts, a median velocity/fast galactic wind is needed in the absence/presence
of secondary CR electrons. 

The FIR-to-radio luminosity ratios of our fiducial model are consistent with observations (within $0.7 \sigma$) for
the local galaxies and the high-z starforming galaxies.
The  model FIR-to-radio luminosity ratios of the high-z starburst galaxies deviate from the observed value by $0.3$~dex 
(or $2 \sigma$).
The $70$~$\mu$m-to-radio luminosity ratios of our fiducial model are consistent with observations (within $1 \sigma$) for
the low-z starbursts, local galaxies, and the total sample.
The production of secondary CR electrons, which decreases $q_{\rm IR}$, is not needed in the framework of our model.
We conclude that our fiducial model of the main sequence starforming galaxies is consistent with the available IR-to-radio luminosity 
ratios determined by observations. The low-z starburst models probably need a galactic wind.
\begin{table*}[!ht]
      \caption{The IR-to-radio ratio.}
         \label{tab:galq}
      \[
         \begin{array}{lcccccc}
           \hline
             &  {\rm local\ spirals} & {\rm low-z\ starbursts} & {\rm local\ galaxies} & {\rm high-z\ SF\ galaxies} & {\rm high-z\ starburst\ galaxies} & {\rm total} \\
           \hline
           {\rm TIR} & & & & & & \\
           \hline
           {\rm observed} & 2.54^{9} \pm 0.01 & 2.67^{1} \pm 0.07 & 2.54 \pm 0.27^{2a,2b} & \sim 2.4-2.6^3 & 2.20^{4a,4b} \pm 0.06 & 2.50^5 \pm 0.24 \\
           {\rm observed} & & & & & 2.56^{4c} \pm 0.05 & \\
           {\rm fiducial} & 2.70 \pm 0.11 & 2.40 \pm 0.06 & 2.51 \pm 0.17 & 2.58 \pm 0.05 & 2.23 \pm 0.02 & 2.56 \pm 0.16 \\
           {\rm wind} & 2.70 \pm 0.12 & 2.51 \pm 0.03 & 2.55 \pm 0.16 & 2.77 \pm 0.08 & 2.29 \pm 0.03 & 2.70 \pm 0.20 \\
           {\rm sec+wind} & 2.52 \pm 0.12 & 2.23 \pm 0.06 & 2.37 \pm 0.21 & 2.59 \pm 0.09 & 2.06 \pm 0.04 & 2.54 \pm 0.24 \\
           {\rm sec+fastwind} & 2.53 \pm 0.12 & 2.47 \pm 0.10 & 2.49 \pm 0.16 & 2.84 \pm 0.11 & 2.25 \pm 0.04 & 2.75 \pm 0.27 \\
           {\rm exp} & 2.61 \pm 0.11 & 2.19 \pm 0.05 & 2.40 \pm 0.24 & 2.36 \pm 0.04 & 1.90 \pm 0.01 & 2.35 \pm 0.21 \\
           {\rm Bsigma} & 2.59 \pm 0.17 & 2.31 \pm 0.06 & 2.45 \pm 0.24 & 2.44 \pm 0.04 & 2.19 \pm 0.05 & 2.43 \pm 0.16 \\
           {\rm Brho} & 2.58 \pm 0.18 & 2.96 \pm 0.10 & 2.81 \pm 0.24 & 2.97 \pm 0.08 & 2.91 \pm 0.09 & 2.92 \pm 0.20 \\
       \noalign{\smallskip}
       \hline
       \noalign{\smallskip}
       {\rm FIR} & & & & & & \\
       \hline
       {\rm observed} &  &  & 2.34^6 \pm 0.01 &  &  &  \\
       {\rm observed} &  &  & 2.35^7 \pm 0.08 & 2.17^7 \pm 0.08 & 2.11^7 \pm 0.08 &  \\
       {\rm fiducial} & 2.40 \pm 0.11 & 2.07 \pm 0.03 & 2.22 \pm 0.19 & 2.21 \pm 0.06 & 1.96 \pm 0.01 & 2.20 \pm 0.15 \\
       {\rm wind} & 2.41 \pm 0.12 & 2.20 \pm 0.08 & 2.24 \pm 0.18 & 2.40 \pm 0.11 & 2.02 \pm 0.02 & 2.34 \pm 0.20 \\
       {\rm sec+wind} & 2.23 \pm 0.13 & 1.96 \pm 0.12 & 2.03 \pm 0.23 & 2.26 \pm 0.11 & 1.77 \pm 0.04 & 2.16 \pm 0.24 \\
       {\rm sec+fastwind} & 2.24 \pm 0.13 & 2.18 \pm 0.14 & 2.18 \pm 0.19 & 2.47 \pm 0.12 & 1.94 \pm 0.06 & 2.38 \pm 0.27 \\
       {\rm exp} & 2.32 \pm 0.11 & 1.90 \pm 0.10 & 2.05 \pm 0.26 & 2.00 \pm 0.07 & 1.65 \pm 0.01 & 1.99 \pm 0.21 \\
       {\rm Bsigma} & 2.29 \pm 0.17 & 1.99 \pm 0.04 & 2.17 \pm 0.26 & 2.09 \pm 0.05 & 1.93 \pm 0.02 & 2.06 \pm 0.17 \\
       {\rm Brho} & 2.31 \pm 0.19 & 2.65 \pm 0.06 & 2.51 \pm 0.23 & 2.62 \pm 0.06 & 2.64 \pm 0.08 & 2.61 \pm 0.17 \\
       \noalign{\smallskip}
       \hline
       \noalign{\smallskip}
       70\ \mu{\rm m} & & & & & & \\
       \hline
       {\rm observed} &  &  & 2.24^6 \pm 0.13  &  &  &  \\
       {\rm observed} &  & 2.29^8 \pm 0.06 & 2.10^{2b} \pm 0.01 &  &  & 2.23^5 \pm 0.25 \\
       {\rm fiducial} & 2.28 \pm 0.13 & 2.33 \pm 0.03 & 2.30 \pm 0.22 & 2.41 \pm 0.06 & 2.19 \pm 0.03 & 2.35 \pm 0.16 \\
       {\rm wind} & 2.30 \pm 0.13 & 2.41 \pm 0.03 & 2.33 \pm 0.24 & 2.59 \pm 0.10 & 2.25 \pm 0.01 & 2.51 \pm 0.23 \\
       {\rm sec+wind} & 2.13 \pm 0.13 & 2.15 \pm 0.07 & 2.13 \pm 0.19 & 2.43 \pm 0.10 & 1.99 \pm 0.03 & 2.35 \pm 0.25 \\
       {\rm sec+fastwind} & 2.17 \pm 0.13 & 2.40 \pm 0.10 & 2.20 \pm 0.24 & 2.65 \pm 0.12 & 2.18 \pm 0.01 & 2.56 \pm 0.31 \\
       {\rm exp} & 2.20 \pm 0.12 & 2.07 \pm 0.05 & 2.13 \pm 0.19 & 2.19 \pm 0.06 & 1.85 \pm 0.04 & 2.16 \pm 0.17 \\
       {\rm Bsigma} & 2.23 \pm 0.20 & 2.24 \pm 0.03 & 2.23 \pm 0.26 & 2.28 \pm 0.06 & 2.15 \pm 0.03 & 2.26 \pm 0.16 \\
       {\rm Brho} & 2.25 \pm 0.23 & 2.90 \pm 0.11 & 2.50 \pm 0.43 & 2.82 \pm 0.08 & 2.86 \pm 0.13 & 2.79 \pm 0.29 \\
       \hline
        \end{array}
      \]
\begin{tablenotes}{}{}
\item $^{1}$: Sargent et al. (2010), $^{2a}$: Jarvis et al. (2010), $^{2b}$: Molnar et al. (2021), $^3$: Delhaize et al. (2017), Delvecchio et al. (2021)
\item $^{4a}$: Algera et al. (2020), $^{4b}$: Thomson et al. (2019), $^{4c}$: Thomson et al. (2014), $^5$: Basu et al. (2015), $^6$: Yun et al. (2001),
\item $^7$: Magnelli et al. (2015), $^8$: Farrah et al. (2003), $^9$: Molnar et al. (2021)
\end{tablenotes}
\end{table*}

\subsection{The SFR-radio correlation \label{sec:sfrradcorr}}

\begin{figure}
  \centering
  \resizebox{\hsize}{!}{\includegraphics{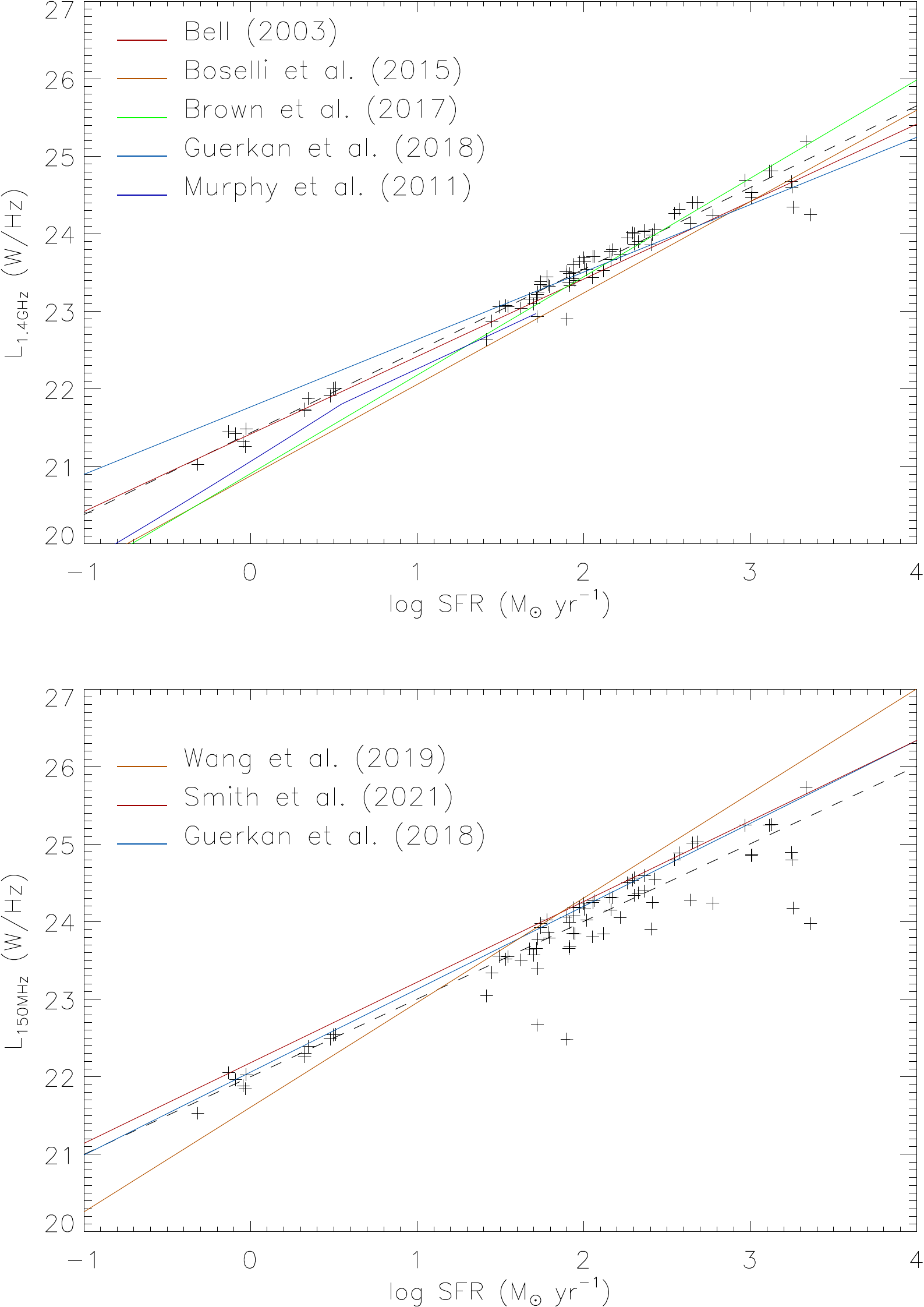}}
  \caption{Upper panel: SFR - $1.4$~GHz correlation. Lower panel: SFR - $150$~MHz correlation.
    Colored lines: observed correlations. Plus signs: model galaxies. 
  \label{fig:galaxies_FRC_vrotDifferentForPhibbs_nosecnowind_5}}
\end{figure}

The model SFR - $1.4$~GHz and SFR - $150$~MHz correlations are presented in Fig.~\ref{fig:galaxies_FRC_vrotDifferentForPhibbs_nosecnowind_5}
together with the observed correlations. The SFRs were derived using different methods and the correlations were derived
for different samples (Table~\ref{tab:slopes}). The exponents of the SFR - radio correlations based on SED
fitting methods are smaller than the exponents based on IR luminosities and extinction-corrected H$\alpha$.
For SFRs derived through extinction-corrected H$\alpha$ and IR luminosities the exponents tend to unity if low-z starbursts with
$\dot{M}_* > 10$~M$_{\odot}$yr$^{-1}$ are included in the sample. It appears that the exponent of the SFR - $150$~MHz correlation
is somewhat steeper than that of the SFR - $1.4$~GHz correlation.
\begin{table*}[!ht]
      \caption{Exponents of the SFR - radio correlation.}
         \label{tab:slopes}
      \[
         \begin{tabular}{llcc}
           \hline
           $\nu=1.4$~GHz & exponent & SFR & sample \\
           \hline
           Bell (2003) & $1.3$ ($L_{\rm rad} < 6.4 \times 10^{21}$~L$_{\odot}$) & ext-corr H$\alpha$ &  \\
           Bell (2003) & $1.0$ ($L_{\rm rad} > 6.4 \times 10^{21}$~L$_{\odot}$) & ext-corr H$\alpha$ &  \\
           Murphy et al. (2011) & $1.0$ & IR & high-z galaxies included \\
           Heesen et al. (2014) & $1.11 \pm 0.08$ & $24$~$\mu$m + FUV & $\dot{M}_* \la 10$~M$_{\odot}$yr$^{-1}$ \\ 
           Boselli et al. (2015) & $1.18$ & ext-corr H$\alpha$ & $\dot{M}_* \la 10$~M$_{\odot}$yr$^{-1}$ \\
           Brown et al. (2017) & $1.27 \pm 0.03$ & ext-corr H$\alpha$ & $\dot{M}_* \la 10$~M$_{\odot}$yr$^{-1}$ \\
           G\"urkan et al. (2018) & $0.87 \pm 0.01$ & SED fitting & $\dot{M}_* \la 10$~M$_{\odot}$yr$^{-1}$ \\
           model (all samples) & $1.05 \pm 0.04$ & \\
           \hline
           $\nu=150$~MHz & & & \\
           \hline
           G\"urkan et al. (2018) & $1.07 \pm 0.01$ & SED fitting & $\dot{M}_* \la 10$~M$_{\odot}$yr$^{-1}$ \\
           Wang et al. (2019) & $1.35 \pm 0.06$ & ext-corr H$\alpha$ & \\
           Smith et al. (2021) & $1.04 \pm 0.01$ & SED fitting & \\
           model (all samples) & $0.99 \pm 0.05$ & \\ 
           \hline
         \end{tabular}
      \]
\end{table*}

We found an exponent of the model SFR - $1.4$~GHz correlation for the combined sample of $1.05 \pm 0.04$. 
The normalization is $\log\left(L_{\rm 1.4GHz}/({\rm W\,Hz}^{-1})\right)=21.43 \pm 0.08$ with an additional systematic uncertainty of 
$\pm 0.15$ stemming from the comparison
between the model and observed IR-radio offsets measured by Yun et al. (2001) and Molnar et al. (2021).
The slope is close to that of Heesen et al. (2014), somewhat steeper than those
of Bell (2003) and Murphy et al. (2011), and shallower than that of Boselli et al. (2015). However, the model radio luminosities
are a factor of two higher than the radio luminosities observed by Boselli et al. (2015). Alternatively, dividing the
SFRs of these authors by a factor of two would make the model and observed correlations identical.
The model slope is close to ($1 \sigma$) that Heesen et al. (2014), lower ($4 \sigma$) than that of Brown et al. (2017), and significantly
higher ($19 \sigma$) than that of  G\"urkan et al. (2018).

The exponent of the model SFR - $150$~MHz correlation for the combined sample is $0.99 \pm 0.05$. This is consistent with the exponents
found by G\"urkan et al. (2018; $1.6 \sigma$) and Smith et al. (2021; $1 \sigma$). 
The slope found by Wang et al. (2019) is higher by $5 \sigma$ than our model slope.
The normalization of the model correlation is $\log\left(L_{\rm 150MHz}/({\rm W\,Hz}^{-1})\right)=21.93 \pm 0.1$.

We conclude that the observed SFR - radio correlation can be reproduced by our fiducial model in a reasonable way (within $\sim 4 \sigma$).
The model exponents are very close to one. 

\section{Discussion \label{sec:discussion}}

The agreement between the slopes of the model and observed IR-radio correlation ($2 \sigma$) is better than that between the 
slopes of the model and observed SFR-radio correlation ($4 \sigma$). We note that, whereas the SFR is an input quantity, 
the IR emission is calculated by the model. The less good agreement for the SFR-radio correlation is at least partly due to the relatively large 
scatter of the different measurements ($1.10 \pm 0.16$ at $1.4$~GHz and $1.15 \pm 0.17$ at $150$~MHz; Table~\ref{tab:slopes}).
This observational scatter should be decreased by about a factor of two before we can think about improvements of the model.
Possible improvements are a better inclusion of radio halos (Eq.~\ref{eq:lnu8}; Krause et al. 2018), a more sophisticated description of the
vertical CR electron diffusion (Eq.~\ref{eq:tdiffe}), the explicit inclusion of CR protons, and a better description of the IR emission of the warm ISM.

G\"urkan et al. (2018), Smith et al. (2021), and Delvecchio et al. (2021) found a mass-dependent IR - radio correlation.
Galaxies of higher masses have lower IR-to-radio luminosity ratios. Unfortunately, the mass range of our model galaxy samples 
is not broad enough to show a mass dependence of $q_{\rm IR}$.

In Sect.~\ref{sec:IRradioseds} it was shown that our fiducial model overpredicts the radio continuum emission of 25\,\% of the low-z starbursts.
Our low-z starburst sample ($\log \left( <L_{\rm TIR}/L_{\odot}> \right)=12$)
has a lower model TIR-to-radio luminosity ratio than the local spiral sample (Sect.~\ref{sec:IRradiocorrelation}).
This is contrary to the observed TIR-to-radio luminosity ratios, which are higher for the low-z starbursts than for the local spiral galaxies.
For the high-z starburst galaxies, the situation is less clear. Whereas our fiducial model overpredicts the radio continuum emission
of 35\,\% of the high-z starburst galaxies, the model IR-to-radio luminosity ratios are significantly higher than those found 
by Thomson et al. (2014) but comparable to those found by Thomson et al. (2019) and Algera et al. (2020) (Table~\ref{tab:galq}). 
The latter authors ascribe the difference with respect to Thomson et al. (2014) to the fact that they used a stacking technique, 
which allowed them to reach lower IR and radio luminosities.
We might observe such a trend in our low-z starburst model compared to observations. 
However, the high-z starburst galaxies of Table~\ref{tab:gbzk} have rather low IR-to-radio 
luminosity ratios ($q_{\rm TIR}=2.2$) despite their high TIR luminosities ($\log \left( <L_{\rm TIR}/L_{\odot}> \right)=13$).
Significantly higher TIR-to-radio luminosity ratios can be achieved via fast galactic winds in the absence of secondary CR electrons.

Models including secondary CR electrons are also viable for low-z starbursts and high-z galaxies but only in the presence of fast
galactic winds (Table~\ref{tab:galq}). The simple prescription of the advection timescale based on the rotation velocity 
(Eq.~\ref{eq:wind}) is not sufficient to yield model radio luminosities comparable to observations. 
We note that the wind velocities measured by Rupke et al. (2002) of $\sim 500$~km\,s$^{-1}$ correspond to a medium 
velocity wind ($v_{\rm wind} \sim \sqrt{2} v_{\rm rot}$). However, the low-z starbursts models with secondary CR electrons
need a ten times higher, fast wind to reproduce observations (Table~\ref{tab:galq}).

As stated in Sect.~\ref{sec:introduction}, non-calorimeter models often have to involve a conspiracy to maintain the 
tightness of the FIR - radio correlation. The fiducial models closest to CR electron calorimeters are
the models of the starburst galaxies (low-z starbursts and high-z starburst galaxies; right panels of Fig.~\ref{fig:average_loss_timescales2}). 
The situation changes if fast winds are added to the models. Lacki et al. (2010) stated that IC cooling alone is very quick in starbursts, 
implying that electrons cannot escape from these galaxies before losing most of their energy (Condon
et al. 1991; Thompson et al. 2006). In our starburst samples (low-z starbursts and high-z starburst galaxies, upper panels of 
Fig.~\ref{fig:average_loss_timescales1} and \ref{fig:average_loss_timescales2}) this is not the case. 
The synchrotron timescale is much smaller than the IC timescale. This is due to equipartition between the
turbulent kinetic and magnetic field energy densities. Not only the density is enhanced in the starburst
galaxies but also the turbulent velocity dispersion (Downes \& Solomon 1998, Genzel et al. 2010, Tacconi et al.  2013, Vollmer et al. 2017).
This is the reason why our model starburst galaxies can be considered as close to CR electron calorimeters.
On the other hand, the model spiral galaxies and high-z starforming galaxies are not CR electron calorimeters.
In both galaxies, energy losses due to bremsstrahlung and IC cooling are important.

\begin{figure*}
  \centering
  \resizebox{\hsize}{!}{\includegraphics{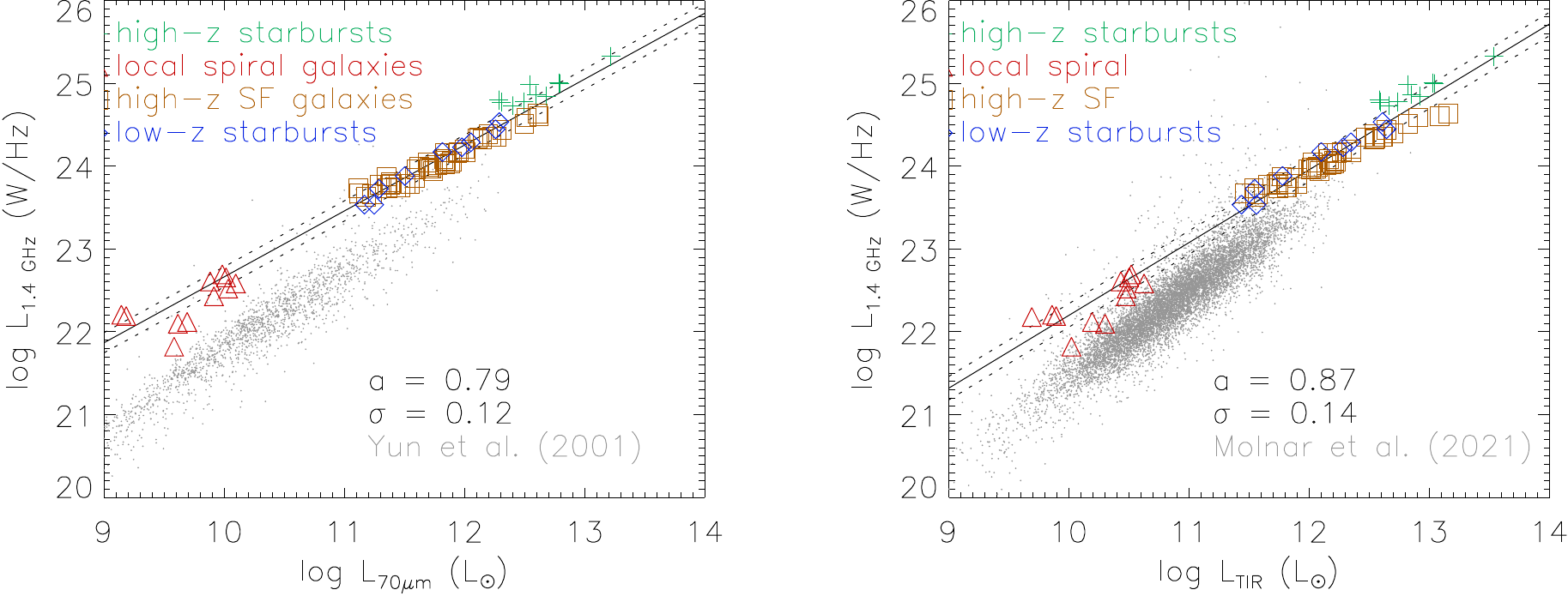}}
  \caption{Calorimeter models. TIR - $1.4$~GHz correlation.
    Colored symbols: model galaxies. Black solid and dotted lines: model linear regression.
    Grey dots: observations.
  \label{fig:galaxies_FRC_vrotDifferentForPhibbs_calorimeter}}
\end{figure*}

To quantify the effect of the different CR electron energy losses, we made model calorimeter calculations
by setting $t_{\rm diff}=t_{\rm wind}=t_{\rm IC}=t_{\rm brems}=t_{\rm ion}=0$. The resulting IR - radio correlations are
shown in Fig.~\ref{fig:galaxies_FRC_vrotDifferentForPhibbs_calorimeter} and can be directly compared to
the upper left and lower right panels of Fig.~\ref{fig:galaxies_FRC_vrotDifferentForPhibbs_nosecnowind_3}.
As expected, the radio continuum luminosities of all galaxies increase, those of the high-z starburst galaxies by $\sim 0.3$~dex,
those of the local spiral galaxies by $\sim 0.8$~dex. Most importantly, the slope of the correlation
flattens ($0.9$ instead of $1.1$ at $70$~$\mu$m and $1.0$ instead of $1.2$ for the TIR). 
We note that the exponent is not unity in our fiducial model because the SFR - FIR correlation is 
slightly superlinear. This slope is significantly different from the observed slope (see Table~\ref{tab:slopes}).
Furthermore, the model radio continuum SEDs of the calorimeter model have much steeper slopes than observed for
the local spiral galaxies, low-z starbursts, and high-z starburst galaxies.

For a further investigation of the influence of the different CRe energy loss times on the TIR-1.4~GHz correlation,
we set all timescales but one to zero. The slopes and normalizations of the resulting TIR-radio correlation
are presented in Table~\ref{tab:modeltimes} and Fig.~\ref{fig:modeltimes}. Advective and ionic energy losses do not play
a role in our models. The slope and normalization of our fiducial model are set by diffusion, bremsstrahlung, and
inverse Compton losses.

Averaged over sufficiently long length- and timescales, the CR distribution may achieve energy equipartition
 with the magnetic field (Beck \& Krause 2005).
It is not known whether most synchrotron sources are in equipartition between the CR particle and
magnetic field energy densities, but radio astronomers often assume so because it is physically plausible
(CRs and magnetic fields have a common source of energy, which are supernova explosions, and CRs are confined by magnetic fields) 
and permits to estimate the relativistic particle energies and the magnetic field strengths of radio sources with measured luminosities and sizes.
However, as stated by Seta et al. (2018), there is no compelling observational or theoretical reason to expect a tight correlation 
between the to particle and the magnetic field energy densities across all scales.
For a recent review on CR - magnetic field equipartition see Seta \& Beck (2019).
Most of the energy of CRs is carried by protons and heavier particles, therefore, the equipartition assumption relies on the assumption 
that relativistic electrons are distributed similarly to the heavier CR particles.
Energy equipartition then can be written as
\begin{equation}
U_{\rm B}=\frac{B^2}{8\,\pi}=(1+\eta)\,U_{\rm e}=(1+\eta)\, \int^{\gamma_1}_{\gamma_2} \gamma m_{\rm e} c^2 \,n_{\rm CRe}(\gamma)\,{\rm d}\gamma\,,
\end{equation}
where $\eta$ is the fraction of energy in heavy CR particles, $m_{\rm e}$ the electron mass, and $c$ the speed of light. 
For strong shocks in non-relativistic gas $\eta \sim 40$ (Beck \& Krause 2005). This corresponds to a proton-to-electron
number ratio of $40$ which is only a factor of $2$ lower  than the observed value (e.g., Yoshida 2008).
We caution the reader that the calculation of the CR electron energy density strongly depends on the assumed
lower energy cutoff $\gamma_2$ and on the exponent of $n_{\rm CRe}(\gamma)$.

\begin{figure*}
  \centering
  \resizebox{\hsize}{!}{\includegraphics{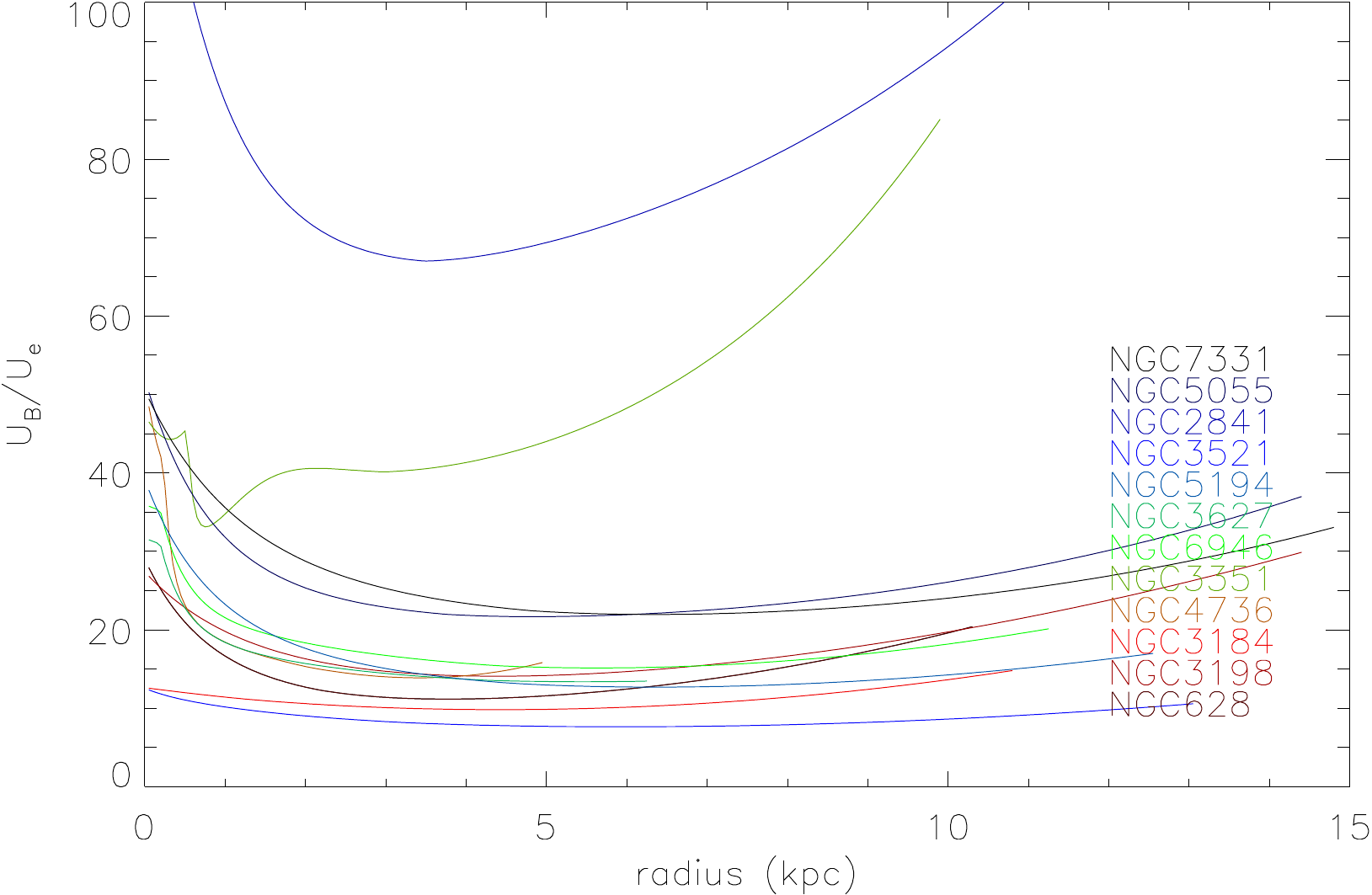}\includegraphics{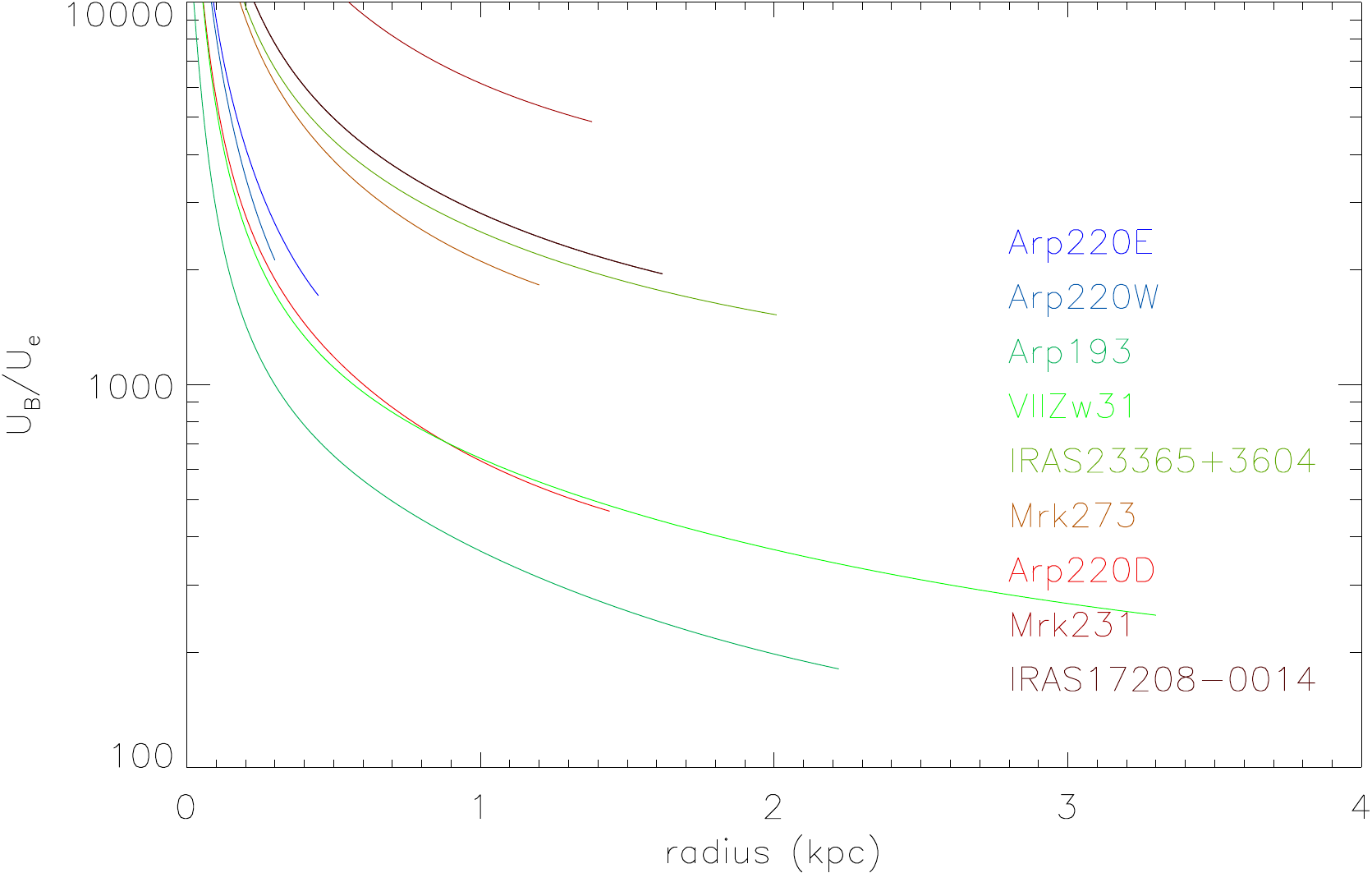}}
  \resizebox{\hsize}{!}{\includegraphics{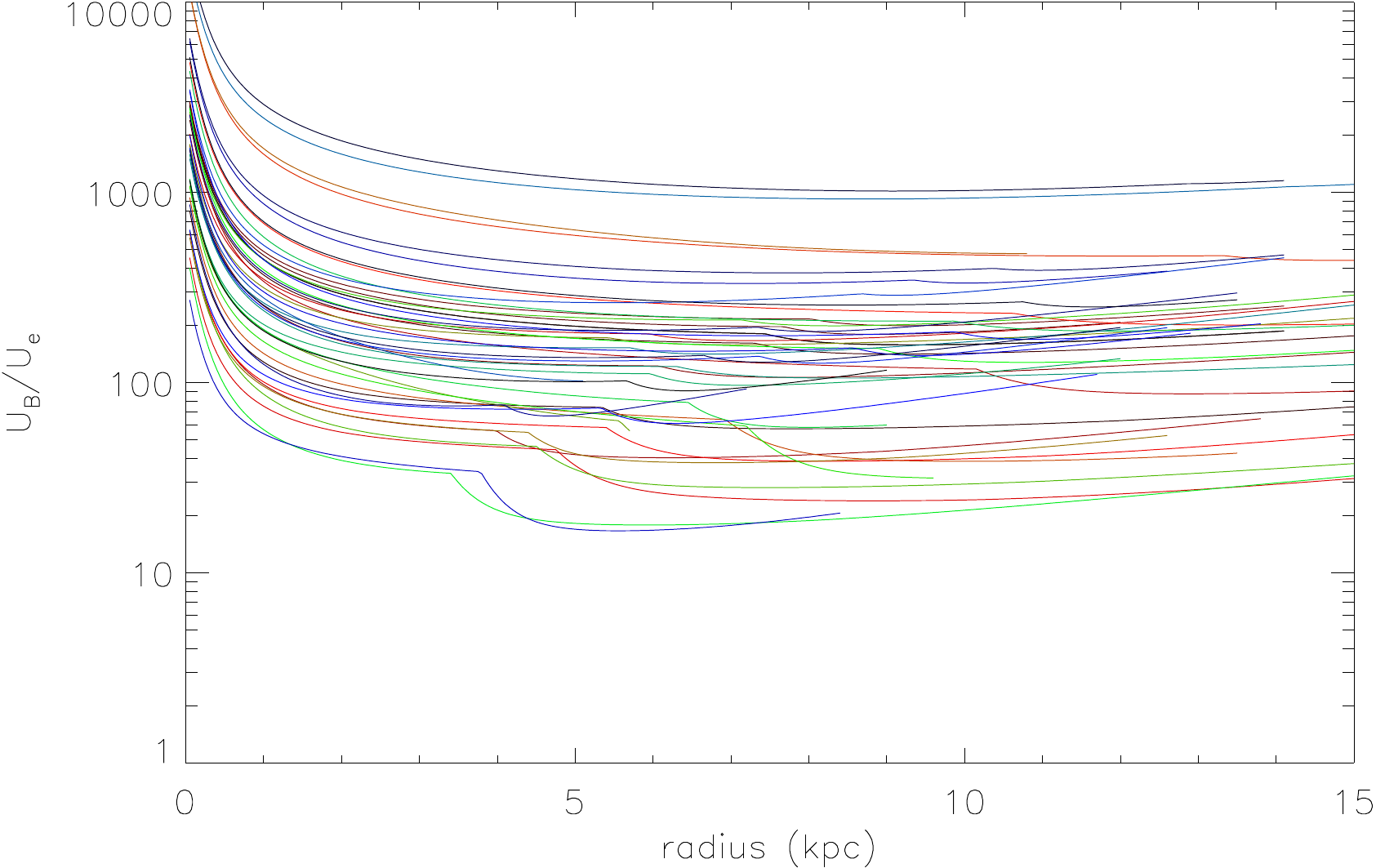}\includegraphics{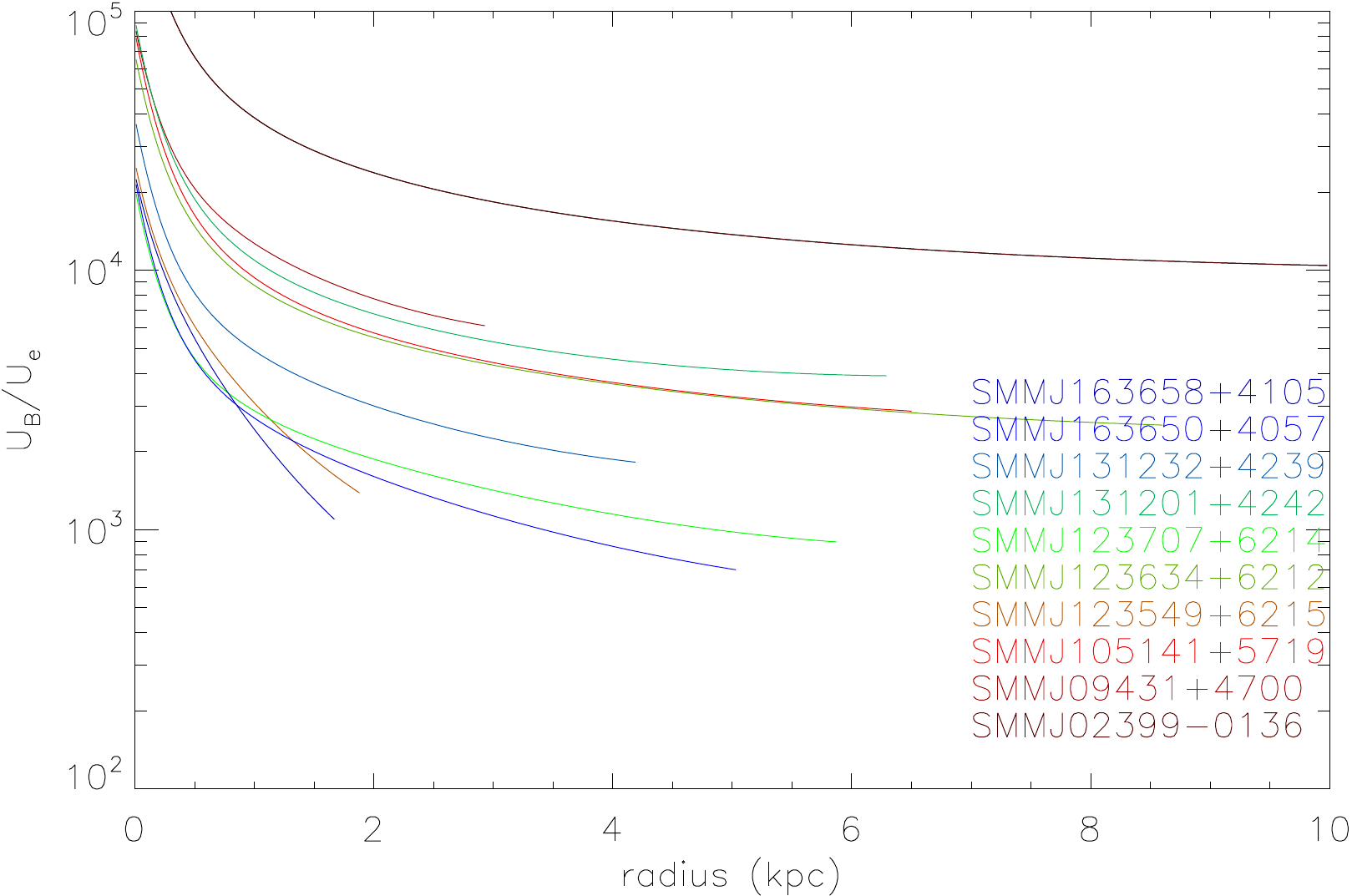}}
  \caption{Ratios between the magnetic and CR electrons energy densities. Upper left: local spiral galaxies.
    Upper right: low-z starbursts. Lower left: high-z starforming galaxies. 
    The kink in the radial profiles is caused by the sudden onset of a galactic wind. 
    Lower right: high-z starburst galaxies.
  \label{fig:equipartition}}
\end{figure*}

The ratios between the magnetic and CR electrons energy densities are presented for the different galaxy samples in
Fig.~\ref{fig:equipartition}.
The ratios $U_{\rm B}/U_{\rm e}$ of most of the local spiral galaxies are between $10$ and $20$, consistent with resonant
scattering of Alfv\'en waves within the turbulent ISM (Beck \& Krause 2005).
The majority of the high-z starforming galaxies have $U_{\rm B}/U_{\rm e} \sim 200$, which is higher than the value
predicted for strong shocks ($U_{\rm B}/U_{\rm e} \sim 40$; Beck \& Krause 2005).
The starburst galaxies have much higher ratios ($U_{\rm B}/U_{\rm e} > 100$).
This is expected because the variation of the magnetic field strengths between the
different samples is much larger than that of the CR electron densities (Fig.~\ref{fig:average_Bmagn_all}). 
Our result is consistent with the findings of Yoast-Hull et al. (2016) who found a significantly larger magnetic field energy density than 
the CR energy density in starburst galaxies. We thus conclude that energy equipartition between the
CR particles and the magnetic field approximately holds in our models of starforming galaxies.

\section{Conclusions\label{sec:conclusions}}

In galaxies not all injected energy of CR electrons is radiated via synchrotron emission meaning that galaxies
can not be treated as electron calorimeters. Multiple energy losses of CR electrons decrease the synchrotron emission:
inverse Compton losses, bremsstrahlung, diffusion of CR electrons into the galactic halo, advection of CR electrons
by galactic wind, and ionic losses. The mixture of these losses shapes the radio continuum SED.
We extended the analytical model of galactic disks of Vollmer et al. (2017) by including a simplified prescription for
the synchrotron emissivity (Eq.~\ref{eq:emfinal}). The galactic gas disks are treated as turbulent clumpy accretion disks.
The different losses are taken into account via their
characteristic timescales. The magnetic field strength is determined by the equipartition between the turbulent
kinetic and the magnetic energy densities. In this way the radio luminosities of the Vollmer et al. (2017) model galaxies were
calculated: local spiral galaxies, low-z starbursts, high-z main sequence starforming and high-z starburst galaxies.
Based on the comparison between our model galaxies and available observations we obtained the following conclusions:
\begin{enumerate}
\item
the exponents of the model log(SFR) - log($70$~$\mu$m) and log(SFR) - log(TIR) correlations are close to one ($1.09 \pm 0.02$ and $0.98 \pm 0.03$; Sect.~\ref{sec:sfrircorr}).
\item
The ratio between the magnetic field strength of the different samples is much larger than the ratio between the
CR electron densities. 
Free-free absorption mainly affects the center of the starburst galaxies (low-z starbursts and high-z starbursts) at frequencies lower than $1$~GHz  (Sect.~\ref{sec:magcr}).
\item
In local spiral galaxies and high-z starforming galaxies IC energy losses and losses due to bremsstrahlung are significant in the outer and inner 
disks, respectively. At low frequencies ($\nu \sim 150$~MHz) ionic losses become important in the inner disks.
The models of the starburst galaxies are close to calorimetric if no fast galactic winds are included (Sect.~\ref{sec:elosstimes}).
\item
The observed radio continuum SEDs of most ($\sim 70$\,\%) of the galaxies are reproduced by the fiducial model in a satisfactory way. 
Except for the local spiral galaxies, fast galactic winds can potentially make conflicting models agree with observations 
(Sect.~\ref{sec:IRradioseds}).
\item
The comparison with data of Yun et al. (2001), Bell (2003), Basu et al. (2015), and Molnar et al. (2021) shows agreement within $2\sigma$
between the model and observed IR - radio correlations.
Our fiducial model is also consistent with the available IR-to-radio luminosity ratios determined by observations. 
Only the low-z starburst models probably need a galactic wind (Sect.~\ref{sec:IRradiocorrelation}).
\item
The observed SFR - radio correlations at $150$~MHz and $1.4$~GHz can be reproduced by our fiducial model within $\sim 4 \sigma$
of the joint uncertainty of model and data for both datasets. The model exponents are $0.99 \pm 0.05$
and $1.05 \pm 0.04$ at $150$~MHz and $1.4$~GHz, respectively (Sect.~\ref{sec:sfrradcorr}).
\item
Advective and ionic energy losses do not play a significant role in our model TIR - 1.4~GHz correlations. 
The slope and normalization of our fiducial model are set by diffusion, bremsstrahlung, and inverse Compton losses.
If a CR electron calorimeter is assumed, the slope of the IR - 1.4~GHz correlation flattens $0.9$ instead of $1.1$ at $70$~$\mu$m and 
$1.0$ instead of $1.2$ for the TIR (Fig.~\ref{fig:galaxies_FRC_vrotDifferentForPhibbs_calorimeter}).
\item
Equipartition between the turbulent kinetic and magnetic field energy densities seems to be realized in the gas disks
of starforming and starburst galaxies.
\item
Energy equipartition between the CR particles and the magnetic field only approximately holds in our models of 
main sequence starforming galaxies (Fig.~\ref{fig:equipartition}).
\item 
IC losses are not dominant in the starburst galaxies because in these galaxies not only the gas density but
also the turbulent velocity dispersion is higher than in normally starforming galaxies. 
Equipartition between the turbulent kinetic and magnetic field energy densities then leads to very high magnetic field strengths
and very short synchrotron timescales.
\end{enumerate}
Our fiducial model reproduces the available IR and radio data in a satisfactory way. However, the role of CR electron
secondaries and galactic winds has still to be elucidated in the framework of our model. In particular, our simple prescription of 
the wind timescale (Eq.~\ref{eq:wind}) is not able reproduce the available data in the presence of CR electron secondaries.
The inclusion of a sample of luminous infrared galaxies bridging parameter space between local spirals and
low-z starburst galaxies will certainly be helpful.

\begin{acknowledgements}
We would like to thank R.~Beck and the anonymous referee for their comments, which helped to significantly improve the article.
\end{acknowledgements}

\clearpage

\begin{appendix}

\section{The large-scale model \label{sec:model1}}

Following Vollmer \& Leroy (2011), the ISM is considered as a single turbulent gas in vertical hydrostatic equilibrium. 
The turbulent pressure is
$p_{\rm turb} = \rho \sigmaturb^2$, where $\sigmaturb= \sqrt{\vturb^2 + \vtherm^2}$ is the total 3D velocity dispersion that 
takes into account both, the turbulent velocity dispersion $\vturb$ and a constant thermal velocity 
$v_{\rm therm} = \vtherm = 6~\rm{km~s^{-1}}$. Following Elmegreen (1989), hydrostatic pressure equilibrium is given by 
\begin{equation} \label{pressure}
p_{\rm turb} = \rho \sigmaturb^2 = \frac{\pi}{2} G \Sigma \left( \Sigma + \Sigma_\star \frac{\sigmaturb}{ \vdisp } \right)\ , 
\end{equation}
where $\Sigma$ is the total gas surface density, $\Sigma_\star$ is the stellar surface density, and $\vdisp$ is the vertical 
stellar velocity dispersion. Given the stellar surface density and the stellar length scale of the disk $l_\star$, the vertical 
stellar velocity dispersion is given according to Kregel et al. (2002)
\begin{equation} \label{vdisp}
    \vdisp = \sqrt{ 2 \pi G \Sigma_\star \frac{l_\star}{7.3}}\ .
\end{equation}

ISM turbulence is mainly maintained by energy input via supernova explosions giving rise to a turbulent driving length scale
$l_{\rm driv}$. The energy per unit time which is dissipated  by turbulence is:
\begin{equation}
    \dot{E} \simeq -\dot{E}_{\rm SN} = - \frac{\rho \nu }{2} \int \frac{\vturbb^2}{\ldriv^2}dV \ ,
\end{equation}
where $\nu$ is the viscosity of the gas defined as $\nu = \vturbb \ldriv$ with the 3D turbulent velocity dispersion 
$\vturbb = \sqrt{3}\,\vturb$. If we define the surface density of the gas as $\Sigma = \rho H$ and assume the integration over the 
volume $\int \rm{d}V = V = AH$, we can connect the energy input into the ISM by SNe directly to the SFR with the assumption of a 
constant initial mass function as:
\begin{equation} 
    \frac{\dot{E}_{SN}}{\Delta A} = \frac{\Sigma \nu }{2} \frac{\vturbb^2}{\ldriv^2} = \xi \dot{\Sigma}_\star \ ,
    \label{eq:sfr2}
\end{equation}
where $\dot{E}_{SN}$ is the energy injected by the supernovae, $\Delta A$ is the unit area and $\sigmastar$ is the star-formation rate. The factor $\xi$ relates the energy injection of supernovae to the star formation rate. It is considered as radially independent and its canonical value was estimated from observations in the Milky Way. 
In the presence of high disk mass accretion rate, the energy injection through the gain of potential energy can be important. 
In this case, Eq.~\ref{eq:sfr2} becomes:
\begin{equation}
  \label{sfr3}
  \frac{\Sigma \nu }{2} \frac{\vturbb^2}{\ldriv^2} = \xi \dot{\Sigma}_\star + \frac{1}{2 \pi} \dot{M} \Omega^2 \ ,
\end{equation}
where $\mdot$ is the mass accretion rate and $\Omega$ is the angular velocity.

The energy released into the ISM per mass turned into stars is 
\begin{equation}
\xi=\frac{\dot{N}_{\rm SN}}{\dot{M}_*}\,E^{\rm kin}_{\rm SN}\ ,
\end{equation}
where $\dot{N}_{\rm SN}$ is the number of SN per time and $E^{\rm kin}_{\rm SN}$ the kinetic energy input from a single SN.
Thornton et al. (1998) have shown by modeling SN explosions
in different environments that the kinetic energy of the remnants is about ten percent of the total SN energy irrespective of the
density and metallicity of the ambient medium. The SN energy
input into the ISM is thus $E^{\rm kin}_{\rm SN} \sim  10^{50}$~ergs. The integrated number of SNe type II in the Galaxy is taken to 
be $\dot{N}_{\rm SN} \sim 1/60$~yr$^{-1}$ (Rozwadowska et al. 2021). The Galactic star
formation rate is taken to be $M_{*}=1.6$~M$_{\odot}$yr$^{-1}$ (Licquia \& Newman 2015). 
With a kinetic to total SN energy fraction of $16$\,\% one obtains  $\xi=  9.2 \times 10^{-8}$~(pc/yr)$^2$,
a factor two higher than the value used by Vollmer \& Beckert (2003) and Vollmer \& Leroy (2011).

The clumpiness of the model implies that the density of a single gas cloud $\rho_{\rm cl}$ depends directly on the average density of the disk $\rho$. In the model, these two quantities are linked by the volume filling factor $\phiv$, such that $\rho_{\rm cl} = \phiv^{-1} \rho$. 
Following Vollmer \& Leroy (2011), the star formation rate per unit volume is given by:
\begin{equation} \label{starformation1}
\rhosfr = \phiv \rho \tauffcl^{-1} 
\end{equation}
For self-gravitating clouds with a Virial parameter of unity, the turbulent crossing time $\tauturbcl$ equals twice the free-fall time $\tauturbcl$ (Vollmer et al. 2021):
\begin{equation} \label{tautau}
    \tauturbcl =\frac{\sqrt{3}}{2}\frac{l_{\rm cl}}{ v_{\rm turb,cl}} = 2 \tauffcl = \sqrt{\frac{3\pi\phiv}{32G\rho}}\ ,
\end{equation}
where $l_{\rm cl}$ and $v_{\rm turb,cl}$ are respectively the size and the turbulent 3D velocity dispersion of a single gas cloud. Following Larson's law (Larson 1981), we can simplify the expression of the turbulent crossing time: 
\begin{equation} \label{larson}
    \frac{\sqrt{3}}{2}\frac{l_{\rm cl}}{ v_{\rm turb,cl}} =  \frac{\sqrt{3}}{2} \frac{\ldriv}{\vturb \sqrt{\delta}}\ ,
\end{equation}
where $\delta$ is the scaling between the driving length scale and the size of the
largest self-gravitating structures, such as $\delta = \ldriv/l_{\rm cl}$. This leads to a star formation rate per volume of 
\begin{equation} \label{starformation}
    \rhosfr = \frac{4\sqrt{\delta}}{\sqrt{3}} \phiv \rho \frac{\vturb}{\ldriv}
\end{equation}
and $\dot{\Sigma}_*=\rhosfr \, l_{\rm driv}$.
This recipe is close to the prescription suggested by Krumholz et al. (2012)
\begin{equation}
\label{eq:krumholz}
\dot{\Sigma}_*=f_{\rm H_2}\,\epsilon_{\rm ff} \frac{\Sigma}{t_{\rm ff}}\ ,
\end{equation}
where $\epsilon_{\rm ff}$ is the star formation efficiency per free-fall time. 
The relevant size scale for the density entering $t_{\rm ff}$ is that corresponding to the outer scale of the turbulence 
that regulates the SFR, which corresponds to $l_{\rm driv}$ in our model.

Turbulent viscosity redistributes the angular momentum within the disk.
Assuming a continuous and non-zero external gas mass accretion $\dot{\Sigma}_{\rm ext}$, the simplified time evolution of the disk surface density is given by:
\begin{equation}
    \frac{\partial \Sigma}{\partial t} \sim \frac{\nu \Sigma}{R^2} - \dot{\Sigma}_\star + \dot{\Sigma}_{\rm ext}\,.
\end{equation} 
The mass accretion rate within the disk is:
\begin{equation}
    \dot{M} = - 2\pi R \Sigma v_{\rm rot} = \frac{1}{v_{\rm rot}} \frac{\partial}{\partial R}\left( 2\pi \Sigma R^3 \frac{d\Omega}{dR}\right) \,.
\end{equation}
With the approximation $\partial/\partial R \sim R$ and $v_{\rm rot} = \Omega R$, one obtains:
\begin{equation}
\label{massacc}
    \nu \Sigma = - \frac{\dot{M}}{2 \pi R}\, ,
\end{equation}
where the viscosity of the gas $\nu$ is defined as:
\begin{equation}
    \nu = \sqrt{3} \vturb \ldriv\ .
\end{equation}

In addition, the model assumes the radial profiles of the Toomre $Q$ parameter derived by Vollmer \& Leroy (2011) with 
\begin{equation}
\label{eq:toomreq}
Q=\frac{\sigmaturb \Omega}{\pi G \Sigma}\ .
\end{equation}
The Toomre $Q$ parameter is used as a measure of the gas content of the disk, with $Q = 1$ for the maximum disk gas mass.

Our large-scale analytical model of a turbulent, star-forming galactic disks is made of Eq.~\ref{pressure}, Eq.~\ref{eq:sfr2}
together with Eq.~\ref{starformation}, Eq.~\ref{massacc}, and Eq.~\ref{eq:toomreq}.

\section{The galaxy samples}

\begin{table*}
\begin{center}
\caption{Local spiral galaxies.\label{tab:gleroy}}
\begin{tabular}{lccccccccc}
\hline
Galaxy & $v_{\rm max}$ & $l_{\rm flat}^{\rm (c)}$ & $l_*$ & $M_*$ & $\dot{M_*}$ & L$_{\rm TIR}$  & $Q^{\rm (b)}$ & $\dot{M}^{\rm (a)}$ & $M_{\rm gas}^{\rm (a)}$ \\
 & (km\,s$^{-1}$) & (kpc) & (kpc) & ($10^{10}$~M$_{\odot}$) & (M$_{\odot}$yr$^{-1}$) & ($10^{10}$~L$_{\odot}$)  & & (M$_{\odot}$yr$^{-1}$) & ($10^{9}$~M$_{\odot}$) \\
\hline
   NGC628 & 217 &  0.8 &  2.2 & 1.26 & 0.81 &  0.8 &   3.0 &  0.2 &  5.4 \\
   NGC3198 & 150 &  2.7 &  3.2 & 1.26 & 0.93 &  1.0 &    2.0 &  0.3 &  8.3 \\
   NGC3184 & 210 &  2.7 &  2.4 & 2.00 & 0.90 &  1.0 &    2.5 &  0.1 &  5.6 \\
   NGC4736 & 156 &  0.2 &  1.1 & 2.00 & 0.48 &  0.6 &    5.0 &  0.1 &  1.2 \\
   NGC3351 & 196 &  0.6 &  2.2 & 2.51 & 0.94 &  0.8 &    6.0 &  0.4 &  4.0 \\
   NGC6946 & 186 &  1.3 &  2.5 & 3.16 & 3.24 &  3.2 &    2.0 &  0.4 &  9.7 \\
   NGC3627 & 192 &  1.2 &  2.7 & 3.98 & 2.22 &  2.5 &    2.0 &  0.3 &  3.3 \\
   NGC5194 & 219 &  0.8 &  2.7 & 3.98 & 3.12 &  0.0 &    2.0 &  0.3 & 11.4 \\
   NGC3521 & 227 &  1.3 &  2.9 & 5.01 & 2.10 &  3.2 &    2.0 &  0.1 &  9.4 \\
   NGC2841 & 302 &  0.6 &  4.0 & 6.31 & 0.74 &  1.3 &    8.0 &  0.3 &  8.0 \\
   NGC5055 & 192 &  0.6 &  3.2 & 6.31 & 2.12 &  2.0 &    3.0 &  0.3 &  8.8 \\
   NGC7331 & 244 &  1.2 &  3.2 & 7.94 & 3.00 &  5.0 &   3.0 &  0.4 & 11.6 \\
\hline
\end{tabular}
\begin{tablenotes}
      \item $^{\rm (a)}$ calculated quantities; the mean CO(1-0)--H$_2$ conversion factor is $\alpha_{\rm CO}=4.7 \pm 1.8$~M$_{\odot}$(K~km\,s$^{-1}$pc$^2$)$^{-1}$ (Vollmer et al. 2017).
        \item$^{\rm (b)}$ assumed quantities; all other columns are input quantities from Leroy et al. (2008).
          \item $^{\rm (c)}$ A rotation curve of the form $v_{\rm rot}=v_{\rm max} (1-\exp(-R/l_{\rm flat}))$ was assumed.
    \end{tablenotes}
\end{center}
\end{table*}

\begin{table*}
\begin{center}
\caption{Ultraluminous infrared galaxies.\label{tab:gulirg}}
\begin{tabular}{lccccccccc}
\hline
Galaxy Name & $v_{\rm max}$ & $l_{\rm flat}^{\rm (b,f)}$ & $l_*$ & $M_*$ & $\dot{M_*}^{\rm (e)}$  & log(L$_{\rm TIR})^{\rm (d)}$ & $Q^{\rm (b)}$ & $\dot{M}^{\rm (a)}$ & $M_{\rm gas}^{\rm (a)}$ \\
 & (km\,s$^{-1}$) & (kpc) & (kpc) & ($10^{10}$~M$_{\odot}$) & (M$_{\odot}$yr$^{-1}$) & L$_{\odot}$) &  & (M$_{\odot}$yr$^{-1}$) & ($10^{9}$~M$_{\odot}$) \\
\hline
IRAS17208-0014 & 260 & 0.02 &  0.5 &  0.8 &  435 &  12.39  &    1.2 &  313.3 & 14.8 \\
        Mrk231 & 345 & 0.02 &  0.4 &  1.3 &  595 & 12.50  &    1.5 &  499.8 & 16.5 \\
       Arp220D & 330 & 0.02 &  0.4 &  1.2 &   52 & 11.49   &    2.5 &   15.8 &  3.8 \\
        Mrk273 & 280 & 0.02 &  0.4 &  0.9 &  253 & 12.21  &    1.5 &  182.4 &  8.4 \\
IRAS23365+3604 & 260 & 0.02 &  0.6 &  1.0 &  258 & 12.13  &    1.5 &  242.9 & 14.4 \\
       VIIZw31 & 290 & 0.02 &  1.1 &  2.2 &  164 & 12.00   &    1.5 &   28.0 & 13.7 \\
        Arp193 & 230 & 0.02 &  0.7 &  0.9 &   81 & 11.73   &    1.5 &   18.8 &  6.5 \\
       Arp220W & 300 & 0.01 &  0.1 &  1.2 &   79 & 11.66   &    2.0 &   34.8 &  1.1 \\
       Arp220E & 350 & 0.01 &  0.1 &  1.9 &   52 & 11.49   &    2.8 &   22.1 &  1.3 \\
\hline
\end{tabular}
\begin{tablenotes}
\item $^{\rm (a)}$ calculated quantities; the mean CO(1-0)--H$_2$ conversion factor is $\alpha_{\rm CO}=1.7 \pm 0.4$~M$_{\odot}$(K~km\,s$^{-1}$pc$^2$)$^{-1}$ (Vollmer et al. 2017).
  \item $^{\rm (b)}$ assumed quantities; all other columns are input quantities from Downes \& Solomon (1998).
    \item $^{\rm (c)}$ Arp220D, Arp220W, and Arp220E refer to the Disk, Western, and Eastern components, respectively.
    \item $^{\rm (d)}$ Garcia-Carpio et al. (2008)
      \item $^{\rm (e)}$ A conversion factor of $\dot{M}_*/L_{\rm TIR}=1.7\ 10^{-10}$~M$_{\odot}$yr$^{-1}$ was assumed.
      \item $^{\rm (f)}$ A rotation curve of the form $v_{\rm rot}=v_{\rm max} (1-exp(-R/l_{\rm flat}))$ was assumed.
    \end{tablenotes}
\end{center}
\end{table*}

\begin{table*}
\begin{center}
\caption{Submillimeter galaxies.\label{tab:gbzk}}
\begin{tabular}{lccccccccc}
\hline
Galaxy Name & $v_{\rm max}$ & $l_{\rm flat}^{\rm (b,d)}$ & $l_*$ & $M_*^{\rm (c)}$ & $\dot{M_*}^{\rm (e)}$ & log(L$_{\rm TIR}^{\rm (f)}$) &  $\dot{M}^{\rm (a)}$ & $M_{\rm gas}^{\rm (a)}$ \\
 & (km\,s$^{-1}$) & (kpc) & (kpc) & ($10^{10}$~M$_{\odot}$) & (M$_{\odot}$yr$^{-1}$) & (L$_{\odot}$) & (M$_{\odot}$yr$^{-1}$) & ($10^{9}$~M$_{\odot}$) \\
\hline
SMM J02399-013 & 590 & 0.10 &  3.5 & 10.0 & 2294 &  13.1;--;13.4;13.0;--  &     1927.0 & 318.5 \\
SMM J09431+470 & 295 & 0.10 &  0.9 & 10.0 & 1746 &  12.9;--;13.0;--;--  &     1117.4 &  39.1 \\
SMM J105141+57 & 457 & 0.10 &  2.1 & 10.0 & 1296 &  12.8;--;--;--;13.1  &      423.4 &  98.2 \\
SMM J123549+62 & 442 & 0.10 &  0.6 & 24.0 & 1794 &  13.0;--;--;--;--  &       71.8 &  17.1 \\
SMM J123634+62 & 343 & 0.10 &  2.8 & 10.0 &  930 &  12.7;--;--;12.7;--  &      737.8 & 117.0 \\
SMM J123707+62 & 317 & 0.10 &  1.9 & 24.0 & 1016 &  12.7;--;--;12.8;--  &      135.5 &  45.6 \\
SMM J131201+42 & 430 & 0.10 &  2.1 & 10.0 & 1340 &  12.8;12.9;--;--;--  &      589.6 &  99.5 \\
SMM J131232+42 & 346 & 0.10 &  1.4 & 10.0 & 1016 &  12.7;--;--;--;--  &      257.4 &  41.5 \\
SMM J163650+40 & 523 & 0.10 &  1.6 & 46.0 & 1772 &  12.9;12.7;--;--;--  &       59.1 &  50.8 \\
SMM J163658+41 & 590 & 0.10 &  0.5 & 52.0 & 2248 &  13.1;12.9;--;--;--  &       28.5 &  16.0 \\
\hline
\end{tabular}
\begin{tablenotes}
      \item $^{\rm (a)}$ calculated quantities; the mean CO(1-0)--H$_2$ conversion factor is $\alpha_{\rm CO}=1.4 \pm 0.7$~M$_{\odot}$(K~km\,s$^{-1}$pc$^2$)$^{-1}$ (Vollmer et al. 2017).
        \item $^{\rm (b)}$ assumed quantities; all other columns are input quantities from Genzel et al. (2010).
        \item $^{\rm (c)}$ we assumed $M_*=10^{11}$~M$_{\odot}$ for galaxies whose mass is not given in Genzel et al. (2010).
            \item $^{\rm (d)}$ A rotation curve of the form $v_{\rm rot}=v_{\rm max} (1-exp(-R/l_{\rm flat}))$ was assumed.
              \item $^{\rm (e)}$ Twice the SFRs from Genzel et al. (2010) who used $\dot{M}_*/L_{\rm TIR}=1.0\ 10^{-10}$~M$_{\odot}$yr$^{-1}$.
                \item $^{\rm (f)}$ Genzel et al. (2010), Kovacs et al. (2006), Valiante et al. (2009), Magnelli et al. (2012), and Chapman et al. (2010).
    \end{tablenotes}
\end{center}
\end{table*}

\begin{table*}
\begin{center}
\caption{High-z star-forming disk galaxies.\label{tab:gphibbs}}
\begin{tabular}{lcccccccccc}
\hline
Galaxy Name & $v_{\rm max}^{\rm (c)}$ & $l_{\rm flat}^{\rm (b,d)}$ & $l_*$ & $M_*$ & $\dot{M_*}$ & log(L$_{\rm TIR}$)$^{\rm (e)}$   & $\dot{M}^{\rm (a)}$ & $M_{\rm gas}^{\rm (a)}$ \\
 & (km\,s$^{-1}$) & (kpc) & (kpc) & ($10^{10}$~M$_{\odot}$) & (M$_{\odot}$yr$^{-1}$) & (L$_{\odot}$) &  (M$_{\odot}$yr$^{-1}$) & ($10^{9}$~M$_{\odot}$) \\
\hline
   EGS12004280 & 230 & 0.10 &  4.7 &  4.1 &  100 & 11.59  &   30.5 &   47.5 \\
   EGS12004754 & 215 & 0.10 &  6.5 &  9.3 &   53 & 11.48  &      5.8 &   39.2 \\
   EGS12007881 & 232 & 0.10 &  5.7 &  5.2 &   94 & 11.74  &     23.5 &   53.8 \\
   EGS12015684 & 233 & 0.10 &  4.0 &  4.6 &  113 & 12.24  &     30.5 &   41.3 \\
   EGS12023832 & 215 & 0.10 &  4.7 &  5.9 &  115 & 11.78  &     36.8 &   47.4 \\
   EGS12024462 & 253 & 0.10 &  8.6 &  6.0 &   78 & 11.99  &     12.9 &   73.5 \\
   EGS12024866 & 221 & 0.10 &  4.6 &  2.5 &   31 & 11.42  &      4.0 &   24.5 \\
   EGS13003805 & 387 & 0.10 &  5.7 & 17.0 &  200 & 12.11  &      9.7 &   70.9 \\
   EGS13004661 & 171 & 0.10 &  5.0 &  3.0 &   60 & 11.91  &     36.3 &   39.4 \\
   EGS13004684 & 295 & 0.10 &  5.0 & 11.0 &   42 & 11.30  &      1.4 &   27.7 \\
   EGS13011148 & 260 & 0.10 &  5.2 & 11.0 &   52 & 11.57  &      2.9 &   31.3 \\
   EGS13011155 & 296 & 0.10 &  7.8 & 12.0 &  201 & 11.77  &     35.2 &  106.5 \\
   EGS13011166 & 363 & 0.10 &  6.5 & 12.0 &  373 & 12.36  &     69.0 &  133.0 \\
   EGS13017614 & 346 & 0.10 &  4.5 & 13.0 &   88 & 11.73  &      2.9 &   36.0 \\
   EGS13017707 & 324 & 0.10 &  3.6 &  7.4 &  351 & 12.25  &     93.0 &   72.2 \\
   EGS13017843 & 227 & 0.10 &  4.2 &  4.0 &   35 & 11.36  &      3.3 &   22.2 \\
   EGS13017973 & 155 & 0.10 &  7.2 &  4.4 &   55 & 11.44  &    36.6 &   52.0 \\
   EGS13018632 & 319 & 0.10 &  1.9 &  5.2 &   82 & 12.01  &     3.9 &   15.1 \\
   EGS13019114 & 327 & 0.10 &  7.2 &  6.6 &   47 & 11.77  &      1.9 &   45.9 \\
   EGS13019128 & 194 & 0.10 &  5.2 &  4.4 &   87 & 11.67  &     39.6 &   48.1 \\
   EGS13026117 & 436 & 0.10 &  3.2 & 13.0 &  113 & 12.27  &      2.3 &   29.9 \\
   EGS13033624 & 301 & 0.10 &  5.3 &  8.9 &  148 & -  &      17.0 &   59.6 \\
   EGS13033731 & 350 & 0.10 &  5.5 &  2.8 &   28 & 11.42   &    0.8 &   29.2 \\
   EGS13034339 & 299 & 0.10 &  3.0 &  6.6 &   86 & 12.28   &     5.4 &   24.4 \\
   EGS13034541 & 330 & 0.10 &  8.0 &  9.3 &  183 & 11.97  &     24.2 &  107.8 \\
   EGS13034542 & 195 & 0.10 &  4.0 &  5.2 &   61 & 11.58  &     11.9 &   26.7 \\
   EGS13035123 & 219 & 0.10 & 11.2 & 15.0 &   87 & 12.04  &     14.8 &   89.3 \\
   EGS13042293 & 167 & 0.10 &  5.2 &  3.9 &   55 & 11.79  &    24.2 &   36.0 \\
      zC406690 & 224 & 0.10 &  6.3 &  4.0 &  480 & -  &     304.8 &  158.1 \\
    Q1623BX599 & 376 & 0.10 &  1.7 &  5.7 &  131 & -  &       5.6 &   17.5 \\
    Q1700BX691 & 260 & 0.10 &  3.9 &  7.6 &   50 & -  &       2.9 &   23.3 \\
    Q2343BX610 & 402 & 0.10 &  4.6 & 10.0 &  212 & -  &      12.7 &   63.2 \\
    Q2343BX442 & 309 & 0.10 &  4.3 & 12.0 &  145 & -  &     10.5 &   44.1 \\
     Q2343MD59 & 371 & 0.10 &  2.8 &  7.6 &   26 & -  &       0.3 &   13.2 \\
  Q2346BX482se & 285 & 0.10 &  2.4 &  0.6 &   34 & -  &       3.9 &   15.8 \\
       BzK4171 & 261 & 0.10 &  4.5 &  4.0 &  101 & -  &      20.2 &   45.4 \\
     BzK210000 & 292 & 0.10 &  4.7 &  7.8 &  231 & -  &      54.3 &   72.2 \\
      BzK16000 & 258 & 0.10 &  4.0 &  4.3 &   82 & -  &      11.9 &   34.4 \\
      BzK17999 & 238 & 0.10 &  4.7 &  3.9 &  450 & -  &     351.0 &  122.8 \\
      BzK12591 & 361 & 0.10 &  4.5 & 11.0 &  267 & -  &      26.7 &   69.4 \\
      BzK25536 & 254 & 0.10 &  3.0 &  3.3 &   62 & -  &       7.4 &   22.2 \\
    J2135-0102 & 381 & 0.10 &  1.5 &  1.7 &  230 & -  &      46.0 &   27.8 \\
\hline
\end{tabular}
\begin{tablenotes}
      \item $^{\rm (a)}$ calculated quantities; the mean CO(1-0)--H$_2$ conversion factor is $\alpha_{\rm CO}=2.6 \pm 0.9$~M$_{\odot}$(K~km\,s$^{-1}$pc$^2$)$^{-1}$ (Vollmer et al. 2017).
        \item $^{\rm (b)}$ assumed quantities; all other columns are input quantities from Tacconi et al. (2013).
        \item $^{\rm (c)}$ if $v_{\rm rot} < \sqrt{(M_{\rm gas}+M_*)\,G/(2\,l_*)}$ the assumed rotation velocity is $v_{\rm rot}=\sqrt{(M_{\rm gas}+M_*)\,G/(2\,l_*)}$.
          \item $^{\rm (d)}$ A rotation curve of the form $v_{\rm rot}=v_{\rm max} (1-exp(-R/l_{\rm flat}))$ was assumed.
          \item $^{\rm (e)}$ From Barro et al. (2011). The mean conversion factor is $\dot{M}_*/L_{\rm TIR}=1.4\ 10^{-10}$~M$_{\odot}$yr$^{-1}$.
    \end{tablenotes}
\end{center}
\end{table*}

\section{SEDs and radio SEDs}

\begin{figure*}
  \centering
  \resizebox{15cm}{!}{\includegraphics{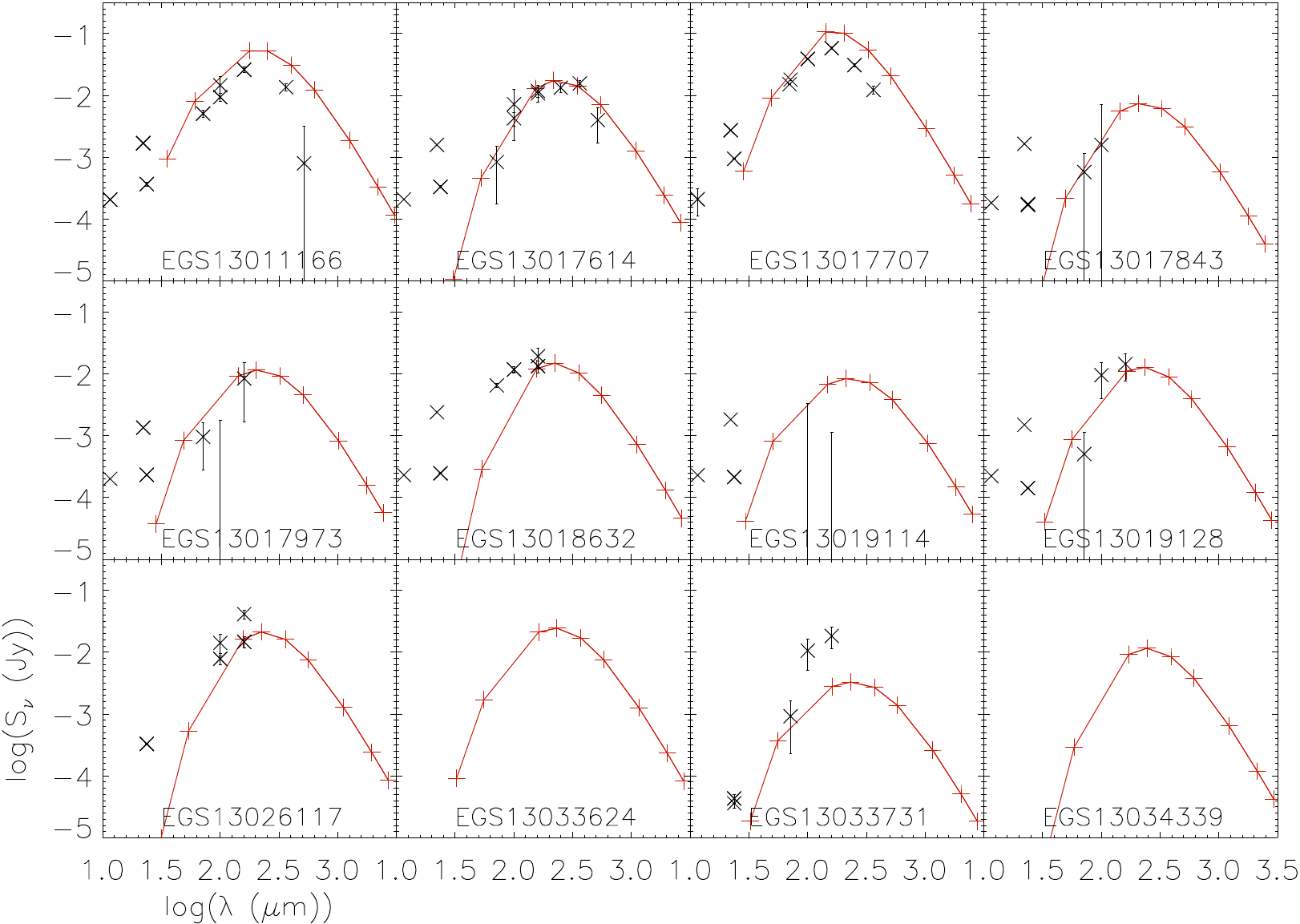}}
  \resizebox{15cm}{!}{\includegraphics{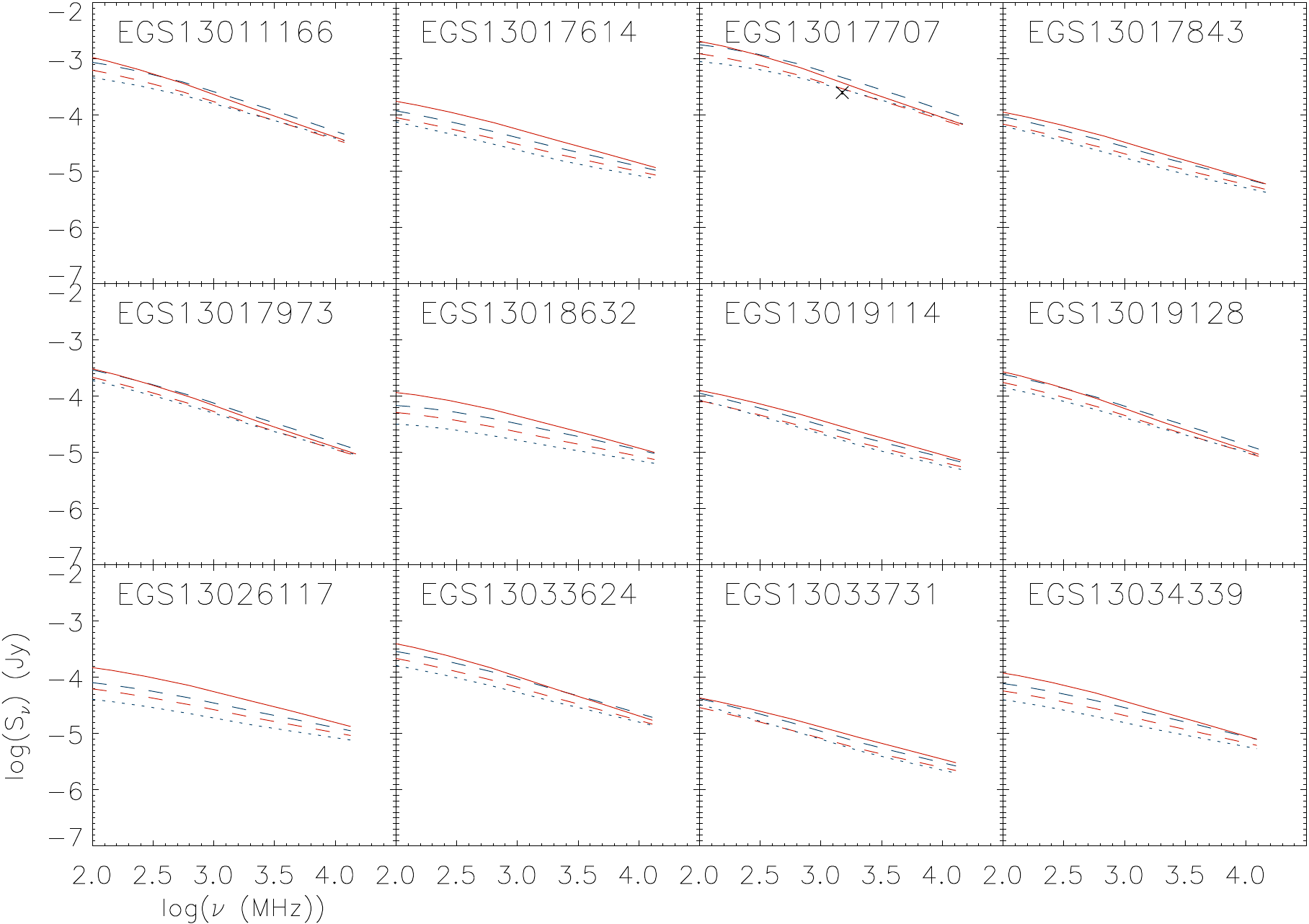}}
  \caption{Same as Fig.~\ref{fig:IRspectra_phibbs}.
  \label{fig:IRspectra_phibbs2}}
\end{figure*}

\begin{figure*}
  \centering
  \resizebox{15cm}{!}{\includegraphics{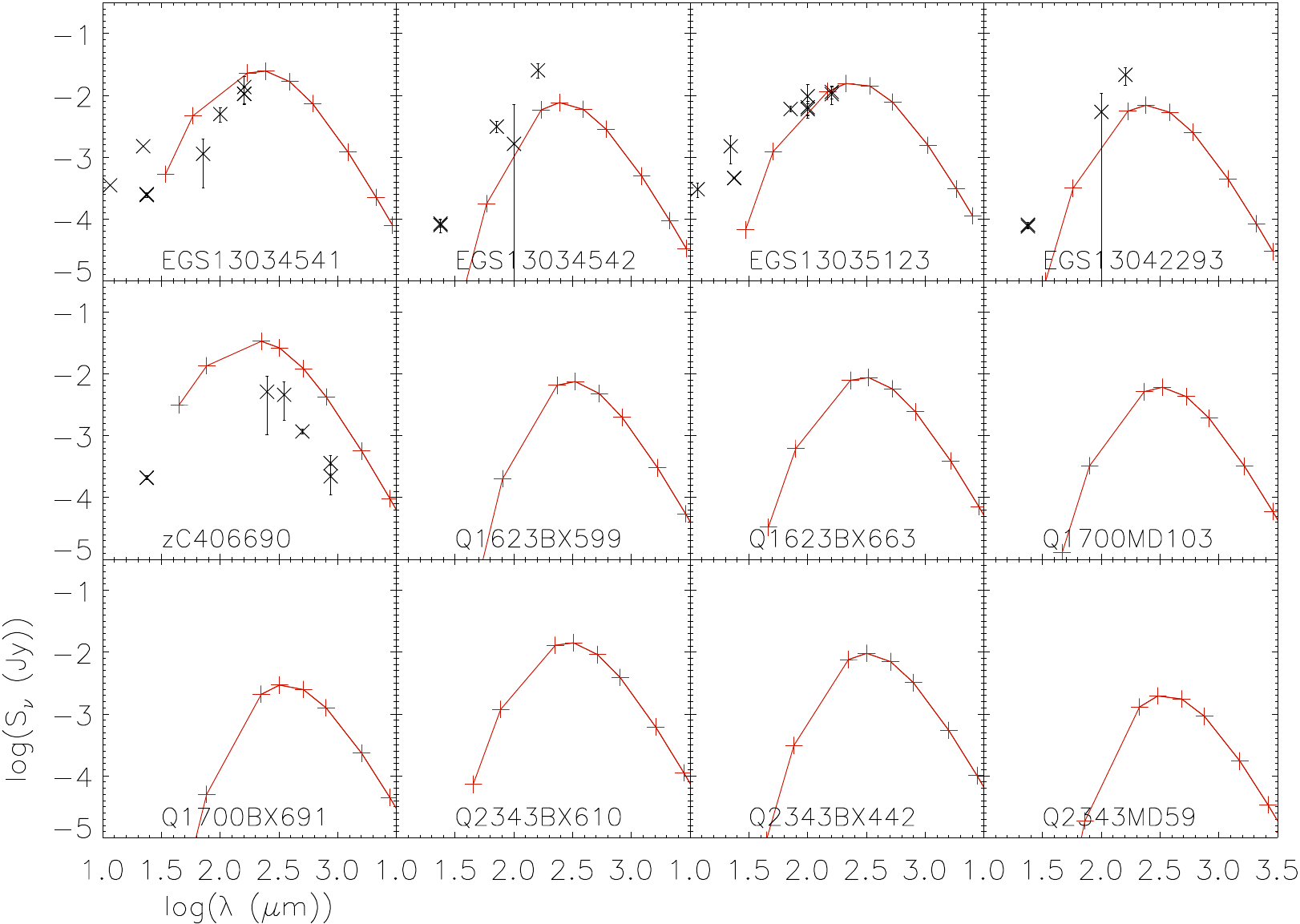}}
  \resizebox{15cm}{!}{\includegraphics{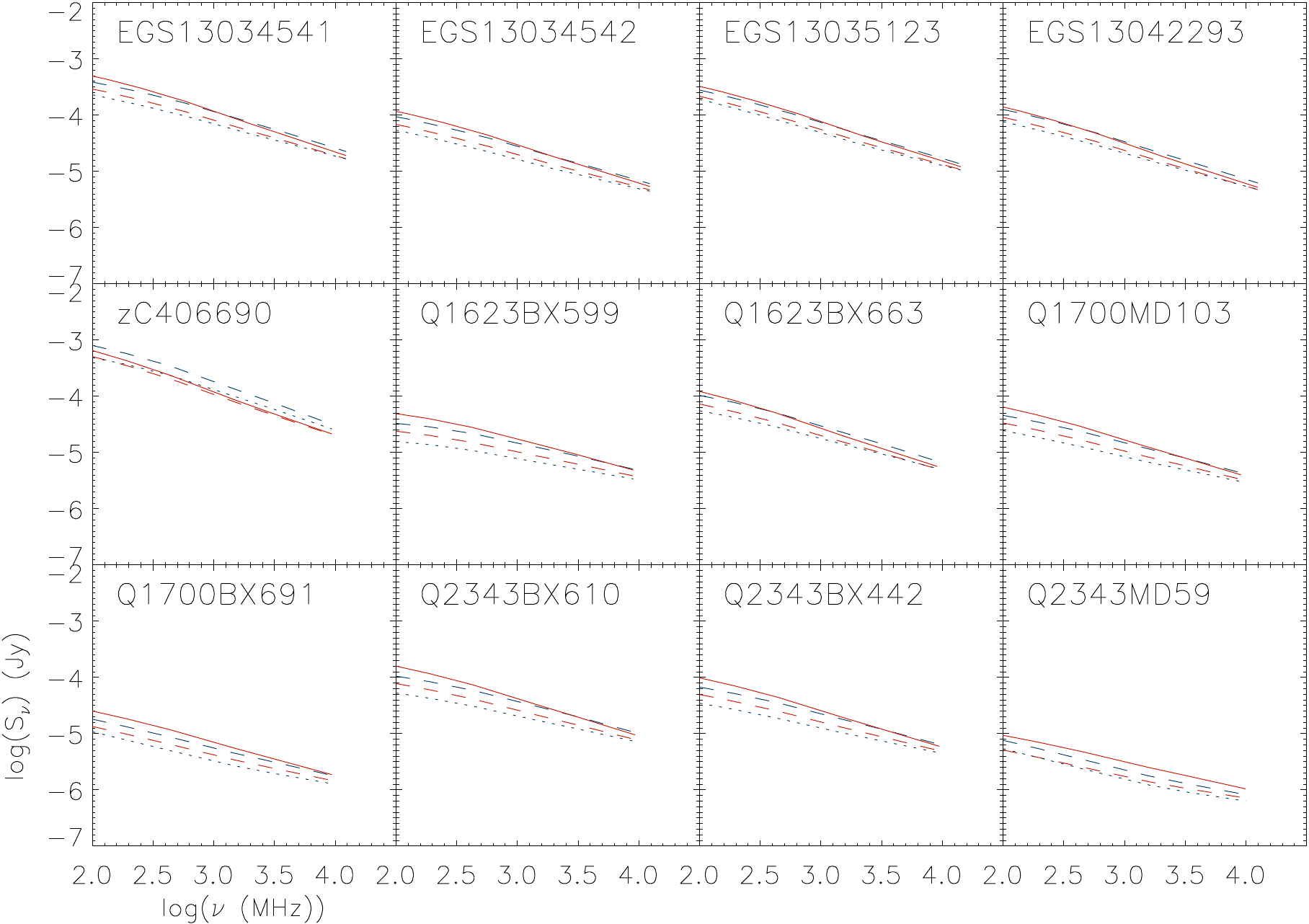}}
  \caption{Same as Fig.~\ref{fig:IRspectra_phibbs}.
  \label{fig:IRspectra_phibbs3}}
\end{figure*}

\begin{figure*}
  \centering
  \resizebox{15cm}{!}{\includegraphics{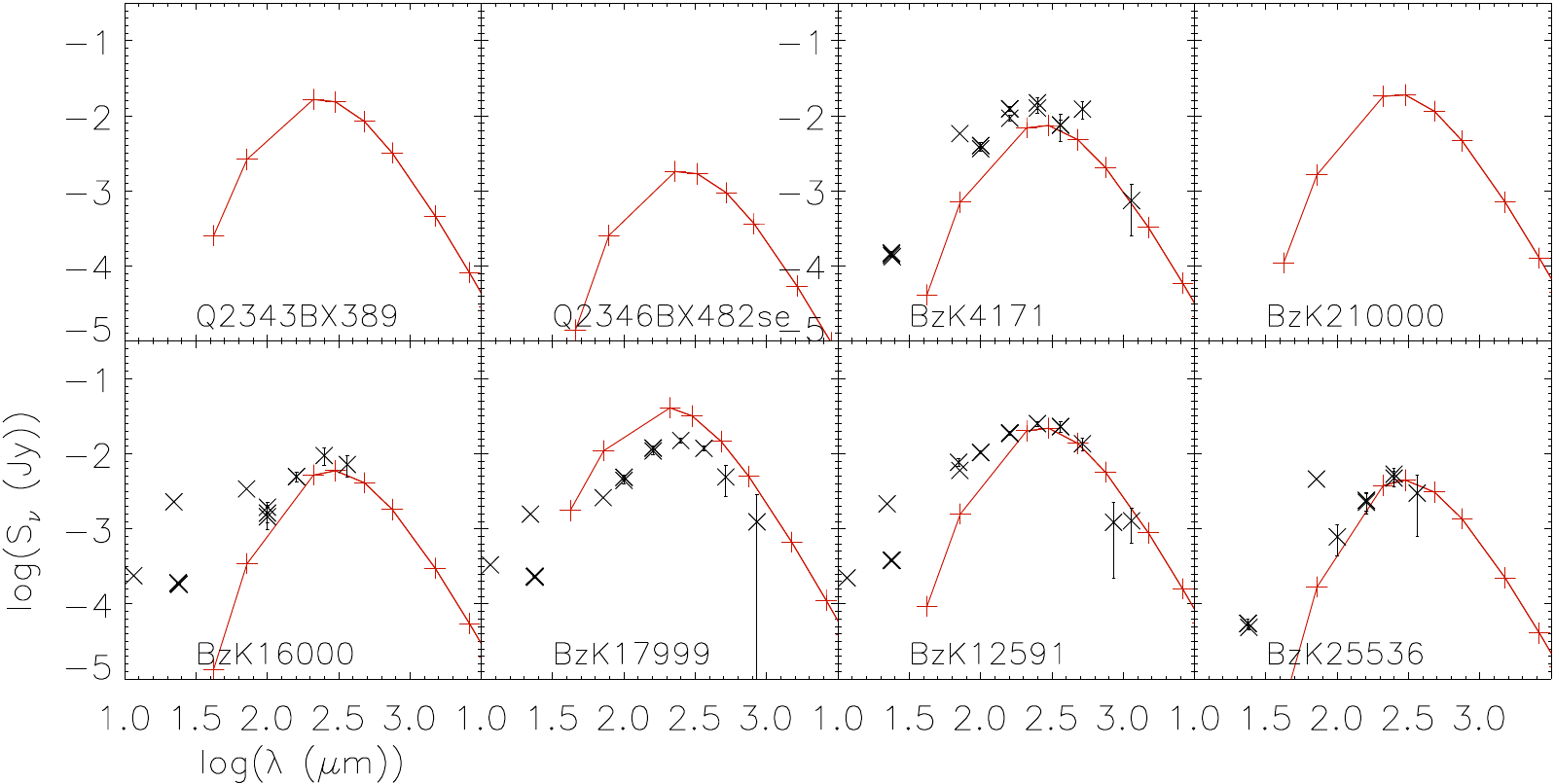}}
  \resizebox{15cm}{!}{\includegraphics{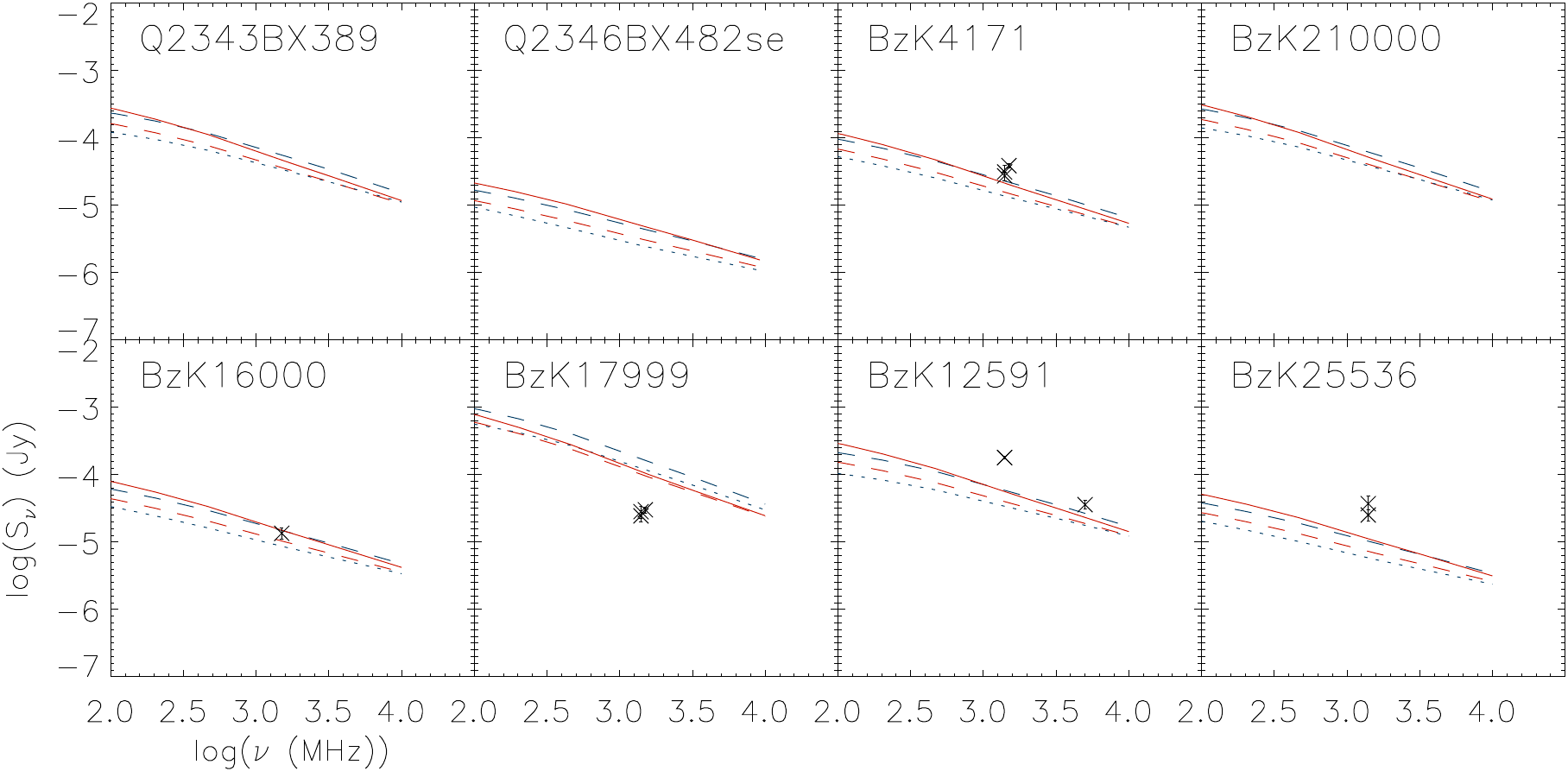}}
  \caption{Same as Fig.~\ref{fig:IRspectra_phibbs}.
  \label{fig:IRspectra_phibbs4}}
\end{figure*}

\section{The influence of different CRe energy loss times on the TIR-radio correlation}

\begin{table}[!ht]
      \caption{TIR-1.4~GHz correlations.}
         \label{tab:modeltimes}
      \[
         \begin{tabular}{lccc}
           \hline
            & involved timscale & slope & offset at $L=10^{10}$~L$_{\odot}$ \\
           \hline
           Molnar et al. (2001) &  & $1.07 \pm 0.01$ & $21.45 \pm 0.01$ \\
           fiducial model  & all & $1.09 \pm 0.05$ & $21.31 \pm 0.09$ \\
           model  & diff & $1.09 \pm 0.05$ & $21.71 \pm 0.09$ \\
           model  & wind & $0.91 \pm 0.04$ & $22.10 \pm 0.09$ \\
           model  & brems & $1.00 \pm 0.04$ & $21.72 \pm 0.09$ \\
           model  & IC & $1.03 \pm 0.05$ & $21.81 \pm 0.09$ \\
           model  & ion & $0.89 \pm 0.04$ & $22.00 \pm 0.09$ \\
           calorimeter model  & none & $0.91 \pm 0.04$ & $22.14 \pm 0.09$ \\
           \hline
         \end{tabular}
      \]
\end{table}

\begin{figure}
  \centering
  \resizebox{\hsize}{!}{\includegraphics{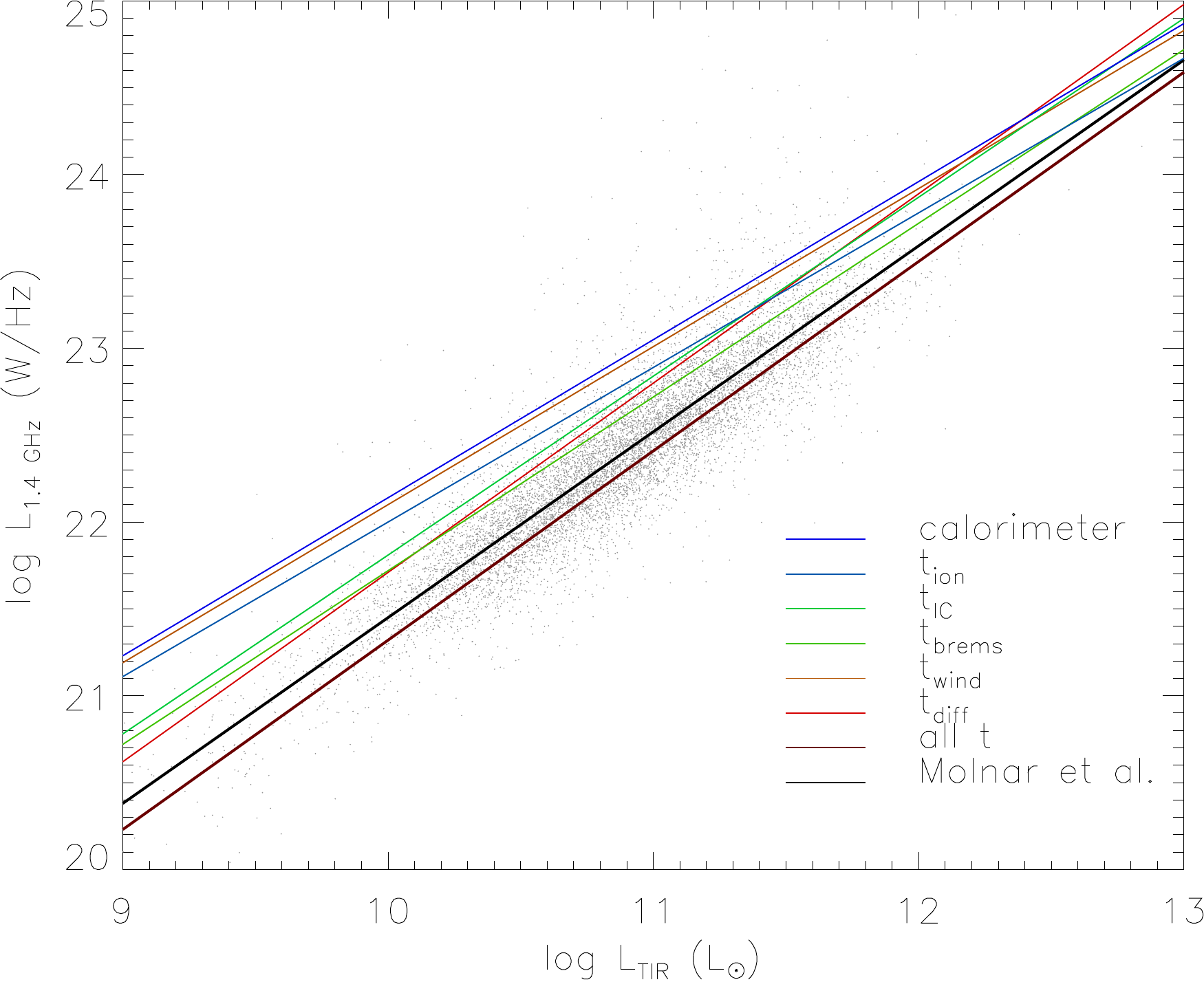}}
  \caption{TIR-1.4~GHz correlation. Grey points: Molnar et al. (2021). Colored lines: models involving different CRe energy loss timescales
    (see Table~\ref{tab:modeltimes}).
  \label{fig:modeltimes}}
\end{figure}

\end{appendix}

\end{document}